\documentclass[12pt,dvipsnames]{article}
\usepackage[top=80pt,bottom=85pt,left=85pt,right=85pt]{geometry}

\pdfoutput=1
\usepackage[utf8]{inputenc}
\usepackage[T1]{fontenc} 
\usepackage[english]{babel} 
\usepackage{braket}
\usepackage[dvipsnames,table]{xcolor}
\usepackage{physics}
\usepackage{tabularx} 
\usepackage{ragged2e}
\usepackage{rotating}
\usepackage{makecell}
\usepackage{multirow} 
\usepackage{comment} 
\usepackage{amsmath}
\usepackage{cite}
\usepackage[all]{xy}

\usepackage{tikz}
\usetikzlibrary{math} 
\usetikzlibrary{fit,backgrounds}
\pgfdeclarelayer{bg}
\pgfsetlayers{bg,main}
\usetikzlibrary{3d} 
\usepackage{tikz-3dplot}
\usepackage{tikz-cd} 
\usepackage{microtype}
\usetikzlibrary{arrows.meta,positioning} 
\usepackage{amsthm}
\usepackage{mathrsfs} 
\usepackage[strict]{changepage}
\newcommand\scalemath[2]{\scalebox{#1}{\mbox{\ensuremath{\displaystyle #2}}}}

\tikzcdset{
  cells={font=\everymath\expandafter{\the\everymath\displaystyle}},
}

\usepackage{amssymb} 
\usepackage{subcaption} 
\DeclareCaptionFormat{custom}

\usepackage[pdftex,colorlinks=true]{hyperref} 
\hypersetup{urlcolor=MidnightBlue, citecolor=red, linkcolor=MidnightBlue}

\usepackage{float} 
\usepackage[textsize=tiny]{todonotes}

\usepackage[capitalise]{cleveref}

\newtheorem{theorem}{Theorem}[section]
\newtheorem{proposition}[theorem]{Proposition}

\numberwithin{equation}{section}

\definecolor{cambridgeblue}{rgb}{0.64, 0.76, 0.68}
\definecolor{caribbeangreen}{rgb}{0.0, 0.8, 0.6}
\definecolor{celadon}{rgb}{0.67, 0.88, 0.69}
\definecolor{champagne}{rgb}{0.97, 0.91, 0.81}
\definecolor{cream}{rgb}{1.0, 0.99, 0.82}
\definecolor{cyan(process)}{rgb}{0.0, 0.72, 0.92}
\definecolor{brilliantlavender}{rgb}{0.96, 0.73, 1.0}
\definecolor{candypink}{rgb}{0.89, 0.44, 0.48}

\usepackage{pdflscape}
\usepackage{paralist}

\usetikzlibrary{positioning}
\usetikzlibrary{arrows}
\usetikzlibrary{decorations.pathreplacing}
\usetikzlibrary{shapes.geometric}
\usetikzlibrary{calc}
\usetikzlibrary{decorations.pathmorphing}
\usetikzlibrary{shapes.misc}

\tikzset{gaugeSU/.style={inner sep=1.7mm,draw=none,fill=yellow,minimum size=2mm,circle, draw}}
\tikzset{flavourSU/.style={draw=none,minimum size=2mm,fill=white, regular polygon,regular polygon sides=4,draw}}
\tikzset{flavour/.style={draw=none,minimum size=0.3mm,fill=white, regular polygon,regular polygon sides=4,draw}}
\tikzset{gaugeBig/.style={inner sep=1mm,draw=none,fill=white,minimum size=2mm,circle, draw}}
\tikzset{bd/.style={circle, draw=black, inner sep=0pt, fill=black, minimum size=2mm}}
\tikzset{wd/.style={circle, draw=black, inner sep=0pt, fill=white, minimum size=2mm}}
\tikzset{Dynkin/.style={circle, draw=black, inner sep=0pt, fill=white, minimum size=2mm}}
\tikzstyle{ligne}=[draw, very thick] 
\tikzstyle{gridline}=[draw, gray] 
\tikzset{gauge/.style={circle, draw,inner sep=2.5pt,fill=white}}
\tikzset{gaugeo/.style={circle, draw,inner sep=2.5pt,fill=orange}}
\tikzset{gaugec/.style={circle, draw,inner sep=2.5pt,fill=cyan}}
\tikzset{gauger/.style={circle, draw,inner sep=2.5pt,fill=red}}
\tikzset{gaugeb/.style={circle, draw,inner sep=2.5pt,fill=blue}}
\tikzset{gaugeg/.style={circle, draw,inner sep=2.5pt,fill=green}}
\tikzset{gaugem/.style={circle, draw,inner sep=2.5pt,fill=magenta}}
\tikzset{gaugey/.style={circle, draw,inner sep=2.5pt,fill=yellow}}
\tikzset{hasse/.style={circle, fill,inner sep=2pt}}
\tikzset{shrinky/.style={circle, fill,inner sep=1pt}}
\tikzset{sized/.style={circle, draw, inner sep=1.5pt}}
\tikzset{seven/.style={circle, draw,inner sep=3pt}}

\tikzset{dotto/.style={circle, orange, draw,inner sep=1.5pt,fill=orange}}
\tikzset{dottp/.style={circle, purple, draw,inner sep=1.5pt,fill=purple}}
\tikzset{dottc/.style={circle, cyan, draw,inner sep=1.5pt,fill=cyan}}
\tikzset{dottr/.style={circle, red, draw,inner sep=1.5pt,fill=red}}
\tikzset{dottb/.style={circle, blue, draw,inner sep=1.5pt,fill=blue}}
\tikzset{dottg/.style={circle, green, draw,inner sep=1.5pt,fill=green}}
\tikzset{dottm/.style={circle, magenta, draw,inner sep=1.5pt,fill=magenta}}
\newcommand{\convexpath}[2]{
  [   
  create hullcoords/.code={
    \global\edef\namelist{#1}
    \foreach [count=\counter] \nodename in \namelist {
      \global\edef\numberofnodes{\counter}
      \coordinate (hullcoord\counter) at (\nodename);
    }
    \coordinate (hullcoord0) at (hullcoord\numberofnodes);
    \pgfmathtruncatemacro\lastnumber{\numberofnodes+1}
    \coordinate (hullcoord\lastnumber) at (hullcoord1);
  },
  create hullcoords
  ]
  ($(hullcoord1)!#2!-90:(hullcoord0)$)
  \foreach [
  evaluate=\currentnode as \previousnode using \currentnode-1,
  evaluate=\currentnode as \nextnode using \currentnode+1
  ] \currentnode in {1,...,\numberofnodes} {
    let \p1 = ($(hullcoord\currentnode) - (hullcoord\previousnode)$),
    \n1 = {atan2(\y1,\x1) + 90},
    \p2 = ($(hullcoord\nextnode) - (hullcoord\currentnode)$),
    \n2 = {atan2(\y2,\x2) + 90},
    \n{delta} = {Mod(\n2-\n1,360) - 360}
    in 
    {arc [start angle=\n1, delta angle=\n{delta}, radius=#2]}
    -- ($(hullcoord\nextnode)!#2!-90:(hullcoord\currentnode)$) 
  }
}

\usepackage{ifthen}
\usepackage{xstring}
\long\def\AtwoWeight#1#2{
    \begin{bmatrix}
        #1&#2\\
    \end{bmatrix}
}

\long\def\AtwoG#1#2{
        \node[gauge] (g1) at (-2,0) {$#1$};
        \node[gauge] (g2) at (-1,0) {$#2$};
        \draw (g1)--(g2);
}
\long\def\AtwoF#1#2{
        \ifthenelse{#1>0}{\node[flavour] (f1) at (-2,-1) {$#1$}; \draw (f1)--(-2,0);}{}
        \ifthenelse{#2>0}{\node[flavour] (f2) at (-1,-1) {$#2$}; \draw (f2)--(-1,0);}{}
}

\newcommand{\AtwoRgens}[1]{%
    \IfEqCase{#1}{%
        {1}{\raisebox{-.5\height}{\begin{tikzpicture}
        \AtwoF{1}{1}
        \AtwoG{1}{1}
    \end{tikzpicture}}}
    {2}{\raisebox{-.5\height}{\begin{tikzpicture}
        \AtwoF{3}{0}
        \AtwoG{2}{1}
    \end{tikzpicture}}}
    {3}{\raisebox{-.5\height}{\begin{tikzpicture}
      \AtwoF{0}{3}
        \AtwoG{1}{2}
    \end{tikzpicture}}}
}}
\long\def\AthreeWeight#1#2#3{
    \begin{bmatrix}
        #1&#2&#3\\
    \end{bmatrix}
}

\long\def\AthreeG#1#2#3{
        \node[gauge] (g1) at (-2,0) {$#1$};
        \node[gauge] (g2) at (-1,0) {$#2$};
        \node[gauge] (g3) at (0,0) {$#3$};
        \draw (g1)--(g2)--(g3);
}
\long\def\AthreeF#1#2#3{
        \ifthenelse{#1>0}{\node[flavour] (f1) at (-2,-1) {$#1$}; \draw (f1)--(-2,0);}{}
        \ifthenelse{#2>0}{\node[flavour] (f2) at (-1,-1) {$#2$}; \draw (f2)--(-1,0);}{}
        \ifthenelse{#3>0}{\node[flavour] (f3) at (0,-1) {$#3$}; \draw (f3)--(0,0);}{}
}

\newcommand{\AthreeRgens}[1]{%
    \IfEqCase{#1}{%
        {1}{\raisebox{-.5\height}{\begin{tikzpicture}
        \AthreeF{1}{0}{1}
        \AthreeG{1}{1}{1}
    \end{tikzpicture}}}
    {2}{\raisebox{-.5\height}{\begin{tikzpicture}
        \AthreeF{0}{2}{0}
        \AthreeG{1}{2}{1}
    \end{tikzpicture}}}
    {3}{\raisebox{-.5\height}{\begin{tikzpicture}
      \AthreeF{2}{1}{0}
        \AthreeG{2}{2}{1}
    \end{tikzpicture}}}
    {4}{\raisebox{-.5\height}{\begin{tikzpicture}
      \AthreeF{0}{1}{2}
        \AthreeG{1}{2}{2}
    \end{tikzpicture}}}
    {5}{\raisebox{-.5\height}{\begin{tikzpicture}
      \AthreeF{4}{0}{0}
        \AthreeG{3}{2}{1}
    \end{tikzpicture}}}
    {6}{\raisebox{-.5\height}{\begin{tikzpicture}
      \AthreeF{0}{0}{4}
        \AthreeG{1}{2}{3}
    \end{tikzpicture}}}
}}
\long\def\AfourWeight#1#2#3#4{
    \begin{bmatrix}
        #1&#2&#3&#4\\
    \end{bmatrix}
}

\long\def\AfourG#1#2#3#4{
        \node[gauge] (g1) at (-2,0) {$#1$};
        \node[gauge] (g2) at (-1,0) {$#2$};
        \node[gauge] (g3) at (0,0) {$#3$};
        \node[gauge] (g4) at (1,0) {$#4$};
        \draw (g1)--(g2)--(g3)--(g4);
}
\long\def\AfourF#1#2#3#4{
        \ifthenelse{#1>0}{\node[flavour] (f1) at (-2,-1) {$#1$}; \draw (f1)--(-2,0);}{}
        \ifthenelse{#2>0}{\node[flavour] (f2) at (-1,-1) {$#2$}; \draw (f2)--(-1,0);}{}
        \ifthenelse{#3>0}{\node[flavour] (f3) at (0,-1) {$#3$}; \draw (f3)--(0,0);}{}
        \ifthenelse{#4>0}{\node[flavour] (f4) at (1,-1) {$#4$}; \draw (f4)--(1,0);}{}
}

\newcommand{\AfourRgens}[1]{%
    \IfEqCase{#1}{%
    {1}{\raisebox{-.5\height}{\begin{tikzpicture}
        \AfourF{1}{0}{0}{1}
        \AfourG{1}{1}{1}{1}
    \end{tikzpicture}}}
    {2}{\raisebox{-.5\height}{\begin{tikzpicture}
        \AfourF{0}{1}{1}{0}
        \AfourG{1}{2}{2}{1}
    \end{tikzpicture}}}
    {3}{\raisebox{-.5\height}{\begin{tikzpicture}
        \AfourF{2}{0}{1}{0}
        \AfourG{2}{2}{2}{1}
    \end{tikzpicture}}}
    {4}{\raisebox{-.5\height}{\begin{tikzpicture}
        \AfourF{1}{2}{0}{0}
        \AfourG{2}{3}{2}{1}
    \end{tikzpicture}}}
    {5}{\raisebox{-.5\height}{\begin{tikzpicture}
        \AfourF{3}{1}{0}{0}
        \AfourG{3}{3}{2}{1}
    \end{tikzpicture}}}
    {6}{\raisebox{-.5\height}{\begin{tikzpicture}
        \AfourF{5}{0}{0}{0}
        \AfourG{4}{3}{2}{1}
    \end{tikzpicture}}}
    {7}{\raisebox{-.5\height}{\begin{tikzpicture}
        \AfourF{0}{3}{0}{1}
        \AfourG{2}{4}{3}{2}
    \end{tikzpicture}}}
    {8}{\raisebox{-.5\height}{\begin{tikzpicture}
        \AfourF{0}{5}{0}{0}
        \AfourG{3}{6}{4}{2}
    \end{tikzpicture}}}
    {9}{\raisebox{-.5\height}{\begin{tikzpicture}
        \AfourF{0}{1}{0}{2}
        \AfourG{1}{2}{2}{2}
    \end{tikzpicture}}}
    {10}{\raisebox{-.5\height}{\begin{tikzpicture}
        \AfourF{0}{0}{2}{1}
        \AfourG{1}{2}{3}{2}
    \end{tikzpicture}}}
    {11}{\raisebox{-.5\height}{\begin{tikzpicture}
        \AfourF{0}{0}{1}{3}
        \AfourG{1}{2}{3}{3}
    \end{tikzpicture}}}
    {12}{\raisebox{-.5\height}{\begin{tikzpicture}
        \AfourF{0}{0}{0}{5}
        \AfourG{1}{2}{3}{4}
    \end{tikzpicture}}}
    {13}{\raisebox{-.5\height}{\begin{tikzpicture}
        \AfourF{1}{0}{3}{0}
        \AfourG{2}{3}{4}{2}
    \end{tikzpicture}}}
    {14}{\raisebox{-.5\height}{\begin{tikzpicture}
        \AfourF{0}{0}{5}{0}
        \AfourG{2}{4}{6}{3}
    \end{tikzpicture}}}
}}
\long\def\DfiveWeight#1#2#3#4#5{
    \begin{bmatrix}
        &&& #5\\
        #1&#2&#3&\\
        &&& #4
    \end{bmatrix}
}

\long\def\DfiveG#1#2#3#4#5{
        \node[gauge] (g1) at (-2,0) {$#1$};
        \node[gauge] (g2) at (-1,0) {$#2$};
        \node[gauge] (g3) at (0,0) {$#3$};
        \node[gauge] (g4) at (1,-1) {$#4$};
        \node[gauge] (g5) at (1,1) {$#5$};
        \draw (g1)--(g2)--(g3)--(g4) (g5)--(g3);
}
\long\def\DfiveF#1#2#3#4#5{
        \ifthenelse{#1>0}{\node[flavour] (f1) at (-2,-1) {$#1$}; \draw (f1)--(-2,0);}{}
        \ifthenelse{#2>0}{\node[flavour] (f2) at (-1,-1) {$#2$}; \draw (f2)--(-1,0);}{}
        \ifthenelse{#3>0}{\node[flavour] (f3) at (0,-1) {$#3$}; \draw (f3)--(0,0);}{}
        \ifthenelse{#4>0}{\node[flavour] (f4) at (2,-1) {$#4$}; \draw (f4)--(1,-1);}{}
        \ifthenelse{#5>0}{\node[flavour] (f5) at (2,1) {$#5$}; \draw (f5)--(1,1);}{}
}

\newcommand{\DfiveRgens}[1]{%
    \IfEqCase{#1}{%
    {1}{\raisebox{-.5\height}{\begin{tikzpicture}
        \DfiveF{0}{1}{0}{0}{0}
        \DfiveG{1}{2}{2}{1}{1}
    \end{tikzpicture}}}
    {2}{\raisebox{-.5\height}{\begin{tikzpicture}
        \DfiveF{2}{0}{0}{0}{0}
        \DfiveG{2}{2}{2}{1}{1}
    \end{tikzpicture}}}
    {3}{\raisebox{-.5\height}{\begin{tikzpicture}
        \DfiveF{1}{0}{1}{0}{0}            \DfiveG{2}{3}{4}{2}{2}
    \end{tikzpicture}}}
    {4}{\raisebox{-.5\height}{\begin{tikzpicture}
        \DfiveF{0}{0}{0}{1}{1}
        \DfiveG{1}{2}{3}{2}{2}
    \end{tikzpicture}}}
    {5}{\raisebox{-.5\height}{\begin{tikzpicture}
        \DfiveF{1}{0}{0}{2}{0}
        \DfiveG{2}{3}{4}{3}{2}
    \end{tikzpicture}}}
    {6}{\raisebox{-.5\height}{\begin{tikzpicture}
        \DfiveF{0}{0}{1}{2}{0}
        \DfiveG{2}{4}{6}{4}{3}
         \end{tikzpicture}}}
        {7}{\raisebox{-.5\height}{\begin{tikzpicture}
        \DfiveF{0}{0}{1}{0}{2}
        \DfiveG{2}{4}{6}{3}{4}
         \end{tikzpicture}}}
        {8}{\raisebox{-.5\height}{\begin{tikzpicture}
        \DfiveF{0}{0}{0}{4}{0}
        \DfiveG{2}{4}{6}{5}{3}
         \end{tikzpicture}}}
        {9}{\raisebox{-.5\height}{\begin{tikzpicture}
        \DfiveF{0}{0}{0}{0}{4}
        \DfiveG{2}{4}{6}{3}{5}
    \end{tikzpicture}}}
     {10}{\raisebox{-.5\height}{\begin{tikzpicture}
        \DfiveF{0}{0}{2}{0}{0}
        \DfiveG{2}{4}{6}{3}{3}
    \end{tikzpicture}}}
     {11}{\raisebox{-.5\height}{\begin{tikzpicture}
        \DfiveF{1}{0}{0}{0}{2}
        \DfiveG{2}{3}{4}{2}{3}
    \end{tikzpicture}}}
}}
\long\def\DSixWeight#1#2#3#4#5#6{
    \begin{bmatrix}
        &&&&#6\\
        #1&#2&#3&#4&\\
        &&&&#5
    \end{bmatrix}
}

\long\def\DSixG#1#2#3#4#5#6{
        \node[gauge] (g1) at (-2,0) {$#1$};
        \node[gauge] (g2) at (-1,0) {$#2$};
        \node[gauge] (g3) at (0,0) {$#3$};
        \node[gauge] (g4) at (1,0) {$#4$};
        \node[gauge] (g5) at (2,1) {$#5$};
        \node[gauge] (g6) at (2,-1) {$#6$};
        \draw (g1)--(g2)--(g3)--(g4)--(g5) (g6)--(g4);
}
\long\def\DSixF#1#2#3#4#5#6{
        \ifthenelse{#1>0}{\node[flavour] (f1) at (-2,-1) {$#1$}; \draw (f1)--(-2,0);}{}
        \ifthenelse{#2>0}{\node[flavour] (f2) at (-1,-1) {$#2$}; \draw (f2)--(-1,0);}{}
        \ifthenelse{#3>0}{\node[flavour] (f3) at (0,-1) {$#3$}; \draw (f3)--(0,0);}{}
        \ifthenelse{#4>0}{\node[flavour] (f4) at (1,-1) {$#4$}; \draw (f4)--(1,0);}{}
        \ifthenelse{#5>0}{\node[flavour] (f5) at (3,-1) {$#5$}; \draw (f5)--(2,-1);}{}
        \ifthenelse{#6>0}{\node[flavour] (f6) at (3,1) {$#6$}; \draw (f6)--(2,1);}{}
}

\newcommand{\DSixRgens}[1]{%
    \IfEqCase{#1}{%
    {1}{\raisebox{-.5\height}{\begin{tikzpicture}
        \DSixF{0}{1}{0}{0}{0}{0}
        \DSixG{1}{2}{2}{2}{1}{1}
    \end{tikzpicture}}}
    {2}{\raisebox{-.5\height}{\begin{tikzpicture}
        \DSixF{2}{0}{0}{0}{0}{0}
        \DSixG{2}{2}{2}{2}{1}{1}
    \end{tikzpicture}}}
    {3}{\raisebox{-.5\height}{\begin{tikzpicture}
        \DSixF{0}{0}{0}{1}{0}{0}
        \DSixG{1}{2}{3}{4}{2}{2}
    \end{tikzpicture}}}
    {4}{\raisebox{-.5\height}{\begin{tikzpicture}
        \DSixF{1}{0}{1}{0}{0}{0}
        \DSixG{2}{3}{4}{4}{2}{2}
    \end{tikzpicture}}}
    {5}{\raisebox{-.5\height}{\begin{tikzpicture}
        \DSixF{0}{0}{0}{0}{2}{0}
        \DSixG{1}{2}{3}{4}{2}{3}
    \end{tikzpicture}}}
    {6}{\raisebox{-.5\height}{\begin{tikzpicture}
        \DSixF{0}{0}{0}{0}{0}{2}
        \DSixG{1}{2}{3}{4}{3}{2}
    \end{tikzpicture}}}
    {7}{\raisebox{-.5\height}{\begin{tikzpicture}
        \DSixF{1}{0}{0}{0}{1}{1}
        \DSixG{2}{3}{4}{5}{3}{3}
    \end{tikzpicture}}}
     {8}{\raisebox{-.5\height}{\begin{tikzpicture}
        \DSixF{0}{0}{2}{0}{0}{0}
        \DSixG{2}{4}{6}{6}{3}{3}
    \end{tikzpicture}}}
     {9}{\raisebox{-.5\height}{\begin{tikzpicture}
        \DSixF{0}{0}{1}{0}{1}{1}
        \DSixG{2}{4}{6}{7}{4}{4}
    \end{tikzpicture}}}
}}
\long\def\eSixWeight#1#2#3#4#5#6{
    \begin{bmatrix}
        &&#6&&\\
        #1&#2&#3&#4&#5
    \end{bmatrix}
}

\long\def\eSixG#1#2#3#4#5#6{
        \node[gauge] (g1) at (-2,0) {$#1$};
        \node[gauge] (g2) at (-1,0) {$#2$};
        \node[gauge] (g3) at (0,0) {$#3$};
        \node[gauge] (g4) at (1,0) {$#4$};
        \node[gauge] (g5) at (2,0) {$#5$};
        \node[gauge] (g6) at (0,1) {$#6$};
        \draw (g1)--(g2)--(g3)--(g4)--(g5) (g3)--(g6);
}
\long\def\eSixF#1#2#3#4#5#6{
        \ifthenelse{#1>0}{\node[flavour] (f1) at (-2,-1) {$#1$}; \draw (f1)--(-2,0);}{}
        \ifthenelse{#2>0}{\node[flavour] (f2) at (-1,-1) {$#2$}; \draw (f2)--(-1,0);}{}
        \ifthenelse{#3>0}{\node[flavour] (f3) at (0,-1) {$#3$}; \draw (f3)--(0,0);}{}
        \ifthenelse{#4>0}{\node[flavour] (f4) at (1,-1) {$#4$}; \draw (f4)--(1,0);}{}
        \ifthenelse{#5>0}{\node[flavour] (f5) at (2,-1) {$#5$}; \draw (f5)--(2,0);}{}
        \ifthenelse{#6>0}{\node[flavour] (f6) at (0,2) {$#6$}; \draw (f6)--(0,1);}{}
}

\newcommand{\EsixRgens}[1]{%
    \IfEqCase{#1}{%
    {3}{\raisebox{-.5\height}{\begin{tikzpicture}
        \eSixF{0}{0}{1}{0}{0}{0}
        \eSixG{2}{4}{6}{4}{2}{3}
    \end{tikzpicture}}}
    {1}{\raisebox{-.5\height}{\begin{tikzpicture}
        \eSixF{0}{0}{0}{0}{0}{1}
        \eSixG{1}{2}{3}{2}{1}{2}
    \end{tikzpicture}}}
    {4}{\raisebox{-.5\height}{\begin{tikzpicture}
        \eSixF{1}{1}{0}{0}{0}{0}
        \eSixG{3}{5}{6}{4}{2}{3}
    \end{tikzpicture}}}
    {2}{\raisebox{-.5\height}{\begin{tikzpicture}
        \eSixF{1}{0}{0}{0}{1}{0}
        \eSixG{2}{3}{4}{3}{2}{2}
    \end{tikzpicture}}}
    {8}{\raisebox{-.5\height}{\begin{tikzpicture}
        \eSixF{0}{1}{0}{1}{0}{0}
        \eSixG{3}{6}{8}{6}{3}{4}
    \end{tikzpicture}}}
    {6}{\raisebox{-.5\height}{\begin{tikzpicture}
        \eSixF{0}{0}{0}{1}{1}{0}
        \eSixG{2}{4}{6}{5}{3}{3}
    \end{tikzpicture}}}
    {5}{\raisebox{-.5\height}{\begin{tikzpicture}
        \eSixF{3}{0}{0}{0}{0}{0}
        \eSixG{4}{5}{6}{4}{2}{3}
    \end{tikzpicture}}}
    {9}{\raisebox{-.5\height}{\begin{tikzpicture}
        \eSixF{2}{0}{0}{1}{0}{0}
        \eSixG{4}{6}{8}{6}{3}{4}
    \end{tikzpicture}}}
    {11}{\raisebox{-.5\height}{\begin{tikzpicture}
        \eSixF{1}{0}{0}{2}{0}{0}
        \eSixG{4}{7}{10}{8}{4}{5}
    \end{tikzpicture}}}
    {14}{\raisebox{-.5\height}{\begin{tikzpicture}
        \eSixF{0}{0}{0}{3}{0}{0}
        \eSixG{4}{8}{12}{10}{5}{6}
    \end{tikzpicture}}}
    {13}{\raisebox{-.5\height}{\begin{tikzpicture}
        \eSixF{0}{3}{0}{0}{0}{0}
        \eSixG{5}{10}{12}{8}{4}{6}
    \end{tikzpicture}}}
    {12}{\raisebox{-.5\height}{\begin{tikzpicture}
        \eSixF{0}{2}{0}{0}{1}{0}
        \eSixG{4}{8}{10}{7}{4}{5}
    \end{tikzpicture}}}
    {10}{\raisebox{-.5\height}{\begin{tikzpicture}
        \eSixF{0}{1}{0}{0}{2}{0}
        \eSixG{3}{6}{8}{6}{4}{4}
    \end{tikzpicture}}}
    {7}{\raisebox{-.5\height}{\begin{tikzpicture}
        \eSixF{0}{0}{0}{0}{3}{0}
        \eSixG{2}{4}{6}{5}{4}{3}
    \end{tikzpicture}}}
}}
\long\def\eSevenWeight#1#2#3#4#5#6#7{
    \begin{bmatrix}
        &&#7&&&\\
        #1&#2&#3&#4&#5&#6
    \end{bmatrix}
}

\long\def\eSevenG#1#2#3#4#5#6#7{
        \node[gauge] (g1) at (-2,0) {$#1$};
        \node[gauge] (g2) at (-1,0) {$#2$};
        \node[gauge] (g3) at (0,0) {$#3$};
        \node[gauge] (g4) at (1,0) {$#4$};
        \node[gauge] (g5) at (2,0) {$#5$};
        \node[gauge] (g6) at (3,0) {$#6$};
        \node[gauge] (g7) at (0,1) {$#7$};
        \draw (g1)--(g2)--(g3)--(g4)--(g5)--(g6) (g3)--(g7);
}
\long\def\eSevenF#1#2#3#4#5#6#7{
        \ifthenelse{#1>0}{\node[flavour] (f1) at (-2,-1) {$#1$}; \draw (f1)--(-2,0);}{}
        \ifthenelse{#2>0}{\node[flavour] (f2) at (-1,-1) {$#2$}; \draw (f2)--(-1,0);}{}
        \ifthenelse{#3>0}{\node[flavour] (f3) at (0,-1) {$#3$}; \draw (f3)--(0,0);}{}
        \ifthenelse{#4>0}{\node[flavour] (f4) at (1,-1) {$#4$}; \draw (f4)--(1,0);}{}
        \ifthenelse{#5>0}{\node[flavour] (f5) at (2,-1) {$#5$}; \draw (f5)--(2,0);}{}
        \ifthenelse{#6>0}{\node[flavour] (f6) at (3,-1) {$#6$}; \draw (f6)--(3,0);}{}
        \ifthenelse{#7>0}{\node[flavour] (f7) at (0,2) {$#7$}; \draw (f7)--(0,1);}{}
}

\newcommand{\EsevenRgens}[1]{%
    \IfEqCase{#1}{%
    {1}{\raisebox{-.5\height}{\begin{tikzpicture}
        \eSevenF{1}{0}{0}{0}{0}{0}{0}
        \eSevenG{2}{3}{4}{3}{2}{1}{2}
    \end{tikzpicture}}}
    {2}{\raisebox{-.5\height}{\begin{tikzpicture}
        \eSevenF{0}{0}{0}{0}{1}{0}{0}
        \eSevenG{2}{4}{6}{5}{4}{2}{3}
    \end{tikzpicture}}}
    {3}{\raisebox{-.5\height}{\begin{tikzpicture}
        \eSevenF{0}{1}{0}{0}{0}{0}{0}
        \eSevenG{3}{6}{8}{6}{4}{2}{4}
    \end{tikzpicture}}}
    {4}{\raisebox{-.5\height}{\begin{tikzpicture}
        \eSevenF{0}{0}{0}{0}{0}{2}{0}
        \eSevenG{2}{4}{6}{5}{4}{3}{3}
    \end{tikzpicture}}}
    {5}{\raisebox{-.5\height}{\begin{tikzpicture}
        \eSevenF{0}{0}{0}{0}{0}{1}{1}
        \eSevenG{3}{6}{9}{7}{5}{3}{5}
    \end{tikzpicture}}}
    {6}{\raisebox{-.5\height}{\begin{tikzpicture}
        \eSevenF{0}{0}{1}{0}{0}{0}{0}
        \eSevenG{4}{8}{12}{9}{6}{3}{6}
    \end{tikzpicture}}}
    {7}{\raisebox{-.5\height}{\begin{tikzpicture}
        \eSevenF{0}{0}{0}{0}{0}{0}{2}
        \eSevenG{4}{8}{12}{9}{6}{3}{7}
    \end{tikzpicture}}}
    {8}{\raisebox{-.5\height}{\begin{tikzpicture}
        \eSevenF{0}{0}{0}{2}{0}{0}{0}
        \eSevenG{6}{12}{18}{15}{10}{5}{9}
    \end{tikzpicture}}}
    {9}{\raisebox{-.5\height}{\begin{tikzpicture}
        \eSevenF{0}{0}{0}{1}{0}{1}{0}
        \eSevenG{4}{8}{12}{10}{7}{4}{6}
    \end{tikzpicture}}}
    {10}{\raisebox{-.5\height}{\begin{tikzpicture}
        \eSevenF{0}{0}{0}{1}{0}{0}{1}
        \eSevenG{5}{10}{15}{12}{8}{4}{8}
    \end{tikzpicture}}}
}}
\long\def\eEightWeight#1#2#3#4#5#6#7#8{
    \begin{bmatrix}
        &&#8&&&&\\
        #1&#2&#3&#4&#5&#6&#7
    \end{bmatrix}
}

\long\def\eEightG#1#2#3#4#5#6#7#8{
        \node[gauge] (g1) at (-2,0) {$#1$};
        \node[gauge] (g2) at (-1,0) {$#2$};
        \node[gauge] (g3) at (0,0) {$#3$};
        \node[gauge] (g4) at (1,0) {$#4$};
        \node[gauge] (g5) at (2,0) {$#5$};
        \node[gauge] (g6) at (3,0) {$#6$};
        \node[gauge] (g7) at (4,0) {$#7$};
        \node[gauge] (g8) at (0,1) {$#8$};
        \draw (g1)--(g2)--(g3)--(g4)--(g5)--(g6)--(g7) (g3)--(g8);
}
\long\def\eEightF#1#2#3#4#5#6#7#8{
        \ifthenelse{#1>0}{\node[flavour] (f1) at (-2,-1) {$#1$}; \draw (f1)--(-2,0);}{}
        \ifthenelse{#2>0}{\node[flavour] (f2) at (-1,-1) {$#2$}; \draw (f2)--(-1,0);}{}
        \ifthenelse{#3>0}{\node[flavour] (f3) at (0,-1) {$#3$}; \draw (f3)--(0,0);}{}
        \ifthenelse{#4>0}{\node[flavour] (f4) at (1,-1) {$#4$}; \draw (f4)--(1,0);}{}
        \ifthenelse{#5>0}{\node[flavour] (f5) at (2,-1) {$#5$}; \draw (f5)--(2,0);}{}
        \ifthenelse{#6>0}{\node[flavour] (f6) at (3,-1) {$#6$}; \draw (f6)--(3,0);}{}
        \ifthenelse{#7>0}{\node[flavour] (f7) at (4,-1) {$#7$}; \draw (f7)--(4,0);}{}
        \ifthenelse{#8>0}{\node[flavour] (f8) at (0,2) {$#8$}; \draw (f8)--(0,1);}{}
}

\newcommand{\EeightRgens}[1]{%
    \IfEqCase{#1}{%
    {1}{\raisebox{-.5\height}{\begin{tikzpicture}
        \eEightF{0}{0}{0}{0}{0}{0}{1}{0}
        \eEightG{2}{4}{6}{5}{4}{3}{2}{3}
    \end{tikzpicture}}}
    {2}{\raisebox{-.5\height}{\begin{tikzpicture}
        \eEightF{1}{0}{0}{0}{0}{0}{0}{0}
        \eEightG{4}{7}{10}{8}{6}{4}{2}{5}
    \end{tikzpicture}}}
    {3}{\raisebox{-.5\height}{\begin{tikzpicture}
        \eEightF{0}{0}{0}{0}{0}{1}{0}{0}
        \eEightG{4}{8}{12}{10}{8}{6}{3}{6}
    \end{tikzpicture}}}
    {4}{\raisebox{-.5\height}{\begin{tikzpicture}
        \eEightF{0}{0}{0}{0}{0}{0}{0}{1}
        \eEightG{5}{10}{15}{12}{9}{6}{3}{8}
    \end{tikzpicture}}}
    {5}{\raisebox{-.5\height}{\begin{tikzpicture}
        \eEightF{0}{0}{0}{0}{1}{0}{0}{0}
        \eEightG{6}{12}{18}{15}{12}{8}{4}{9}
    \end{tikzpicture}}}
    {6}{\raisebox{-.5\height}{\begin{tikzpicture}
        \eEightF{0}{1}{0}{0}{0}{0}{0}{0}
        \eEightG{7}{14}{20}{16}{12}{8}{4}{10}
    \end{tikzpicture}}}
    {7}{\raisebox{-.5\height}{\begin{tikzpicture}
        \eEightF{0}{0}{0}{1}{0}{0}{0}{0}
        \eEightG{8}{16}{24}{20}{15}{10}{5}{12}
    \end{tikzpicture}}}
    {8}{\raisebox{-.5\height}{\begin{tikzpicture}
        \eEightF{0}{0}{1}{0}{0}{0}{0}{0}
        \eEightG{10}{20}{30}{24}{18}{12}{6}{15}
    \end{tikzpicture}}}
}}

\begin{document}

\begin{titlepage}

\phantom{wowiezowie}

\vspace{-1cm}

\begin{center}

{\Huge {\bf From Quivers to Geometry:\\
\vspace{0.2cm}
5d Conformal Matter}}


\vspace{1cm}
{\Large Antoine Bourget,$^{\circ}$ Mario De Marco,$^{\ast}$ Michele Del Zotto,$^{\dagger\ddagger\star}$}\\ 

\medskip

{\Large  Julius F. Grimminger,$^{\sharp}$ and Andrea Sangiovanni $^{\dagger\ddagger\star}$}\\

\vspace{1cm}

{\it
{\fontsize{9pt}{10pt}\selectfont
\renewcommand{\arraystretch}{2.5}
\noindent\hspace*{-2.7cm}
\begin{tabular}{cc}
\makecell{$^{\circ}$ \text{Institut de Physique Théorique,  CEA, CNRS,}   \\ \text{Université Paris-Saclay, 91191 Gif-sur-Yvette, France}} & \makecell{$^\dagger$ \text{Mathematics Institute, Uppsala University,} \\ \text{Box 480, SE-75106 Uppsala, Sweden}}\\
\makecell{$^\star$ \text{Centre for Geometry and Physics, Uppsala University,} \\ \text{Box 480, SE-75106 Uppsala, Sweden}\\} & \makecell{$^\ddagger$ \text{Department of Physics and Astronomy, Uppsala University,}\\ \text{Box 516, SE-75120 Uppsala, Sweden}\\}\\
\makecell{$^{\sharp}$ \text{Mathematical Institute, University of Oxford,}\\\text{Andrew Wiles Building, Woodstock Road, Oxford, OX2 6GG, UK}\\} & \makecell{$^{\ast}$ \text{Physique Th\'eorique et Math\'ematique and International Solvay Institutes,}\\
\text{Universit\'e Libre de Bruxelles, C.P. 231, 1050 Brussels, Belgium}}
\end{tabular}
}}

\vskip .5cm
{\footnotesize \tt antoine.bourget@ipht.fr \hspace{1cm} mario.de.marco@ulb.be \hspace{1cm} michele.delzotto@math.uu.se } \\
{\footnotesize \tt    julius.grimminger@maths.ox.ac.uk \hspace{1cm} andrea.sangiovanni@math.uu.se}

\vskip 1cm
     	{\bf Abstract }
\vskip .1in

\end{center}

\noindent 
We show that all 5d balanced (ADE-shaped) special unitary quivers with no Chern-Simons level admit a UV completion which is a 5d conformal matter SCFT. We give explicit local Calabi-Yau threefolds realizing each of these models in M-theory. This unified description enables a systematic exploration of their physical properties, such as their Higgs Branch, as well as connections to class-$\mathcal{S}$ constructions and the affine Grassmannian.
 
\noindent 

\eject

\end{titlepage}

\tableofcontents

\section{Introduction}\label{sec: summary}

Devising a classification of interacting superconformal fixed points in more than 4 spacetime dimensions has been a highly coveted prize since their discovery in the late nineties \cite{Witten:1995ex,Strominger:1995ac,Witten:1995em,Ganor:1996mu,Seiberg:1996qx,Seiberg1996,Morrison_1997,Douglas:1996xp}. The main strategy laid down to tackle this problem relies on the \textit{geometric engineering} of superconformal field theories (SCFTs), both via pure geometric M-theory/F-theory/String theory backgrounds, or via suitable brane configurations, or by combinations of the two. Oftentimes the geometric and the brane picture are related via a chain of dualities. For 6d $\mathcal{N}=(2,0)$ and $\mathcal{N}=(1,0)$ SCFTs, this tactic has met tremendous success, respectively employing the dimensional reduction of Type IIB string theory on Du Val singularities \cite{Witten:1995ex,Strominger:1995ac}, and of F-theory on elliptically fibered singular threefolds \cite{Heckman:2013pva,DelZotto:2014hpa,Heckman:2015bfa}. Despite some remaining subtleties due to ``frozen singularities'' \cite{Bhardwaj:2018jgp}, this framework substantially exhausts the 6d SCFTs that can be constructed \textit{within} the realm of geometric engineering.\\
\indent On the contrary, in 5 spacetime dimensions the picture is murkier, as no general organizing principle has yet been put forward (in no small part due to the lack of the organizing logic provided by elliptically fibered threefolds). On the one hand, it has been proposed that all 5d SCFTs possess a 6d origin, e.g.\ via dimensional reduction on a circle (including possible holonomies along said circle) \cite{Jefferson_2018}. Indeed, a plethora of examples relying on F-theory constructions has been realized in \cite{DelZotto:2017pti,Bhardwaj:2018yhy,Bhardwaj:2018vuu,Bhardwaj:2019xeg,Bhardwaj:2020gyu}. Alternatively, one can employ bottom-up constructions of smooth geometries satisfying suitable consistency conditions to engineer low-rank 5d SCFTs \cite{Jefferson_2018}.\footnote{See however \cite{Collinucci:2025rrh} for a recent result that questions the completeness of the rank-one list proposed in \cite{Jefferson_2018}.} The main drawback of this method is that it quickly grows in complexity once higher-rank 5d SCFTs are probed, making a concrete classification scheme impractical. From the Type IIB perspective, 5-brane webs \cite{Aharony:1997ju,Aharony:1997bh,DeWolfe:1999hj} have been employed to great effect to inspect large swaths of the 5d SCFT landscape \cite{Benini:2009gi,Bergman:2013aca,Zafrir:2014ywa,Hayashi:2015zka,Hayashi:2015fsa,Bergman:2015dpa,Hayashi:2018lyv,Hayashi:2018bkd,Hayashi:2019yxj,Hayashi_2020,Bergman:2020myx}. It is well-known that some classes of 5d SCFTs engineered from 5-brane webs have a dual description in terms of an M-theory setup on a singular non-compact Calabi-Yau threefold (CY3): the duality is explicitly known in brane webs without orientifold planes, with or without white dots \cite{leung1997branes}\footnote{See e.g. \cite{Bourget:2023wlb} for recent work on constructing the singular CY3 corresponding to 5-brane webs with white dots on one external side of the corresponding generalized toric polygon. \cite{Alexeev:2024bko} deals with white dots on multiple sides.}. In all cases, the Calabi-Yau threefolds at hand are either toric, or non-toric deformations of toric threefolds. For all the remaining 5-brane webs, no explicit dual CY3 is known. The last and arguably least explored method relies on a pure geometric background involving the reduction of M-theory on a non-compact CY3 with \textit{canonical singularities}. This approach has been first phrased in such terms in \cite{Xie_2017}, and it naturally encompasses the CY3's that are dual to brane-webs. It also comprises a broad set of canonical threefolds which are not toric singularities (or obvious deformations thereof), and that hence lack a known 5-brane web description.  A classification of isolated canonical hypersurface CY3 admitting a quasi-homogeneous action is known thanks to the work of \cite{yau2003classification}\footnote{This set comprises notable subclasses such as     quasi-homogeneous compound Du Val singularities.}, as well as a list of canonical Calabi-Yau threefolds constructed as orbifolds of type $\mathbb{C}^3/\Gamma$, with $\Gamma \in SL(3,\mathbb{C})$ \cite{Tian:2021cif}. Some collections of 5d SCFTs engineered from hypersurface or complete intersection canonical isolated singularities have been further explored in \cite{Closset:2020scj,
Closset:2020afy,
Closset:2021lwy,
Collinucci:2021ofd,Mu:2023uws}, and examples of 5d SCFTs engineered by non-toric non-complete intersections have been constructed in \cite{DeMarcoDelZottoGraffeoSangiovanni:WIP, Collinucci:2025rrh, Dramburg:2025tlb}.\\

\indent In this work, we aim at employing M-theory geometric engineering on canonical singularities as our chief tool to chart a specific class of 5d SCFTs, proposing an ``atomic'' classification scheme. Such a program entails identifying a finite set of indecomposable SCFTs, that act as building blocks that can be gauged together to form a family (usually with infinitely many members) of 5d SCFTs that share some essential features. This is in the same vein as the exploration of 4d $\mathcal{N}=2$ theories in terms of class-$\mathcal{S}$ fixtures (that act as indecomposable building blocks) and their conformal gaugings \cite{Chacaltana:2010ks,Chacaltana:2011ze,Chacaltana:2012zy,Chacaltana:2012ch,Chacaltana:2013oka,Chacaltana:2014jba,Chacaltana:2015bna,Chacaltana:2016shw,Chacaltana:2017boe,Chacaltana:2018vhp}. 

We wish to apply this philosophy to a class of 5d SCFTs that are particularly amenable to be organized into an atomic classification scheme. Namely, we focus on \textit{bifundamental} 5d SCFTs, where ``bifundamental'' means that the theory sports at least a $\mathfrak{g}\oplus\mathfrak{g}$ flavor symmetry, with $\mathfrak{g}\in ADE$. These theories are dubbed ``5d conformal matter'' SCFTs (5d CM) in analogy with their 6d parents \cite{DelZotto:2014hpa}.\\
\indent We tread our first steps starting from the work \cite{DeMarco:2023irn}, that recently introduced a new class of non-toric canonical CY3 with non-isolated singularities, motivated by the search for interacting 5d CM theories. In \cite{DeMarco:2023irn}, a set of 5d SCFTs that play the role of indecomposable building blocks (``atoms'') have been identified, along with their fusion rules: via gauging operations, that correspond to gluings of the associated CY3, the atoms can be fused to obtain novel 5d CM SCFTs (``molecules'').
The Higgs Branch of such 5d SCFTs and the connection to class-$\mathcal{S}$ constructions were explored in \cite{DeMarco:2025ugw}. All the 5d CM theories considered in \cite{DeMarco:2023irn} and \cite{DeMarco:2025ugw} admit an IR phase which is a balanced special unitary quiver of shape $\mathfrak{g}\in ADE$, with no Chern-Simons level. Crucially in \cite{DeMarco:2023irn} \textit{not all quiver gauge theories with such features are obtained as IR phases of the discovered 5d CM theories.} It is natural to ask whether these further quiver theories admit a UV completion in terms of a 5d SCFT and an associated singular threefold geometry.\\
\indent In the present work, we answer this question in the affirmative, showing that: \\

\textit{All 5d special unitary quiver gauge theories with trivial Chern-Simons levels and with balanced nodes, arranged in the shape of the Dynkin diagram of $\mathfrak{g}$, with $\mathfrak{g}\in ADE$, admit a UV completion in terms of a 5d CM SCFT, with flavor symmetry at least $\mathfrak{g}\oplus\mathfrak{g}$. The UV fixed points can be explicitly engineered via M-theory on a singular Calabi-Yau threefold.}\\

This result comes with its own unique features:
\begin{itemize}
    \item We show that rigorous Lie-theoretic criteria form the backbone for the classification of 5d CM theories.
    \item We identify the set of ``distinguished'' CY3 that systematically reproduce the CY3 geometries needed to engineer \textit{all} the 5d CM SCFTs admitting a low-energy description in terms of a balanced quiver gauge theory, with gauge nodes arranged like the $\mathfrak{g}$ Dynkin diagram, and vanishing Chern-Simons levels. We unveil precise connections of specific classes of 5d CM theories with class-$\mathcal{S}$ constructions.
    \item As a welcome bonus, we study the deformation theory of 5d CM theories, and show that they all descend from 6d CM theories.
\end{itemize}
In the next Section we summarize our results in a self-contained fashion.

\subsection{Summary of results}
5d bifundamental conformal matter theories are SCFTs with flavor symmetry containing \textit{at least} a $\mathfrak{g}\oplus\mathfrak{g}$ subalgebra, with $\mathfrak{g} \in ADE$. As such, in order to engineer them it is reasonable to employ Calabi-Yau threefolds that display two non-compact lines of singularities of ADE type, also commonly referred to as Du Val singularities. We denote Du Val singularities as $\mathsf{Y}_{\mathfrak{g}}$, where $\mathfrak{g}$ specifies the ADE type.\\
\indent The canonical Calabi-Yau threefolds that engineer 5d CM theories of type $\mathfrak{g}\oplus\mathfrak{g}$ are of the form:
\begin{equation}\label{CY3 CM summary}
    \begin{cases}
       Y_{\mathfrak{g}}(x,y,z) =  0\\
        uv = P(x,y,z)\\
    \end{cases},
\end{equation}
where $Y_{\mathfrak{g}}(x,y,z)$ is the polynomial defining the Du Val singularity, and $P(x,y,z)$ is an arbitrary polynomial. We will show how a choice of $\mathfrak{g}$ and $P(x,y,z)$ uniquely fixes a dominant coweight $\mu$ in the Lie algebra $\mathfrak{g}$. In turn, each dominant coweight dictates a special unitary quiver $\mathsf{Q}_{\mu}$, which is balanced and with no Chern-Simons levels. The quiver gauge theory associated to $\mathsf{Q}_{\mu}$ corresponds to a specific low-energy limit of the SCFT engineered by \eqref{CY3 CM summary}. \textit{In this work we construct the 5d CM theories associated to all dominant coweights, i.e.\ that admit a IR quiver gauge theory phase  $\mathsf{Q}_{\mu}$}. Geometrically, this phase is reached via an appropriate partial crepant resolution of \eqref{CY3 CM summary}. The theorem illustrated in \cite{Yonekura:2015ksa} shows that all the quiver gauge theories $\mathsf{Q}_{\mu}$ flow in the UV to SCFTs with \textit{at least} $\mathfrak{g}\oplus\mathfrak{g}$ flavor symmetry, confirming the intuition drawn from the singular geometry also from the field-theoretic perspective. Conversely, for every choice of $P(x,y,z)$ in \eqref{CY3 CM summary} we provide an algorithm to compute the 5d CM theory corresponding to it: each of these will admit a low-energy quiver phase encoded by a dominant coweight. Pictorially we have:
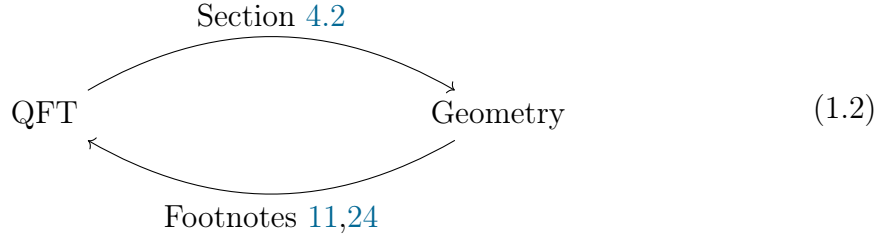
\begin{equation}\raisebox{-.5\height}{
\begin{tikzpicture}
    \node (1) at (0,0) {QFT};
    \node (2) at (6,0) {Geometry};
    \draw[->] (1) to [bend left=30] node[above, sloped, pos=0.5] {Section \ref{sec:formal algorithm}} (2);
    \draw[->] (2) to [bend left=30] node[below, sloped, pos=0.5] {Footnotes \ref{footnote ideals},\ref{footnote ideals 2}} (1);
\end{tikzpicture}}
\end{equation}
We also attach in an ancillary file a Macaulay2 code reproducing the algorithm corresponding to the right to left arrow.\\

\indent The chief tools to explore the features of 5d CM SCFTs are gauging and Higgsing. By gauging we mean the operation of gauging a diagonal $\mathfrak{g}$ subalgebra of two 5d CM theories of $\mathfrak{g}\oplus\mathfrak{g}$ type: this procedure can be repeated iteratively to fuse 5d CM theories to produce arbitrarily long ``molecules'', which are still 5d CM SCFTs. Geometrically, this amounts to gluing the corresponding Calabi-Yau threefolds along non-compact singular lines, making sure that the Calabi-Yau condition is upheld\footnote{We remark here, as shown in \cite{DeMarco:2023irn}, that only bivalent gaugings (i.e.\ gaugings of the diagonal subalgebra between at most two flavor algebras) preserve the Calabi-Yau condition.}. Higgsing refers to turning on a dynamical complex structure deformation in the threefold, that field-theoretically translates into triggering a RG flow from a given 5d CM theory to another. Three different types of 5d CM theories can be identified, employing their properties under gauging and Higgsing:
\begin{itemize}
    \item 5d CM \textit{molecules}, which are all the 5d CM theories that can be obtained via gauging of at least two other 5d CM theories.
    \item 5d CM \textit{atoms}, that cannot be obtained via gauging of other 5d CM SCFTs. Higgsing 5d CM atoms exclusively produces further 5d CM atoms.
    \item 5d CM \textit{hybrids}, that cannot be obtained via gauging of other 5d CM SCFTs. Higgsing 5d CM hybrids produces further 5d CM atoms, and \textit{at least} a 5d CM molecule. This is the key difference between 5d CM atoms and hybrids. This type of 5d CM SCFT is a novel addition with respect to \cite{DeMarco:2023irn}.
\end{itemize}
We identify the singular CY3 geometries, of the form \eqref{CY3 CM summary}, corresponding to all 5d CM atoms, hybrids and molecules. To this end, for each $\mathfrak{g}\in ADE$, we list a finite set of CY3 geometries, that we term \textit{distinguished CY3 geometries}, that are sufficient to retrieve $P(x,y,z)$ for a generic choice of $\mu$ in the dominant coweight lattice. This is achieved via an algorithm that takes as input $\mathfrak{g}$ and $\mu$, relates $\mu$ to an appropriate combination of coweights related to distinguished geometries, and outputs $P(x,y,z)$ in \eqref{CY3 CM summary}, corresponding to $\mathsf{Q}_{\mu}$. The reverse version of the algorithm associates a dominant coweight to any singular threefold specified by a choice of $P(x,y,z)$. This shows the power of the spirit of atomic classifications, as only a handful of building blocks are needed to characterize a class of 5d SCFTs with infinitely many elements.\\

\indent The physical properties of 5d CM SCFTs possess a direct counterpart in Lie-theoretic language: 
\begin{itemize}
    \item 5d CM molecules, whose low-energy quiver gauge theory phases $\mathsf{Q}_{\mu}$ are in one-to-one correspondence with \textit{decomposable} dominant coweights\footnote{We review Lie-theoretic definitions in more details in Section \ref{sec:atomeshybridsmolecules}.}. As we have mentioned, 5d CM molecules are obtained from appropriate gaugings of atoms and hybrids. Lie-theoretically, gauging is interpreted as summing indecomposable coweights (i.e.\ atoms and hybrids) in order to produce decomposable coweights (molecules).
    \item 5d CM atoms admit low-energy quiver gauge theory phases $\mathsf{Q}_{\mu}$ that are in one-to-one correspondence with \textit{small dominant coweights} of $\mathfrak{g}$. In particular, small dominant coweights are indecomposable (i.e.\ they cannot be obtained as a sum of dominant coweights).
    \item 5d CM hybrids, whose low-energy quiver gauge theory phases $\mathsf{Q}_{\mu}$ are in one-to-one correspondence with indecomposable dominant coweights which are \textit{not} small. 5d CM hybrids can be Higgsed to molecules via an appropriate deformation. Namely, indecomposable coweights which are not small are greater, in the sense of the partial ordering, than at least one decomposable coweight.
\end{itemize}
\indent Dominant coweights admit a partial ordering that organizes them into a Hasse diagram. The Hasse diagram of dominant coweights encodes a subspace in the (but not the full) Higgs Branch of 5d CM theories. Higgsing from a given 5d CM theory to another is possible if the corresponding coweights are ordered in the Hasse diagram \cite{Bourget:2019aer}. This observation is directly confirmed by the analysis of \textit{dynamical complex structure deformations} of the threefolds associated to 5d CM theories, that we explore in detail relying on the techniques introduced by \cite{Collinucci:2020jqd}.\\
\indent For 5d CM atoms, the Hasse diagram has a further interpretation in terms of class-$\mathcal{S}$ theories. Via circle reduction, 5d CM atoms descend to 4d $\mathcal{N}=2$ theories captured by a class-$\mathcal{S}$ configuration on a sphere with three regular punctures: two of the punctures are full, accounting for the $\mathfrak{g}\oplus\mathfrak{g}$ flavor symmetry; the third puncture is denoted by $\mathcal{O}_{\mu}$, as it depends on the coweight $\mu$ that corresponds to the atom. Higgsing from an atom to another amounts to a (partial) closure of the puncture $\mathcal{O}_{\mu}$. It has been shown in \cite{DeMarco:2025ugw} that molecules composed of atoms admit a mass deformation that descends via circle reduction to a class-$\mathcal{S}$ theory with regular punctures. Here we extend this result to all 5d CM exclusively composed of atoms. Any 5d CM hybrid, or any molecule obtained via gauging of at least one hybrid theory, does not admit a straightforward class-$\mathcal{S}$ interpretation.

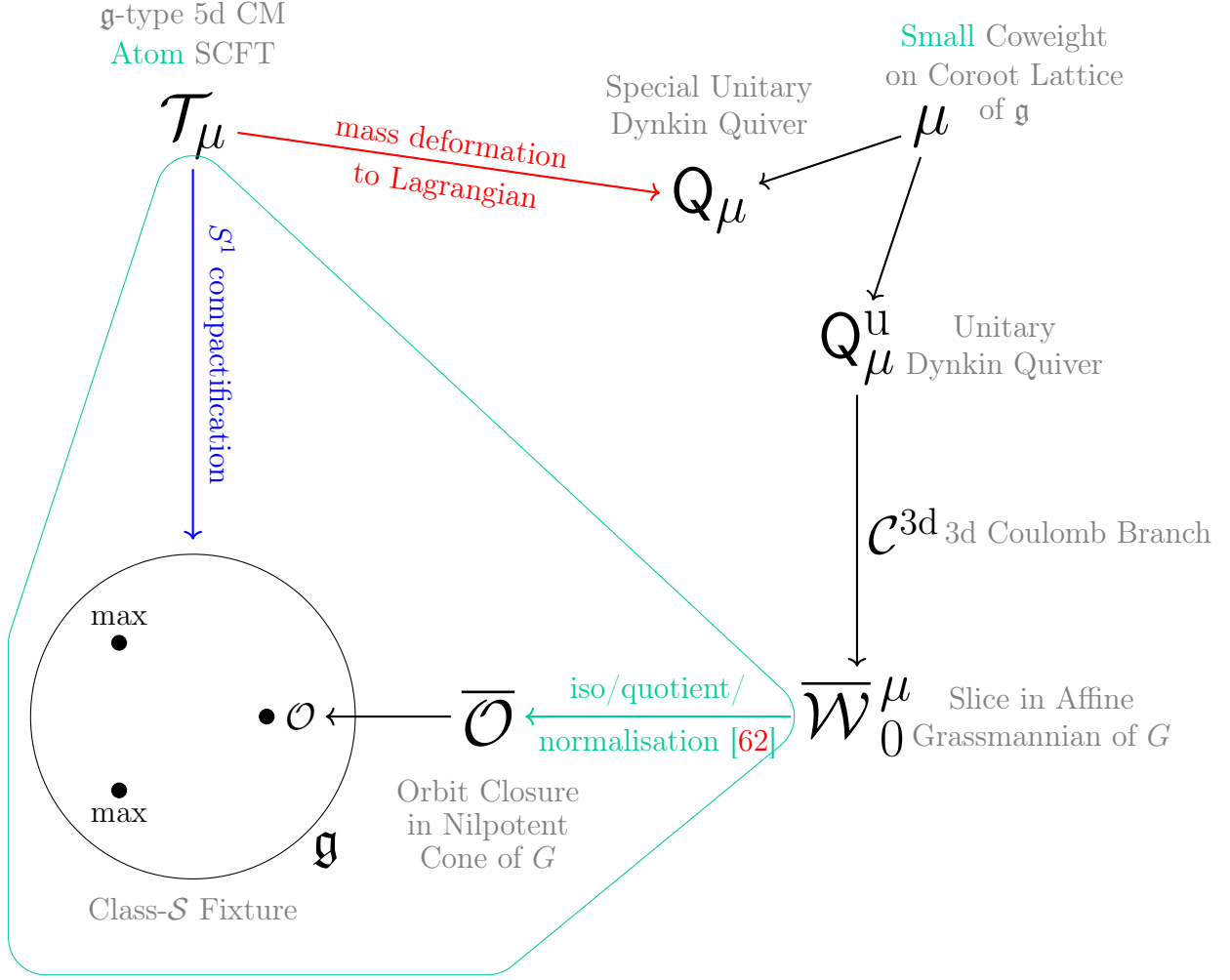
\begin{figure}
    \makebox[\textwidth][c]{\begin{tikzpicture}
        \node (a) at (0,0) {\Huge$\mathcal{T}_{\mu}$};
        \node[gray] at (0,1) {{\color{caribbeangreen}Atom} SCFT};
        \node[gray] at (0,1.5) {$\mathfrak{g}$-type 5d CM};
        \node (b) at (7,-1) {\Huge$\mathsf{Q}_{\mu}$};
        \node[gray] at (7,0) {Dynkin Quiver};
        \node[gray] at (7,0.5) {Special Unitary};
        \draw[->,thick,red] (a)--(b) node[midway, sloped,above] {mass deformation} node[midway, sloped,below] {to Lagrangian};
        \node (c) at (10,0) {\Huge$\mu$};
        \node[gray] at (11,0.2) {of $\mathfrak{g}$};
        \node[gray] at (11,0.7) {on Coroot Lattice};
        \node[gray] at (11,1.2) {{\color{caribbeangreen}Small} Coweight};
        \draw[->,thick] (c)--(b);
        \node (d) at (9,-3) {\Huge$\mathsf{Q}^{\mathrm{u}}_{\mu}$};
        \node[gray] at (11,-2.75) {Unitary};
        \node[gray] at (11,-3.25) {Dynkin Quiver};
        \draw[->,thick] (c)--(d);
        \node (e) at (9,-8) {\Huge$\overline{\mathcal{W}}^{\,\mu}_{\,0}$};
        \node[gray] at (11.5,-7.75) {Slice in Affine};
        \node[gray] at (11.5,-8.25) {Grassmannian of $G$};
        \draw[->,thick] (d)--(e) node[midway,right] {\LARGE$\mathcal{C}^{3\mathrm{d}}$};
        \node[gray] at (12,-5.5) {3d Coulomb Branch};
        \node (f) at (4,-8) {\Huge$\overline{\mathcal{O}}$};
        \node[gray] at (4,-9) {Orbit Closure};
        \node[gray] at (4,-9.5) {in Nilpotent};
        \node[gray] at (4,-9.9) {Cone of $G$};
        \draw[->,thick,caribbeangreen] (e)--(f) node[midway,above] {iso/quotient/} node[midway,below] {normalisation \cite{achar2013geometric}};
        \node[bd,label=right:{\large$\mathcal{O}$}] (g) at (1,-8) {};
        \draw[->,thick] (f)--(1.8,-8);
        \node[bd,label=above:{$\mathrm{max}$}] at (-1,-7) {};
        \node[bd,label=below:{$\mathrm{max}$}] at (-1,-9) {};
        \draw (0,-8) circle (2.2);
        \node at (1.8,-9.8) {\LARGE$\mathfrak{g}$};
        \node[gray] at (0,-10.6) {Class-$\mathcal{S}$ Fixture};
        \draw[->,thick,blue] (a)--(0,-5.6) node[midway,sloped,above] {$S^1$ compactification};
        \node (1) at (0,-0.9) {};
        \node (2) at (7.65,-8) {};
        \node (3) at (4,-11) {};
        \node (4) at (-2,-11) {};
        \node (5) at (-2,-7) {};
        \draw[caribbeangreen] \convexpath{1,2,3,4,5}{0.5cm};
    \end{tikzpicture}}
    \caption{A $\mathfrak{g}$-type 5d CM {\color{caribbeangreen}atom} SCFT $\mathcal{T}_{\mu}$ -- defined by a {\color{caribbeangreen}small} coweight $\mu$ of $\mathfrak{g}$ -- flows to a $\mathfrak{g}$-type class-$\mathcal{S}$ fixture (defined by a sphere with three regular punctures) {\color{blue}upon $S^1$ compactification}. Two punctures are maximal, and the third puncture is determined by $\mu$ via an intricate relationship: The 5d atom SCFT can be {\color{red}mass deformed to a Lagrangian} $\mathfrak{g}$-type Dynkin quiver phase with special unitary gauge nodes, $\mathsf{Q}_{\mu}$, whose gauge and flavour labels are specified by $\mu$. The unitary version of the Dynkin quiver, $\mathsf{Q}^{\mathrm{u}}_{\mu}$, has a 3d Coulomb branch, $\mathcal{C}^{\mathrm{3d}}\left(\mathsf{Q}^{\mathrm{u}}_{\mu}\right)=\overline{\mathcal{W}}^{\,\mu}_{\,0}$, which is a slice in the affine Grassmannian of $G$ (with $\mathrm{Lie}(G)=\mathfrak{g}$). As shown in \cite{achar2013geometric}, and reviewed in Section \ref{sec:class S}, if and only if $\mu$ is a {\color{caribbeangreen}small} coweight of $\mathfrak{g}$, there is a {\color{caribbeangreen}map} from $\overline{\mathcal{W}}^{\,\mu}_{\,0}$ to a nilpotent orbit closure $\overline{\mathcal{O}}$ in the nilpotent cone of $G$. This map may be an isomorphism, a discrete quotient, or a normalisation map. The nilpotent orbit $\mathcal{O}$ specifies the third puncture in the fixture.\\
    If $\mu$ is a \emph{non-small} coweight on the coroot lattice of $\mathfrak{g}$, then the part of the construction circled in {\color{caribbeangreen}green} does not exist. The corresponding 5d CM SCFT is either a hybrid or a molecule, which do not flow to a class-$\mathcal{S}$ fixture upon circle reduction.}
    \label{fig:SmallRepAtomFixture}
\end{figure}

The Lie-theoretic structure underpinning the construction of 5d CM theories bears an additional remarkable fruit. As we have mentioned, to each 5d CM theory there corresponds a IR special unitary quiver gauge theory phase encoded by $\mathsf{Q}_{\mu}$. The unitary version of this quiver, denoted by $\mathsf{Q}_{\mu}^{\mathsf{u}}$, can be interpreted as a 3d $\mathcal{N}=4$ theory, with Coulomb branch $\mathcal{C}^{3d}$. As shown in \cite{Braverman:2016pwk,Bullimore:2015lsa,Bourget:2021siw}, $\mathcal{C}^{3d}$ is a slice $\overline{\mathcal{W}}_0^{\mu}$ in the affine Grassmannian of $\mathfrak{g}$. Strikingly, if and only if $\mu$ is a small coweight, i.e.\ for 5d CM atoms, there exists a map from $\overline{\mathcal{W}}_0^{\mu}$ to a nilpotent orbit closure $\overline{\mathcal{O}}_{\mu}$, in the nilpotent cone of $\mathfrak{g}$. This map may be an isomorphism, a discrete quotient, or a normalisation map. The nilpotent orbit $\mathcal{O}_{\mu}$ specifies the third puncture in the class-$\mathcal{S}$ setup which is the circle reduction of the 5d atom. This state of affairs is schematically summarized in Figure \ref{fig:SmallRepAtomFixture}.\\

Finally, we describe the possible HB and extended CB (ECB) RG flows connecting the various theories presented in this work. In particular, this allows us to show that all the 5d theories we list can be obtained from a 6d theory, via a circle reduction followed by HB RG flow.

\subsection{Outlook}
This work opens up several promising avenues for future research. Among the most direct applications, the study of dynamical deformations of the class of non-toric threefolds introduced in this work, explored in Section \ref{sec: Higgs branch}, drives us closer to an understanding of the Higgs branch of 5d SCFTs lacking a toric description. The 5-brane-web interpretation of some of our singularities, fleshed out in detail in \cite{DeMarco:2025ugw}, strongly hints towards a bridge between geometric engineering of non-toric singularities and Type IIB descriptions. Therefore, building on the results for generalized toric polygons of \cite{Bourget:2023wlb}, and the upcoming work \cite{HiggsRGupcoming}, we anticipate that a fully geometric understanding of key features of the Higgs Branch, such as magnetic quivers, is within reach. Along the same line of thinking, it would be tempting to achieve a classification of 5d conformal matter bifundamentals for non-simply-laced flavor groups.\\
\indent On a different note, the technology we present in Section \ref{sec: CB RG flow} clearly encourages the analysis of more general 5d SCFTs, admitting a low-energy phase which is a 5d quiver with \textit{non-vanishing} Chern-Simons levels. This suggests that a systematic classification of these theories may also be viable.\\
\indent Moreover, given the low-energy quiver phases amply investigated here, we plan on examining more 5d dualities among seemingly different IR quiver theories, of the type studied in \cite{DeMarco:2023irn}.\\
\indent Finally, the approach to the exploration of the landscape of 5d SCFTs in terms of indecomposable constituents (``atoms'' and ``hybrids''), presents new opportunities for computing topological string partition functions in these threefold backgrounds, offering a natural generalization to the frameworks of \cite{Hayashi:2017jze} and \cite{Hayashi:2018iqb}.

\subsection{Outline}
We start in Section \ref{sec:atomeshybridsmolecules} by reviewing key notions of ADE algebras that serve as the organizing principle for our construction of 5d CM SCFTs. We proceed in Section \ref{sec: 5d CM geometry} by reviewing the singular Calabi-Yau threefold geometries that engineer 5d conformal matter theories, recalling how to compute the IR low-energy quiver phase from the singular threefold. In Section \ref{sec: new results} we present our main result, exhibiting the singular CY3 that engineer 5d CM theories encoded by all dominant coweights. We also lay down an algorithm to produce all 5d CM geometries from a minimal set of distinguished threefolds. In Section \ref{sec:class S} we discuss the circle reduction of 5d CM theories, showing the connection with class-$\mathcal{S}$ theories and affine Grassmannians. Section \ref{sec: Higgs branch} is devoted to the study of the Higgs Branch of 5d CM theories, which is probed by dynamical complex structure deformations of the corresponding threefold, that we compute explicitly. In Section \ref{sec: CB RG flow} we discuss extended Coulomb branch RG flows and their implications for the Chern-Simons levels of the 5d CM theories. Finally, in Section \ref{sec: 6d origin} we describe how to connect the various CM theories presented in this paper by HB and extended CB RG flows, showing  that all 5d CM theories can be obtained from 6d CM SCFTs. In appendix \ref{app:DuValres} we include background material concerning the crepant resolution of Du Val singularities in toric ambient space. In appendix \ref{appendix:proofdistinguished} we prove that our list of distinguished CY3 geometries is complete. 
In appendix \ref{sec: tables} we collect Tables that list all the 5d CM atoms and hybrids for the algebras $A_2,A_3,A_4,D_5,D_6,E_6,E_7,E_8$, as well as the exhaustive list of indecomposable coweights for the $D_n$ algebras. Finally, appendix \ref{app:acharhend} collects Hasse diagrams for symplectic leaves related to small coweights in the $D$ and $E$ algebras.

\section{Atomic, hybrid and distinguished coweights}
\label{sec:atomeshybridsmolecules}

In this section we set up our notations and introduce the notions of atomic (or small), hybrid and distinguished coweights. 

\paragraph{Coweights and Quivers. }
Let $\mathfrak{g}$ be a simple complex Lie algebra of rank $r$ and $\mathfrak{h} \subset \mathfrak{g}$ be a Cartan subalgebra.  We assume $\mathfrak{g}$ is of ADE type, or equivalently that the Dynkin diagram is simply laced. We pick once and for all a basis of simple roots $(\alpha_1 , \dots , \alpha_r)$ in such a way that the labels the nodes of the Dynkin diagram follow the conventions of LieART \cite{Feger:2019tvk}. The Cartan matrix is denoted $C_{\mathfrak{g}}$. We call $\Gamma_{\text{r}} \subset \Gamma_{\text{w}} \subset \mathfrak{h}$ the coroot and coweight lattices, and $\Lambda_{\text{r}} \subset \Lambda_{\text{w}} \subset \mathfrak{h}^\ast$ the root and weight lattices. 

The coweight lattice $\Gamma_{\text{w}}$ is a partially ordered set: we say that $\lambda \leq \lambda '$ for $\lambda , \lambda ' \in \Gamma_w$ if $\alpha_i ( \lambda ' - \lambda) \geq 0$ for all $i = 1 , \dots , r$. A coweight $\lambda \geq 0$ is called \emph{nonnegative}, and a nonnegative and nonzero coweight is called \emph{positive}. The set of nonnegative (respectively positive) coweights is called $\Gamma_{\text{w}}^{\geq 0}$ (resp. $\Gamma_{\text{w}} ^{>0}$), with similar notations with $\Gamma_{\text{r}}$. 

Now we focus on nonnegative coweights lying on the coroot lattice, i.e.\ elements of $\Gamma_{\text{r}}^{\geq 0}$. Our motivation is that balanced quivers of the ADE type of $\mathfrak{g}$ are in one-to-one correspondence with $\Gamma_{\text{r}}^{\geq 0}$. 
To any $\lambda \in \Gamma_{\text{r}}^{\geq 0}$ one associates the quiver $\mathsf{Q}_{\lambda}$ as follows: the quiver has special unitary gauge nodes forming the shape of the Dynkin diagram of $\mathfrak{g}$, flavour labels $\lambda$ and gauge labels  \footnote{The components  of the simple roots in the basis of fundamental weights are exactly the columns of the Cartan matrix. Hence, the gauge labels are the coefficients of the expansion of the coweight on the simple coroots.} 
\begin{equation}
\label{eq:gRLDWdef}
    \mathbf{r}=C_{\mathfrak{g}}^{-1}\cdot\lambda. 
\end{equation} 
For instance, if $\mathfrak g = E_6$ and $\lambda = \tiny{\begin{bmatrix}
            &&0&&\\
            1&1&0&0&0
        \end{bmatrix}}$, the quiver is 
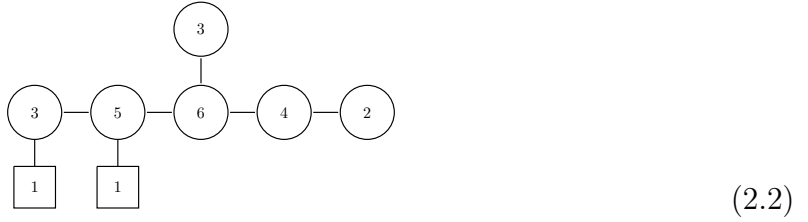
\begin{equation}\label{E6 first quiver}
     \scalebox{0.55}{
\begin{tikzpicture}
        \draw[thick] (0,0) circle (0.65);
        \node at (0,0) {\small$6$};

        \draw[thick] (0.7,0)--(1.3,0);
        \draw[thick] (2,0) circle (0.65);
        \node at (2,0) {\small$4$};

        \draw[thick] (-2.7,0)--(-3.3,0);
        \draw[thick] (-0.7,0)--(-1.3,0);

        \draw[thick] (-4,0) circle (0.65);
        \node at (-4,0) {\small$3$};

        \draw[thick] (-2,0) circle (0.65);
        \node at (-2,0) {\small$5$};

        \draw[thick] (0,0.7)--(0,1.3);
        \draw[thick] (0,2) circle (0.65);
        \node at (0,2) {\small$3$};

        \draw[thick] (2.7,0)--(3.3,0);
        \draw[thick] (4,0) circle (0.65);
        \node at (4,0) {\small$2$};

        \draw[thick] (-4.5,-1.3)--(-3.5,-1.3)--(-3.5,-2.3)--(-4.5,-2.3)--(-4.5,-1.3);
        \draw[thick] (-4,-0.65)--(-4,-1.3);
        \node at (-4,-1.8) {\small$1$};

        \draw[thick] (-2.5,-1.3)--(-1.5,-1.3)--(-1.5,-2.3)--(-2.5,-2.3)--(-2.5,-1.3);
        \draw[thick] (-2,-0.65)--(-2,-1.3);
        \node at (-2,-1.8) {\small$1$};

\end{tikzpicture}
        }
\end{equation}

\renewcommand{\arraystretch}{1.2}

 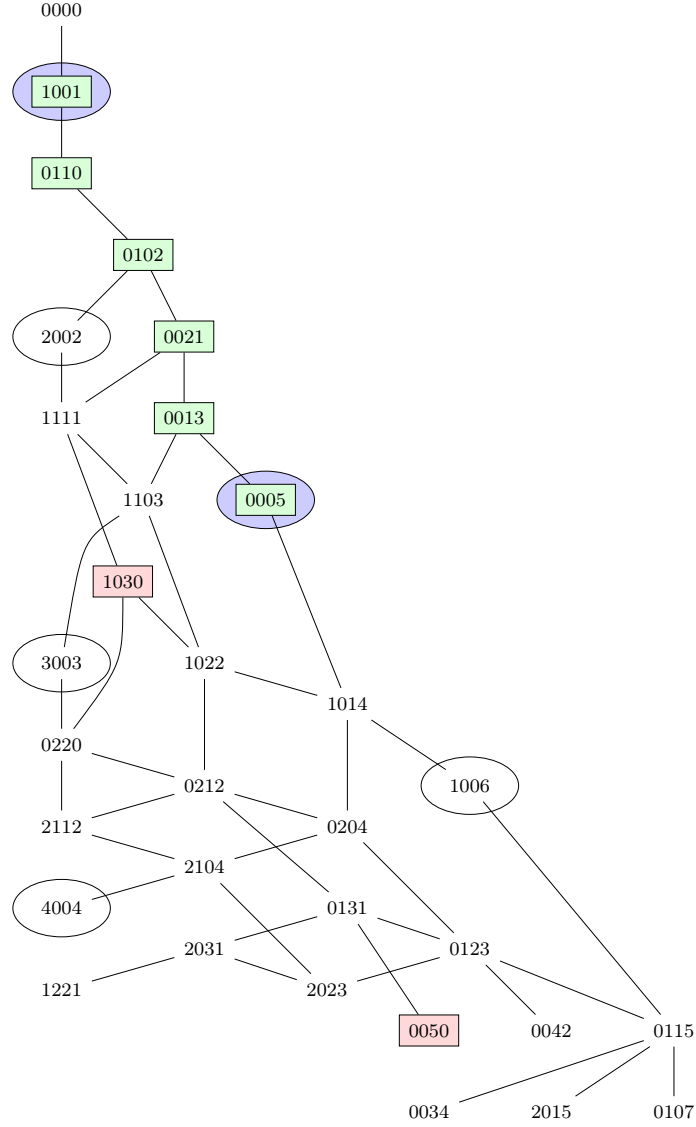
\begin{figure} 
    \centering
   \scalemath{0.9}{ \begin{tikzpicture}[xscale=0.6,yscale=0.6]
        \node (0000) at (0,0) {{\scriptsize 0000}};
        \node[draw,fill=green!15] (1001) at (0,-2) {{\scriptsize 1001}};
        \node[draw,fill=green!15] (0110) at (0,-4) {{\scriptsize 0110}};
        \node[draw,fill=green!15] (0102) at (2,-6) {{\scriptsize 0102}};
        \node (2002) at (0,-8) {{{\scriptsize 2002}}};
        \node[draw,fill=green!15] (0021) at (3,-8) {{\scriptsize 0021}};
        \node (1111) at (0,-10) {{\scriptsize 1111}};
        \node[draw,fill=green!15] (0013) at (3,-10) {{\scriptsize 0013}};
        \node[draw,fill=green!15] (0005) at (5,-12) {{{\scriptsize 0005}}};
        \node (1103) at (2,-12) {{\scriptsize 1103}};
        \node[draw,fill=red!15] (1030) at (1.5,-14) {{\scriptsize 1030}};
         \node (3003) at (0,-16) {{{\scriptsize 3003}}};
         \node (1014) at (7,-17) {{\scriptsize 1014}};
        \node (1022) at (3.5,-16) {{\scriptsize 1022}};
        \node (0220) at (0,-18) {{\scriptsize 0220}};
        \node (2112) at (0,-20) {{\scriptsize 2112}};
        \node (4004) at (0,-22) {{{\scriptsize 4004}}};
        \node (1221) at (0,-24) {{\scriptsize 1221}};
        \node (0212) at (3.5,-19) {{\scriptsize 0212}};
        \node (0204) at (7,-20) {{\scriptsize 0204}};
         \node (2104) at (3.5,-21) {{\scriptsize 2104}};
         \node (2031) at (3.5,-23) {{\scriptsize 2031}};
         \node (1006) at (10,-19) {{{\scriptsize 1006}}};
         \node (0123) at (10,-23) {{\scriptsize 0123}};
         \node (0131) at (7,-22) {{\scriptsize 0131}};
        \node (2023) at (6.5,-24) {{\scriptsize 2023}}; 
        \node[draw,fill=red!15] (0050) at (9,-25) {{\scriptsize 0050}};
        \node (0042) at (12,-25) {{\scriptsize 0042}}; 
        \node (0115) at (15,-25) {{\scriptsize 0115}}; 
        \node (0034) at (9,-27) {{\scriptsize 0034}}; 
        \node (2015) at (12,-27) {{\scriptsize 2015}}; 
        \node (0107) at (15,-27) {{\scriptsize 0107}};

\begin{pgfonlayer}{bg}
        \node[draw, ellipse, fit=(1001), fill=blue!20, inner sep=1.8pt] {};
        \node[draw, ellipse, fit=(2002), inner sep=1.8pt] {};
        \node[draw, ellipse, fit=(3003), inner sep=1.8pt] {};
        \node[draw, ellipse, fit=(4004), inner sep=1.8pt] {};
        \node[draw, ellipse, fit=(0005),fill=blue!20, inner sep=1.8pt] {};
        \node[draw, ellipse, fit=(1006), inner sep=1.8pt] {};
\end{pgfonlayer}
        \draw  (0000)--(1001);
        \draw  (1001)--(0110);
        \draw  (0110)--(0102);
        \draw  (0102)--(2002);
        \draw  (0102)--(0021);
        \draw  (0021)--(1111);
        \draw  (2002)--(1111);
        \draw  (0021)--(0013);
        \draw  (0013)--(0005);
        \draw  (0013)--(1103);
        \draw  (1111)--(1103);
        \draw  (1111)--(1030);
        \draw  (0005)--(1014);
        \draw  (1103)--(1022);
        \draw  (1030) .. controls (1.5,-16) .. (0220);
        \draw  (1103) .. controls (0.5,-13) .. (3003);
        \draw  (3003)--(0220);
        \draw  (0220)--(2112);
        \draw  (1030)--(1022);
        \draw  (1022)--(1014);
        \draw  (1022)--(0212);
        \draw  (1014)--(0204);
        \draw  (1014)--(1006);
        \draw  (0220)--(0212);
        \draw  (0212)--(0204);
        \draw  (0204)--(0123);
        \draw  (0204)--(2104);
        \draw  (2112)--(2104);
        \draw  (0212)--(0131);
        \draw  (0212)--(2112);
        \draw  (2104)--(4004);
        \draw  (2104)--(2023);
        \draw  (1022)--(0212);
        \draw  (2031)--(2023);
        \draw  (2031)--(1221);
        \draw  (0131)--(2031);
        \draw  (0131)--(0123);
        \draw  (0131)--(0050);
        \draw  (0115)--(0034);
        \draw  (0115)--(2015);
        \draw  (0115)--(0107);
        \draw  (1006)--(0115);
        \draw  (0123)--(0115);
        \draw  (0123)--(0042);
        \draw  (0123)--(2023);
    \end{tikzpicture}}
    \caption{Hasse diagram for dominant coweights of the $A_4$ algebra. We draw all coweights for which the shortest path that connects them to the trivial weight has length at most 10. For the sake of visual clarity, we draw only half of the Hasse diagram: the complete one can be obtained by adding the mirror image along the vertical axis passing through the left-right symmetric coweights. The boxed coweights are the indecomposable coweights (we show only 8, the other 6 are obtained by left-right symmetry). The green ones are atomic, the red ones are hybrid. The coweights inside ellipses are those which have only one parent. The set of such coweights is generated by the \textit{distinguished} coweights, which are painted in blue (only 2 are drawn, the third one is obtained by symmetry).}
    \label{fig:summaryWeightsA4}
\end{figure}

\begin{table}[]
\begin{center}
   \begin{tabular}{c|c|c|c|c}
    Algebra & Indecomposable & Small $\equiv$ Atomic & Hybrid & Distinguished \\ \hline 
    $A_{n}$ & A096337  &   & & 3 \\ 
    $D_{2n}$ & $\frac{n^2 + 3n}{2}$ & $2n+1$ & $\frac{(n-2)(n+1)}{2}$  & 4\\ 
    $D_{2n+1}$ & $\frac{n^2 + 7n + 4}{2}$ & $2n+2$ &  $\frac{n(n+3)}{2}$ & 6 \\ 
    $E_6$ & 14& 7 & 7 & 4\\
    $E_7$ & 10 & 5 & 5 & 4\\
    $E_8$ & 8 & 4 & 4  & 3\\
\end{tabular} 
\end{center} 
    \caption{Numbers of indecomposable, atomic and hybrid coweights for each ADE algebra. For $A_n$, we give the OEIS label \cite{oeis}, and we are not aware of a counting of the number of small coweights. The distinguished coweights are listed in Table \ref{table:generators}.    }
    \label{tab:numbers}
\end{table}

\begin{table}
\makebox[\linewidth][c]{
\centering

\resizebox{3.5cm}{!}{
$\scalemath{0.9}{
\begin{array}{|l|}
\hline
 \multicolumn{1}{|c|}{\boldsymbol{A_n}}\\
 \hline
 \hline
\hspace{1cm} \boldsymbol{ \mu_i^{\mathfrak{g}}}  \\
 \hline
\scriptsize  \begin{bmatrix}
  n+1&0 &\dots&& 0\\
\end{bmatrix}\\
\scriptsize\begin{bmatrix}
  0& &\dots&0 & n+1\\
\end{bmatrix}\\
\scriptsize\begin{bmatrix}
  1&0 &\dots&0 & 1\\
\end{bmatrix}  \\
\hline
\end{array}
 }
$ 
}

\resizebox{3.5cm}{!}{
$\scalemath{0.9}{
\begin{array}{|l|}
\hline
 \multicolumn{1}{|c|}{\boldsymbol{D_{2j}}}\\
 \hline
 \hline
 \hspace{1cm} \boldsymbol{\mu_i^{\mathfrak{g}}} \\
 \hline
\scriptsize\begin{bmatrix}
    &&&& 1\\
    1 & 0 &\ldots & 0 &\\
      &   &       &   & 1 \\
\end{bmatrix}\\
\scriptsize\begin{bmatrix}
    &&&& 0\\
    2 & 0 &\ldots & 0 &\\
      &   &       &   & 0 \\
\end{bmatrix}\\
\scriptsize \begin{bmatrix}
    &&&& 2\\
    0 & 0 &\ldots & 0 &\\
      &   &       &   & 0 \\
\end{bmatrix}\\
\scriptsize \begin{bmatrix}
    &&&& 0\\
    0 & 0 &\ldots & 0 &\\
      &   &       &   & 2 \\
\end{bmatrix}\\
\hline
\end{array}
 }
$ 
}

\resizebox{3.5cm}{!}{
$\scalemath{0.9}{
\begin{array}{|l|}
\hline
 \multicolumn{1}{|c|}{\boldsymbol{D_{2j+1}}}\\
 \hline
 \hline
\hspace{1cm}\boldsymbol{\mu_i^{\mathfrak{g}}} \\
 \hline

\scriptsize\begin{bmatrix}
    &&&& 1\\
    0 & 0 &\ldots & 0 &\\
      &   &       &   & 1 \\
\end{bmatrix}\\
\scriptsize\begin{bmatrix}
    &&&& 0\\
    2 & 0 &\ldots & 0 &\\
      &   &       &   & 0 \\
\end{bmatrix}\\
\scriptsize\begin{bmatrix}
    &&&& 0\\
    1 & 0 &\ldots & 0 &\\
      &   &       &   & 2 \\
\end{bmatrix}\\
\scriptsize\begin{bmatrix}
    &&&& 2\\
    1 & 0 &\ldots & 0 &\\
      &   &       &   & 0 \\
\end{bmatrix}\\
\scriptsize\begin{bmatrix}
    &&&& 0\\
    0 & 0 &\ldots & 0 &\\
      &   &       &   & 4 \\
\end{bmatrix}
\\
\scriptsize\begin{bmatrix}
    &&&& 4\\
    0 & 0 &\ldots & 0 &\\
      &   &       &   & 0 \\
\end{bmatrix}
\\
\hline
\end{array}
 }
$ 
}

}
\makebox[\linewidth][c]{
\centering

\resizebox{3.5cm}{!}{
$\scalemath{0.9}{
\begin{array}{|l|}
\hline
 \multicolumn{1}{|c|}{\boldsymbol{E_6}}\\
 \hline
 \hline
 \hspace{1cm} \boldsymbol{\mu_i^{\mathfrak{g}}} \\
 \hline

  \scriptsize\eSixWeight{1}{0}{0}{0}{1}{0}\\
   \scriptsize\eSixWeight{0}{0}{0}{0}{0}{1} \\
  \scriptsize\eSixWeight{3}{0}{0}{0}{0}{0}\\
  \scriptsize\eSixWeight{0}{0}{0}{0}{3}{0} \\
\hline
\end{array}
 }
$ 
}

\resizebox{3.5cm}{!}{
$\scalemath{0.9}{
\begin{array}{|l|}
\hline
 \multicolumn{1}{|c|}{\boldsymbol{E_7}}\\
 \hline
 \hline
 \hspace{1.2cm} \boldsymbol{\mu_i^{\mathfrak{g}}} \\
 \hline
 \scriptsize\eSevenWeight{0}{0}{0}{0}{0}{1}{1}  \\
 \scriptsize\eSevenWeight{0}{0}{0}{0}{0}{2}{0}   \\
 \scriptsize\eSevenWeight{1}{0}{0}{0}{0}{0}{0}  \\
 \scriptsize\eSevenWeight{0}{0}{0}{0}{0}{0}{2}  \\
\hline
\end{array}
 }
$ 
}

\resizebox{3.5cm}{!}{
$\scalemath{0.9}{
\begin{array}{|l|}
\hline
\multicolumn{1}{|c|}{\boldsymbol{E_8}}\\
\hline
\hline
 \hspace{1.4cm} \boldsymbol{\mu_i^{\mathfrak{g}}} \\
\hline
\scriptsize\eEightWeight{0}{0}{0}{0}{0}{0}{0}{1} \\
\scriptsize\eEightWeight{1}{0}{0}{0}{0}{0}{0}{0} \\
\scriptsize\eEightWeight{0}{0}{0}{0}{0}{0}{1}{0} \\
\hline
\end{array}
 }$ }}
  \caption{Distinguished coweights (which form by definition a Hilbert basis $(\mu_1 , \dots , \mu_s)$ of $\tilde{\Gamma}$) for each simple $\mathfrak{g}$. }
    \label{table:generators}
\end{table}

A coweight $\lambda \in \Gamma_{\text{r}}^{\geq 0}$ which cannot be written as $\lambda= \lambda_1 + \lambda_2$ for $\lambda_1 , \lambda_2 \in \Gamma_r^{> 0}$ is called \emph{indecomposable}. The indecomposable coweights in $\Gamma_{\text{r}}^{\geq 0}$ seen as a semigroup are its Hilbert basis; this proves that there are finitely many. Their numbers are given in the second column of Table \ref{tab:numbers}. 
An indecomposable coweight $\lambda$ is either 
\begin{compactenum}
      \item[-] \emph{atomic} if every $\mu  \in \Gamma_{\text{r}}^{> 0}$ such that $\mu \leq \lambda$ is indecomposable; 
    \item[-] \emph{hybrid} if it is not atomic.   
\end{compactenum}
It turns out that a coweight $\lambda \in \Gamma_{\text{r}}^{\geq 0}$ is atomic if and only if $\lambda$ is not larger or equal to twice the longest coroot. In the math literature, such a coweight is called a \emph{small coweight} \cite{broer1995sum}.

\paragraph{Poset structure. }
The partial order in $\Gamma_{\text{r}}^{\geq 0}$ can be represented using a Hasse diagram. Formally, this means there is a binary relation $\rightsquigarrow$ such that $\lambda \rightsquigarrow \mu$ holds if and only if $\lambda > \mu$ and there is no $\nu \in \Gamma_{\text{r}}^{\geq 0}$ such that $\lambda > \nu > \mu$. The Hasse diagram is a depiction of this relation. An example is depicted in Figure \ref{fig:summaryWeightsA4}. 

We introduce the set of non-zero coweights having \emph{only one} parent: 
\begin{equation}\label{coweights one parent}
    \tilde{\Gamma} \doteq \{ \mu \in \Gamma_{\text{r}}^{> 0} \, | \, \exists \, \textbf{!}  \, \lambda \in \Gamma_{\text{r}}^{> 0}  \, : \,  \lambda \rightsquigarrow \mu \} \, . 
\end{equation}
This is an infinite subset of $\Gamma_{\text{r}}^{ \geq 0}$. It admits a finite Hilbert basis, which we denote $(\mu_1 , \mu_2 , \dots , \mu_s )$. We tabulate these bases for each $\mathfrak{g}$ in Table \ref{table:generators}, and we  prove in \Cref{appendix:proofdistinguished} that these are indeed all the generators of \eqref{coweights one parent}. We call the coweights $(\mu_1 , \mu_2 , \dots , \mu_s )$ the \emph{distinguished coweights}; they play a fundamental role in the definition of the threefolds in Section \ref{sec: 5d CM geometry}.

\section{M-theory geometric engineering of 5d conformal matter: review}\label{sec: 5d CM geometry}
In the previous section we have reviewed the classification of dominant coweights, which can be decomposable or indecomposable. Indecomposable coweights are either atomic or hybrid. To each coweight $\mu$ there corresponds a special unitary balanced quiver gauge theory $\mathsf{Q}_{\mu}$, with gauge nodes arranged as the Dynkin diagram of $\mathfrak{g}$, with $\mathfrak{g}\in ADE$, and no Chern-Simons levels. The coweight encodes the flavor nodes, and the balancing condition uniquely fixes the gauge nodes. For a given choice of Lie algebra, one can then introduce a finite set of quiver gauge theories, in correspondence with atomic and hybrid coweights. The gauge and flavor labels of all the remaining quiver gauge theories (i.e.\ encoded by decomposable coweights) can be obtained as linear combinations of the gauge and flavor labels of atoms and hybrids. We collect all the atomic and hybrid quivers in Appendix \ref{sec: tables}, for each $\mathfrak{g}\in ADE$.
The quivers identified in such fashion correspond to a putative low-energy phase of a 5d conformal matter SCFT with flavor algebra at least $\mathfrak{g}\oplus\mathfrak{g}$, thanks to the argument by \cite{Yonekura:2015ksa}. It remains to be checked whether the 5d fixed points actually exist. 

In this Section we review the construction of the Calabi-Yau threefold geometries corresponding to 5d CM SCFTs. The singular geometric phase corresponds to the SCFT point, whereas performing a (partial) resolution of the singularity amounts to a motion along the extended Coulomb branch of the theory: in \cref{sec:5dCMthreefolds} we introduce the singular threefolds that engineer 5d conformal matter theories. In Section \ref{sec: special unitary} we exhibit a preferred choice of resolution of these threefolds, such that the corresponding 5d SCFT flows precisely to the 5d quivers of the type reviewed in Section \ref{sec:atomeshybridsmolecules}.

\subsection{5d conformal matter threefolds}
\label{sec:5dCMthreefolds}
All the 5d low-energy quiver phases identified in Section \ref{sec:atomeshybridsmolecules} flow from  a putative 5d SCFT with flavor symmetry \textit{at least} $\mathfrak{g}\oplus\mathfrak{g}$, thanks to the symmetry-enhancement argument introduced in \cite{Tachikawa}, and recalled in \cite{Yonekura:2015ksa}. According to the well-established M-theoretic dictionary, that translates the geometric features of a canonical CY3 geometry\footnote{As first established by \cite{Xie:2017pfl}, we require the threefold to be canonical,  so that the singular phase can be reached from its smoothing at finite distance in the K\"ahler moduli space. Notice that all Calabi-Yau threefolds are, by definition, also canonical.} into 5d SCFT data, manifest flavor symmetry factors arise from non-compact singular lines, that necessarily support Du Val singularities. For our purposes, we thus wish to construct a CY3 geometry that displays at least two non-compact singular lines, each supporting a Du Val singularity of type $\mathfrak{g}\in ADE$. In this Section we review the technique introduced in \cite{DeMarco:2023irn},  and generalize it. 

The Du Val singularities $\mathsf{Y}_{\mathfrak{g}}$ are defined as hypersurfaces in $\mathbb{C}^3$ by the following polynomials $Y_{\mathfrak{g}}$:\footnote{Notice that the presentation of the $A_k$ singularities in this work differs from what was employed in \cite{DeMarco:2023irn}, i.e. $\widetilde{x}^2+\tilde{y}^2=z^{k+1}$. A linear redefinition of the variables 
\begin{equation*}
    \widetilde{x} \rightarrow ix-\frac{iy}{4},\quad \tilde{y} \rightarrow x+\frac{y}{4},
\end{equation*}
takes from the form of \cite{DeMarco:2023irn} to the one in this work.}
\begin{equation}\label{ADE sing}
 \begin{cases}
Y_{A_k}(x,y,z) = xy+z^{k+1}, \\
Y_{D_k}(x,y,z) = x^2 +zy^2+z^{k-1}, \\
Y_{E_6}(x,y,z) =x^2+y^3+z^4, \\
Y_{E_7}(x,y,z) = x^2+y^3+yz^3, \\
Y_{E_8}(x,y,z) = x^2+y^3+z^5.
\end{cases}
\end{equation}
We call $\widetilde{\mathsf{Y}}_{\mathfrak{g}}$ the standard resolution of $\mathsf{Y}_{\mathfrak{g}}$. The preimage of the origin of $\mathsf{Y}_{\mathfrak{g}}$ under the resolution map is a collection of rank$(\mathfrak{g})$ $\mathbb{P}^1$'s, arranged as the $\mathfrak{g}$ Dynkin diagram. In general, by standard Du Val theory, we have 
\begin{equation}
\label{eq:rootlatticeascompacthomology}
    H_{2}(\mathsf{Y}_{\mathfrak{g}},\mathbb Z) =  \Lambda_{\text{r}},\qquad  H_{2}(\mathsf{Y}_{\mathfrak{g}},\partial \mathsf{Y}_{\mathfrak{g}}) = \Lambda_{\text{w}}. 
\end{equation}
and since the closed cohomology (resp. cohomology) is the dual of the homology with compact support (resp. of relative homology), we have\footnote{Other Lie-theoretic quantities are reproduced by the geometry:  It is well-known that minus the intersection form on the $H_2(Y,\mathbb Z)$ is the Cartan matrix, hence the Poincar\'e duality is identified with the map 
\begin{equation*}
    \alpha \in H_2(Y,\mathbb Z) \cong \Lambda_r \to \omega_{\alpha} = -2\frac{\alpha^a H_a}{\alpha^2} \in \Gamma_r \cong H^2_{\text{compact}}(Y,\mathbb Z).
\end{equation*}
} 
\begin{equation}
\label{eq:coweightlatticeasrelativehomology}
H^{2}(\mathsf{Y}_{\mathfrak{g}},\mathbb Z) =  \Gamma_{\text{w}}, \qquad H^{2}_{\text{compact}}(\mathsf{Y}_{\mathfrak{g}},\mathbb Z) =  \Gamma_{\text{r}}.
\end{equation}
We also have the isomorphism\footnote{We recall that, for a given variety $X$, the Picard group $\text{Pic}(X) = \text{Div}(X)/\text{Prin(X)}$, with $\text{Prin}(X)$ the principal divisors of $X$, is the group that captures the isomorphism classes of line bundles on $X$.} $H^2(\mathsf{Y}_{\mathfrak{g}},\mathbb Z) \cong \text{Pic}(\mathsf{Y}_{\mathfrak{g}})$, that associates to each line bundle its first Chern class.  In particular, to each element of the  coroot lattice corresponds one line bundle over $\mathsf{Y}_{\mathfrak{g}}$. We will expand on this identification between line bundles on the Du Val singularity and the coweights in \Cref{sec:genericgeomertryCSAn}.\\

Consider first the non-compact canonical threefold $\mathbb{C}\times \mathsf{Y}_{\mathfrak{g}}$. 
The fibration of the exceptional $\mathbb{P}^1$'s along the non-compact line spanned by the $\mathbb{C}$ factor is a set of non-compact 4-cycles. Reducing the M-theory 3-form on the Poincaré duals of these 4-cycles provides the Cartan subalgebra of the background vector fields corresponding to a $\mathfrak{g}$ flavor symmetry factor. These are non-dynamical vector fields since the volume of the 4-cycle is infinite. 

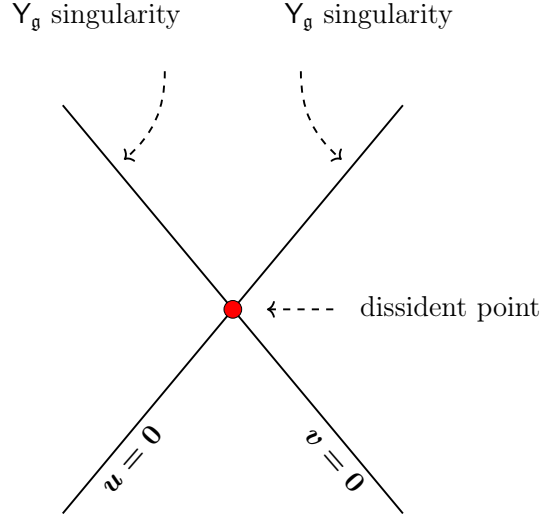
\begin{figure}[H]
\centering
 \scalebox{0.9}{
    \begin{tikzpicture}
        \draw[thick] (-2.5,-3)--(2.5,3);
        \draw[thick] (-2.5,3)--(2.5,-3);

        \draw[thick,dashed,->] (-1,3.5) to [out=270, in = 40] (-1.6,2.2);
        \draw[thick,dashed,->] (1,3.5) to [out=270, in = 140] (1.6,2.2);
        \draw[thick,dashed,<-] (0.5,0)--(1.5,0);
        \draw[fill=red] (0,0) circle (0.13);
        \node at (-2.0,4.3) {$\mathsf{Y}_{\mathfrak{g}} \text{ singularity}$};
    
        \node at (2,4.3) {$\mathsf{Y}_{\mathfrak{g}} \text{ singularity}$};
       
          \node at (3.2,0) {$\text{dissident point}$};
       
        \node[rotate=49] at (-1.5,-2.2) {$\boldsymbol{u=0}$};
        \node[rotate=-49] at (1.5,-2.2) {$\boldsymbol{v=0}$};
        \end{tikzpicture}}
    \caption{Schematic representation of the singular locus of the threefolds \eqref{CY3 CM}.}
     \label{fig:CY3 CM}
\end{figure}

We can swiftly obtain a natural candidate for a 5d bifundamental CM SCFT by transversally intersecting two non-compact lines supporting a Du Val singularity of type $\mathfrak{g}$. Away from the intersection point, the CY3 locally looks like $\mathbb{C}\times \mathsf{Y}_{\mathfrak{g}}$. The intersection point is a \textit{dissident point} \cite{reid1980canonical}, where the singularity enhances, and hence the threefold cannot be locally written as a Cartesian product. The specific features of the enhanced singularity specifies the data of the 5d SCFT supported in the directions transverse to the CY3.
Thanks to the presence of the two non-compact lines of Du Val singularity of type $\mathfrak{g}$, we expect to detect at least a $\mathfrak{g}\times\mathfrak{g}$ flavor symmetry in the SCFT. This statement holds provided the resulting singularity is canonical. The philosophy underlying the construction we have just outlined is visualized in Figure \ref{fig:CY3 CM}.\\

Concretely, an ansatz for a CY3 with the sought-after features to realize 5d CM SCFTs is 
\begin{equation}\label{CY3 CM}
  X : \quad  \begin{cases}
       Y_{\mathfrak{g}}(x,y,z) =  0\\
        uv = w(x,y,z)\\
    \end{cases} ,
\end{equation}
with $w(x,y,z)$ a polynomial. We wish to prove that the threefold $X$ is Calabi-Yau. 

To show that \eqref{CY3 CM} has trivial canonical bundle, one can proceed as follows: we first consider a partially-resolved phase of \eqref{CY3 CM} (that we review in detail in Section \ref{sec: special unitary}), and we notice that the complete intersection defining the threefold is a section of the anticanonical bundle of a toric ambient space. This, by adjunction, is enough to ensure that the canonical bundle of the partially resolved  threefold is trivial. Then, we go back to the singular phase, inferring the triviality of the canonical bundle of \eqref{CY3 CM} from the fact that the blow-down map is crepant. 

Having settled this point, notice that \eqref{CY3 CM} is singular along the lines
\begin{equation}
    x=y=z =u = 0,\quad\quad x=y=z = v = 0.
\end{equation}
The fibers along the singular lines are Du Val singularities of type $\mathfrak{g}$.  
Consequently, we expect the singularities \eqref{CY3 CM} to engineer a 5d SCFT with at least $\mathfrak{g}\oplus\mathfrak{g}$ flavor symmetry. Finally, the intersection between the two lines, $x=y=z = u=v = 0$, is a point where the singularity enhances. In the next section we lay down the recipe to perform our preferred choice of resolution, showing the appropriate phase in the extended CB that coincides with the balanced quivers $\mathsf{Q}_{\mu}$.

\subsection{Special unitary quiver phase of 5d CM}\label{sec: special unitary}

We wish to present a preferred choice of crepant resolution of the singular threefolds \eqref{CY3 CM}: physically, it corresponds to turning on mass deformations that produce an IR Lagrangian 5d gauge theory, with special unitary gauge nodes.

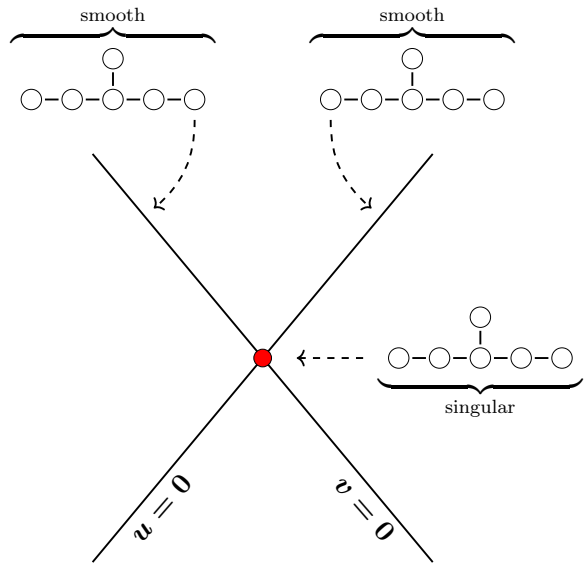
\begin{figure}[H]
\centering
 \scalebox{0.9}{
    \begin{tikzpicture}
        \draw[thick] (-2.5,-3)--(2.5,3);
        \draw[thick] (-2.5,3)--(2.5,-3);

        \draw[thick,dashed,->] (-1,3.5) to [out=270, in = 40] (-1.6,2.2);
        \draw[thick,dashed,->] (1,3.5) to [out=270, in = 140] (1.6,2.2);
        \draw[thick,dashed,<-] (0.5,0)--(1.5,0);
        \draw[fill=red] (0,0) circle (0.13);
        \node at (-2.2,4.9) {$\overbrace{\hspace{3cm}}^{\text{smooth}}$};
        \draw (-3.4,3.8) circle (0.15);
        \draw[thick] (-3.2,3.8)--(-3.0,3.8);
        \draw (-2.8,3.8) circle (0.15);
        \draw[thick] (-2.6,3.8)--(-2.4,3.8);
        \draw (-2.2,3.8) circle (0.15);
        \draw[thick] (-2,3.8)--(-1.8,3.8);
        \draw (-1.6,3.8) circle (0.15);
        \draw[thick] (-1.4,3.8)--(-1.2,3.8);
        \draw (-1,3.8) circle (0.15);
        \draw[thick] (-2.2,4.0)--(-2.2,4.2);
        \draw (-2.2,4.4) circle (0.15);
        \node at (2.2,4.9) {$\overbrace{\hspace{3cm}}^{\text{smooth}}$};
        \draw (3.4,3.8) circle (0.15);
        \draw[thick] (3.2,3.8)--(3.0,3.8);
        \draw (2.8,3.8) circle (0.15);
        \draw[thick] (2.6,3.8)--(2.4,3.8);
        \draw (2.2,3.8) circle (0.15);
        \draw[thick] (2,3.8)--(1.8,3.8);
        \draw (1.6,3.8) circle (0.15);
        \draw[thick] (1.4,3.8)--(1.2,3.8);
        \draw (1,3.8) circle (0.15);
        \draw[thick] (2.2,4.0)--(2.2,4.2);
        \draw (2.2,4.4) circle (0.15);
          \node at (3.2,-0.6) {$\underbrace{\hspace{3cm}}_{\text{singular}}$};
        \draw (4.4,0) circle (0.15);
        \draw[thick] (4.2,0)--(4,0);
        \draw (3.8,0) circle (0.15);
        \draw[thick] (3.6,0)--(3.4,0);
        \draw (3.2,0) circle (0.15);
        \draw[thick] (3,0)--(2.8,0);
        \draw (2.6,0) circle (0.15);
        \draw[thick] (2.4,0)--(2.2,0);
        \draw (2,0) circle (0.15);
        \draw[thick] (3.2,0.2)--(3.2,0.4);
        \draw (3.2,0.6) circle (0.15);
        \node[rotate=49] at (-1.5,-2.2) {$\boldsymbol{u=0}$};
        \node[rotate=-49] at (1.5,-2.2) {$\boldsymbol{v=0}$};
        \end{tikzpicture}}
    \caption{Partially resolved phase of \eqref{CY3 CM}, for the case $\mathfrak{g}=E_6$.}
     \label{fig:CY3 CM partial}
\end{figure}

The resolution consists of two steps:
\begin{equation}
    \widetilde{\widetilde{X}} \xrightarrow{\epsilon} \widetilde{X} \xrightarrow{\pi} X.
\end{equation}
The map $\pi$ is nothing but the usual resolution of a Du Val singularity. It acts as:
\begin{equation}\label{pi resolution}
    \pi: \hspace{0.2cm} \left(x(d_i),y(d_i),z(d_i),u,v\right) \longrightarrow (x,y,z,u,v),
\end{equation}
where the $d_i, i=1,\ldots,k$ are coordinates in a $k$-dimensional toric ambient space subjected to $k-4$ suitable $\mathbb{C}^*$-actions.
The partially resolved threefold $\widetilde{X}$ hence is 
\begin{equation}\label{partial resolution}
    \widetilde{X} : \quad \begin{cases}
    \widetilde{Y}_{\mathfrak{g}}(x,y,z)\big|_{x = x(d_i),y = y(d_i),z = z(d_i)} = 0,\\
        uv = w(x,y,z) \big|_{x = x(d_i),y = y(d_i),z = z(d_i)},
    \end{cases}
\end{equation}
where we denote by $\widetilde{Y}_{\mathfrak{g}}$ the proper transform of $Y$ (we refer to Appendix \ref{app:DuValres} for more details on the resolution procedure of Du Val singularities). 
Expression \eqref{partial resolution} is a polynomial in the variables $d_i$, and it engineers a IR Lagrangian quiver phase with special unitary nodes. This can be seen as follows: the coordinates $(u,v)$ admit a $\mathbb{C}^*$-action of the form:
\begin{equation}
\label{eq:cstarreductionfirst}
  (u,v) \rightarrow (\lambda u, \lambda^{-1}v), \quad \lambda \in \mathbb{C}^*.  
\end{equation}
The dimensional reduction on the $S^1 \subset \mathbb{C}^*$ repackages the M-theory setup into a Type IIA description in terms of D6-branes. The D6-branes lie exactly on the locus $\Delta_{D6}$ where the $\mathbb{C}^*$-action degenerates, namely on $uv = 0$, \textit{up to the first equation of \eqref{eq:cstarreductionfirst}}.
Hence we can rewrite \eqref{partial resolution} to emphasize the brane locus $\Delta(d_i)$:
\begin{equation}\label{partial resolution 2}
    \widetilde{X}: \quad \begin{cases}
    \widetilde{Y}_{\mathfrak{g}}(x,y,z)\big|_{x = x(d_i),y = y(d_i),z = z(d_i)} = 0,\\
        uv = \Delta(d_i).
    \end{cases}
\end{equation}
The partially resolved phase $\widetilde{X}$ in \eqref{partial resolution 2} hence describes a collection of $\mathbb{P}^1_j$'s, $j=1,\ldots \text{rank}(\mathfrak{g})$, arranged as the $\mathfrak{g}$ Dynkin diagram (this is the outcome of the resolution map $\pi$ in \eqref{pi resolution}). Generically, the $\mathbb{P}^1_j$'s support a singularity of type $A_{r_j-1}$, which means that $r_j$ D6-branes wrap said $\mathbb{P}^1$. The $r_j$'s are given by the vanishing order of $\Delta(d_i)=0$ on the curve $\mathbb{P}^1_j$, up to the first equation of \eqref{partial resolution 2}. In physical terms, the phase $\widetilde{X}$ is therefore describing a low-energy special unitary quiver description of the 5d SCFT engineered by M-theory on $X$. The gauge nodes of the quiver correspond to the $\mathbb{P}^1_j$, and the dimension vectors of the flavor and gauge symmetries are dictated by the choice of $w(x,y,z)$ and hence by the brane locus\footnote{We can rephrase the correspondence between partially resolved threefold and IR quiver phase in terms of commutative algebra. Denote by $C_j$ the prime ideal corresponding to the curve $\mathbb{P}^1_j$, and define the ring $R\equiv \mathbb{C}[d_1,\ldots d_k]/\tilde{Y}$. The ideals $C_j$ are explicitly collected in Appendix \ref{app:DuValres}, for various ADE algebras. We wish to evaluate the vanishing order of $\Delta_{D6}(d_i)$ on the curves related to $C_j$, in the ring $R$. To this end, one can take the localization of $R$ on the curve $C_j$, denoted by $R_{C_j}$. $R_{C_j}$ possesses a unique maximal ideal $M_j\equiv C_jR_{C_j}$, which is the extension of $C_j$ to $R_{C_j}$. We then define the vanishing order of $\Delta_{D6}(d_i)$ on the curve $\mathbb{P}^1_j$ as the maximum positive integer $p_j$ such that $\Delta_{D6}(d_i) \in M_j^{p_j}$. This construction can be neatly translated into code for a computer algebra software. As a result, one can directly draw the quiver corresponding to $\Delta_{D6}(d_i)$, where the nodes corresponding to $C_j$ are labeled by the integers $p_j$. See the ancillary Macaulay2 code for an implementation. \label{footnote ideals}}. We schematically represent the resolved phase $\widetilde{X}$ in Figure \ref{fig:CY3 CM partial}. The two non-compact lines of type $\mathfrak{g}$ have been completely resolved, except for their intersection point, where a set of $\mathbb{P}^1$'s arranged as the Dynkin diagram of $\mathfrak{g}$ still supports singularities of the threefold. In a patch that captures the geometry near the intersection between the $j$-th and $j+1$-th $\mathbb{P}^1$'s, the threefold looks like:
\begin{equation}
    uv = a_i^{r_j}b_i^{r_{j+1}},
\end{equation}
where $(u,v,a_i,b_i)$ are the coordinates on the patch, and $r_j$, $r_{j+1}$ are the D6-brane multiplicities introduced above.
The map $\epsilon$ takes care of resolving these leftover singularities, giving rise to a set of compact and non-compact divisors, as dictated by the Lagrangian quiver. This step completely resolves all the singularities. For further details, we refer the reader to \cite{DeMarco:2023irn}, where this operation is fully spelled out.

\section{M-theory geometric engineering of 5d conformal matter: new results}\label{sec: new results}
In this Section we lay down the core new result of this work, building on the material reviewed in Section \ref{sec: 5d CM geometry}. Namely, we explicitly construct the singular CY3 corresponding to any given balanced special unitary Lagrangian quiver $\mathsf{Q}_{\mu}$, which is specified by a dominant coweight $\mu$ in the appropriate Lie algebra. \textit{Physically, this proves that all quivers $\mathsf{Q}_{\mu}$ admit a UV completion in terms of a 5d CM SCFT}. We recall that we have collected $\mathsf{Q}_{\mu}$ for atomic and hybrid coweights in Appendix \ref{sec: tables}.\\
\indent We start in Section \ref{sec:coweighttothreefold} by writing down the singular threefolds associated to the distinguished coweights (for each $\mathfrak g$). Then, in Section \ref{sec:formal algorithm} we present the algorithm to explicitly write down, starting from the geometries related to distinguished coweights, and without performing any resolution or blowup, the threefold associated with any dominant coweight $\mu$. We then show it at work in a concrete example in Section \ref{sec:algorithm}.

\subsection{Distinguished coweights and singular CY3}
\label{sec:coweighttothreefold}

Consider the singular CY3 \eqref{CY3 CM}. 
As we have outlined, we wish to identify the expressions $w(x,y,z)$ that produce the singular CY3's corresponding (in an appropriate chamber of the extended CB of the associated 5d SCFT) to the special unitary quivers encoded by all dominant coweights, that can be retrieved from Appendix \ref{sec: tables}.\\
In order to construct such CY3's, we only need to identify the singular geometries related to a \textit{finite} set of coweights, that we have called \textit{distinguished} in Section \ref{sec:atomeshybridsmolecules}. These act as elementary generators for all the other singular geometries.

In Table \ref{table:generators2} we list the singular geometries corresponding to the distinguished coweights $\mu_i^{\mathfrak g}$ for all $\mathfrak{g}\in ADE$, encoded by a choice of $w(x,y,z) = w_{i}^{\mathfrak g}$. Employing the preferred resolution reviewed in Section \ref{sec: special unitary}, it can be explicitly checked that the geometries defined by the $w_{i}^{\mathfrak g}$ admit a IR phase which is precisely $\mathsf{Q}_{\mu_i^{\mathfrak g}}$\footnote{A more pictorial rationale underpinning the explicit form of the singular geometries in Table \ref{table:generators2} can be garnered from the point of view of the resolution of Du Val singularities. Indeed, the $w_i^{\mathfrak{g}}$ correspond to monomials (or polynomials) that define the ideals of the exceptional curves of the corresponding resolved Du Val singularity. These can be found in \eqref{curves A},\eqref{curves D5},\eqref{curves D6},\eqref{curves E6},\eqref{curves E7},\eqref{curves E8}, and reproduce the generators $w_i^{\mathfrak{g}}$ once appropriately pulled back to the coordinates on the singular phase via the blowdown maps \eqref{An blowdown},\eqref{D5 blowdown},\eqref{D6 blowdown},\eqref{E6 blowdown},\eqref{E7 blowdown},\eqref{E8 blowdown}. Alternatively, we can proceed even faster as follows. We first identify the three weights corresponding to $x$, $y$, $z$.  Then we look at the number of fundamental weights needed to write them. If one needs two, we look at the ADE equation and "factorize" accordingly. For instance, for $E_6$, we find the weight associated to $y$ is the sum of two fundamentals. Since $-y^3 = x^2 + z^4 = (x+iz^2)(x-iz^2)$, we need to add two more weights, corresponding to three times the fundamental weights appearing in $y$, and with those polynomials. }. 

\renewcommand{\arraystretch}{1.2}
\begin{table}
\makebox[\linewidth][c]{
\centering

\resizebox{5cm}{!}{
$\scalemath{0.9}{
\begin{array}{|l|l|}
\hline
 \multicolumn{2}{|c|}{\boldsymbol{A_n}}\\
 \hline
 \hline
\hspace{1cm} \boldsymbol{ \mu_i^{\mathfrak{g}}} & \hspace{0.5cm} \boldsymbol{w^{\mathfrak{g}}_i} \\
 \hline
\scriptsize  \begin{bmatrix}
  n+1&0 &\dots&& 0\\
\end{bmatrix} &    w^{A_n}_1 = x \\
\scriptsize\begin{bmatrix}
  0& &\dots&0 & n+1\\
\end{bmatrix} &    w^{A_n}_2 = y \\
\scriptsize\begin{bmatrix}
  1&0 &\dots&0 & 1\\
\end{bmatrix} &    w^{A_n}_3 = z \\
\hline
\end{array}
 }
$ 
}

\resizebox{5cm}{!}{
$\scalemath{0.9}{
\begin{array}{|l|l|}
\hline
 \multicolumn{2}{|c|}{\boldsymbol{D_{2j}}}\\
 \hline
 \hline
 \hspace{1cm} \boldsymbol{\mu_i^{\mathfrak{g}}} & \hspace{1cm} \boldsymbol{w^{\mathfrak{g}}_i} \\
 \hline
\scriptsize\begin{bmatrix}
    &&&& 1\\
    1 & 0 &\ldots & 0 &\\
      &   &       &   & 1 \\
\end{bmatrix}&  w^{D_{2j}}_1 = x \\
\scriptsize\begin{bmatrix}
    &&&& 0\\
    2 & 0 &\ldots & 0 &\\
      &   &       &   & 0 \\
\end{bmatrix} &     w^{D_{2j}}_2 = z \\
\scriptsize \begin{bmatrix}
    &&&& 2\\
    0 & 0 &\ldots & 0 &\\
      &   &       &   & 0 \\
\end{bmatrix}&      w^{D_{2j}}_3 =y+ i z^{j-1}\\
\scriptsize \begin{bmatrix}
    &&&& 0\\
    0 & 0 &\ldots & 0 &\\
      &   &       &   & 2 \\
\end{bmatrix}&      w^{D_{2j}}_4 =y- i z^{j-1}\\
\hline
\end{array}
 }
$ 
}

\resizebox{5cm}{!}{
$\scalemath{0.9}{
\begin{array}{|l|l|}
\hline
 \multicolumn{2}{|c|}{\boldsymbol{D_{2j+1}}}\\
 \hline
 \hline
\hspace{1cm}\boldsymbol{\mu_i^{\mathfrak{g}}} & \hspace{1cm} \boldsymbol{w^{\mathfrak{g}}_i}\\
 \hline

\scriptsize\begin{bmatrix}
    &&&& 1\\
    0 & 0 &\ldots & 0 &\\
      &   &       &   & 1 \\
\end{bmatrix}&     w^{D_{2j+1}}_1 = y \\
\scriptsize\begin{bmatrix}
    &&&& 0\\
    2 & 0 &\ldots & 0 &\\
      &   &       &   & 0 \\
\end{bmatrix}&     w^{D_{2j+1}}_2 = z \\
\scriptsize\begin{bmatrix}
    &&&& 0\\
    1 & 0 &\ldots & 0 &\\
      &   &       &   & 2 \\
\end{bmatrix}&      w^{D_{2j+1}}_3 =x-i z^j\\
\scriptsize\begin{bmatrix}
    &&&& 2\\
    1 & 0 &\ldots & 0 &\\
      &   &       &   & 0 \\
\end{bmatrix}&      w^{D_{2j+1}}_4 =x+ i z^j\\
\scriptsize\begin{bmatrix}
    &&&& 0\\
    0 & 0 &\ldots & 0 &\\
      &   &       &   & 4 \\
\end{bmatrix}
&      w^{D_{2j+1}}_5 =y^2-2iz(x-iz^j)\\
\scriptsize\begin{bmatrix}
    &&&& 4\\
    0 & 0 &\ldots & 0 &\\
      &   &       &   & 0 \\
\end{bmatrix}
&      w^{D_{2j+1}}_6 =y^2-2iz(x+iz^j)\\
\hline
\end{array}
 }
$ 
}

}
\makebox[\linewidth][c]{
\centering

\resizebox{5cm}{!}{
$\scalemath{0.9}{
\begin{array}{|l|l|}
\hline
 \multicolumn{2}{|c|}{\boldsymbol{E_6}}\\
 \hline
 \hline
 \hspace{1cm} \boldsymbol{\mu_i^{\mathfrak{g}}} & \hspace{1cm} \boldsymbol{w^{\mathfrak{g}}_i} \\
 \hline
  \scriptsize\eSixWeight{1}{0}{0}{0}{1}{0} &   w^{E_6}_1 = y \\
   \scriptsize\eSixWeight{0}{0}{0}{0}{0}{1} &   w^{E_6}_2 = z \\
  \scriptsize\eSixWeight{3}{0}{0}{0}{0}{0}  &   w^{E_6}_3=x+ i z^2\\
  \scriptsize\eSixWeight{0}{0}{0}{0}{3}{0}  &   w^{E_6}_4=x- i z^2\\
\hline
\end{array}
 }
$ 
}

\resizebox{5cm}{!}{
$\scalemath{0.9}{
\begin{array}{|l|l|}
\hline
 \multicolumn{2}{|c|}{\boldsymbol{E_7}}\\
 \hline
 \hline
 \hspace{1.2cm} \boldsymbol{\mu_i^{\mathfrak{g}}} & \hspace{1cm} \boldsymbol{w^{\mathfrak{g}}_i} \\
 \hline
 \scriptsize\eSevenWeight{0}{0}{0}{0}{0}{1}{1} &     w^{E_7}_1 = x \\
 \scriptsize\eSevenWeight{0}{0}{0}{0}{0}{2}{0}  &    w^{E_7}_2 = y \\
 \scriptsize\eSevenWeight{1}{0}{0}{0}{0}{0}{0}  &    w^{E_7}_3 = z \\
 \scriptsize\eSevenWeight{0}{0}{0}{0}{0}{0}{2}  &   w^{E_7}_4=y^2+z^3\\
\hline
\end{array}
 }
$ 
}

\resizebox{5cm}{!}{
$\scalemath{0.9}{
\begin{array}{|l|l|}
\hline
\multicolumn{2}{|c|}{\boldsymbol{E_8}}\\
\hline
\hline
 \hspace{1.4cm} \boldsymbol{\mu_i^{\mathfrak{g}}} & \hspace{0.4cm} \boldsymbol{w^{\mathfrak{g}}_i} \\
\hline
\scriptsize\eEightWeight{0}{0}{0}{0}{0}{0}{0}{1} &  w^{E_8}_1 = x \\
\scriptsize\eEightWeight{1}{0}{0}{0}{0}{0}{0}{0} &    w^{E_8}_2 = y \\
\scriptsize\eEightWeight{0}{0}{0}{0}{0}{0}{1}{0}&    w^{E_8}_3 = z \\
\hline
\end{array}
 }$ }}
  \caption{Distinguished coweights (which form by definition a Hilbert basis $(\mu_1 , \dots , \mu_s)$ of $\tilde{\Gamma}$ introduced in Section \ref{sec:atomeshybridsmolecules}) for each simple $\mathfrak{g}$ and the corresponding polynomials $(w_1 , \dots , w_s)$. }
    \label{table:generators2}
\end{table}

Given the finite set of geometries in Table \ref{table:generators2}, our algorithm predicts $w(x,y,z)$, and therefore the singular CY3, for all the (infinite) remaining set of dominant coweights.\\
\indent As we have illustrated via Lie-theoretic arguments in the preceding sections, the 5d SCFTs corresponding to dominant coweights are organized into three classes:
\begin{itemize}
    \item \textit{atoms}, whose low-energy Lagrangian quiver phase is encoded by a \textit{small dominant coweight};
    \item \textit{hybrids}, whose low-energy Lagrangian quiver phase is encoded by a \textit{lattice generating (but not small) dominant coweight};
    \item \textit{molecules}, whose low-energy Lagrangian quiver phase is encoded by a \textit{dominant coweight that is not lattice generating}.
\end{itemize}
This structure can be sharply detected also via M-theory geometric engineering on singular CY3. In particular\footnote{In the following classification, we assume that $w(x,y,z)$ has been stripped of all the monomials $m_i(x,y,z)$ that do not affect the D6-brane multiplicities $r_j$ introduced in Section \ref{sec: special unitary}. Namely, we define $w'(x,y,z) = w(x,y,z)-\sum_im_i$, which has the same brane multiplicities as $w(x,y,z)$. Of course the set of such $m_i$'s could be empty. The monomials $m_i$ can be interpreted as irrelevant deformations of a ``core'' SCFT engineered by $w'(x,y,z)$.}:
\begin{itemize}
    \item $w(x,y,z)$ in \eqref{CY3 CM} has at least one linear monomial, and it is not factorizable in terms of at least two non-trivial polynomials $ \longrightarrow$  the geometry \eqref{CY3 CM} engineers a 5d CM \textit{atom}.
      \item $w(x,y,z)$ in \eqref{CY3 CM} has no linear monomials, and it is \textit{not} factorizable in terms of at least two non-trivial polynomials $ \longrightarrow$  the geometry \eqref{CY3 CM} engineers a 5d CM \textit{hybrid}. 
    \item $w(x,y,z)$ in \eqref{CY3 CM} has no linear monomials, but it is factorizable in terms of at least two non-trivial polynomials $ \longrightarrow$ the geometry \eqref{CY3 CM} engineers a 5d CM \textit{molecule}.
\end{itemize}
Notice that in the case of molecules, each factor appearing in $w(x,y,z)$ corresponds to an atom (if the factor contains at least a linear monomial) or a hybrid (if the factor contains \textit{no} linear monomials). This is the geometric footprint of the fact that 5d CM molecules are obtained as gaugings of 5d CM atoms and hybrids. Namely, given two 5d CM SCFTs defined as:
\begin{equation}\label{gauging addends}
    \begin{cases}
        Y_{\mathfrak{g}}(x,y,z) = 0\\
        uv = w_1(x,y,z)\\
    \end{cases},
    \quad\quad
    \begin{cases}
        Y_{\mathfrak{g}}(x,y,z) = 0\\
        uv = w_2(x,y,z)\\
    \end{cases},
\end{equation}
the 5d SCFT produced by gauging a common $\mathfrak{g}$ flavor factor is geometrically encoded by the singular threefold:
\begin{equation}\label{molecule}
    \begin{cases}
        Y_{\mathfrak{g}}(x,y,z) = 0\\
        uv = w_1(x,y,z)w_2(x,y,z)\\
    \end{cases}.
\end{equation}
The IR balanced quiver gauge theory phase corresponding to \eqref{molecule} is then obtained by summing the occupation vectors for the gauge and flavor nodes of the two starting 5d CM SCFTs in \eqref{gauging addends}.
For more details on the gauging operation we refer to \cite{DeMarco:2023irn}.
Of course this procedure can be iterated to produce molecules that are gaugings of all possible combinations of atoms and hybrids: in this way the whole space of dominant coweights explored in Section \ref{sec:atomeshybridsmolecules} is reproduced from the M-theoretic point of view.\\

We now turn to illustrating the algorithm to compute the singular CY3 for all dominant coweights in terms of Lie-algebraic data.

\subsection{Algorithm relating dominant coweights and singular threefolds}\label{sec:formal algorithm}

\paragraph{From the coweight to the threefold} Consider the ring $\mathbb{C}[x,y,z]$. To each $\mu_i^{\mathfrak{g}}$ in the Hilbert basis of $\tilde{\Gamma}$ (namely, to each distinguished coweight) we have associated a polynomial $w_i^{\mathfrak{g}} \in \mathbb C[x,y,z]$, as explained in \Cref{sec:coweighttothreefold}. From now on we drop the superscript $\mathfrak{g}$, for the sake of visual ease. Then for each $\mu \in \tilde{\Gamma}$, we associate an ideal $I_\mu$ of $\mathbb C[x,y,z]$ as follows:
\begin{equation}
    I_\mu = (w_1^{n_1} \cdots w_s^{n_s}) + (w_1^{n'_1} \cdots w_s^{n'_s}) + \dots
\end{equation}
where $\mu = n_1 \mu_1 + \dots + n_s \mu_s = n'_1 \mu_1 + \dots + n'_s \mu_s = \dots $ are all the decompositions of $\mu$ in terms of the distinguished coweights.  
Consider now an arbitrary dominant coweight $\lambda \in \Gamma_{\text{r}}^{ \geq 0}$. The set 
\begin{equation}\label{red weights algorithm}
    M_\lambda \doteq \min \{ \mu \in \tilde{\Gamma} | \mu \geq \lambda \}
\end{equation}
is \emph{finite} -- but it can contain several coweights. Consider the ideal 
\begin{equation}\label{red ideal algorithm}
    I_{\lambda} = \sum\limits_{\mu \in M_{\lambda}} I_\mu \, 
\end{equation}
then, the threefold associated with the coweight $\lambda$ is \eqref{CY3 CM}, with $w \in I_{\lambda}$.

\paragraph{From the threefold to the coweight}

To each polynomial $P \in R$, one can associate $\lambda (P) \in \Gamma_{\text{r}}^{ \geq 0}$ by\footnote{In principle, the maximum might not be unique; however we have observed that it is unique in all cases considered in this paper. It would be nice to have a formal proof of this uniqueness. } 
\begin{equation}
    \lambda (P) \doteq \max \{ \lambda \in \Gamma_{\text{r}}^{ \geq 0} | P \in \widetilde{I}_\lambda \}, \quad \text{with: } \widetilde{I}_\lambda =   \sum_{\mu \geq \lambda} I_{\mu}.
\end{equation}

We expect that the ideal $\widetilde{I}_{\lambda}$ can be identified with the ideal $I_{\lambda}$ in \eqref{red ideal algorithm} by passing to the appropriate localization of the ambient space coordinate ring\footnote{Indeed, an element $\widetilde{p}$ of $\widetilde{I}_{\lambda}$ can be written as 
\begin{equation*}
   \widetilde{p} = p + \epsilon, 
\end{equation*}
with $p \in I_{\lambda}$ and $\epsilon$ a term that is subleading with respect to $p$ in an appropriate analytic open neighbourhood of all the $\mathbb P^1$s resolving $\mathrm{Y}_{\mathfrak g}$. This means that if we zoom-in on the exceptional locus, by including in the ambient space ring all the power-series that are smooth in such neighbourhood, the ratio $u =  \frac{\widetilde{p}}{p}$ is a unit of the ambient space coordinate ring, and hence any ideal containing $p$ also contains $\widetilde{p}$.}.

\begin{proposition}\label{prop}
For each $P$, the geometry
\begin{equation}\label{CY3 CM intro}
    \begin{cases}
       Y_{\mathfrak{g}}(x,y,z) =  0\\
        uv = P(x,y,z)\\
    \end{cases},
\end{equation}
corresponds to the quiver $\mathsf{Q}_{\lambda (P)}$.
\end{proposition} 
Physically, proposition \ref{prop} shows that there exists a canonical CY3 associated to each quiver $\mathsf{Q}_{\lambda(P)}$ encoded by a dominant coweight. Therefore, \textit{all quiver gauge theories $\mathsf{Q}_{\lambda(P)}$ admit a UV completion which is a 5d conformal matter SCFT, engineered by M-theory on \eqref{CY3 CM intro}.}

\subsubsection{Example: $E_6$ conformal matter}
\label{sec:algorithm}

We now present an explicit application of the algorithm reviewed in the previous Section.
Pick $\mathfrak g = E_6$ and $\lambda = \tiny{\begin{bmatrix}
            &&0&&\\
            1&1&0&0&0
        \end{bmatrix}}$.
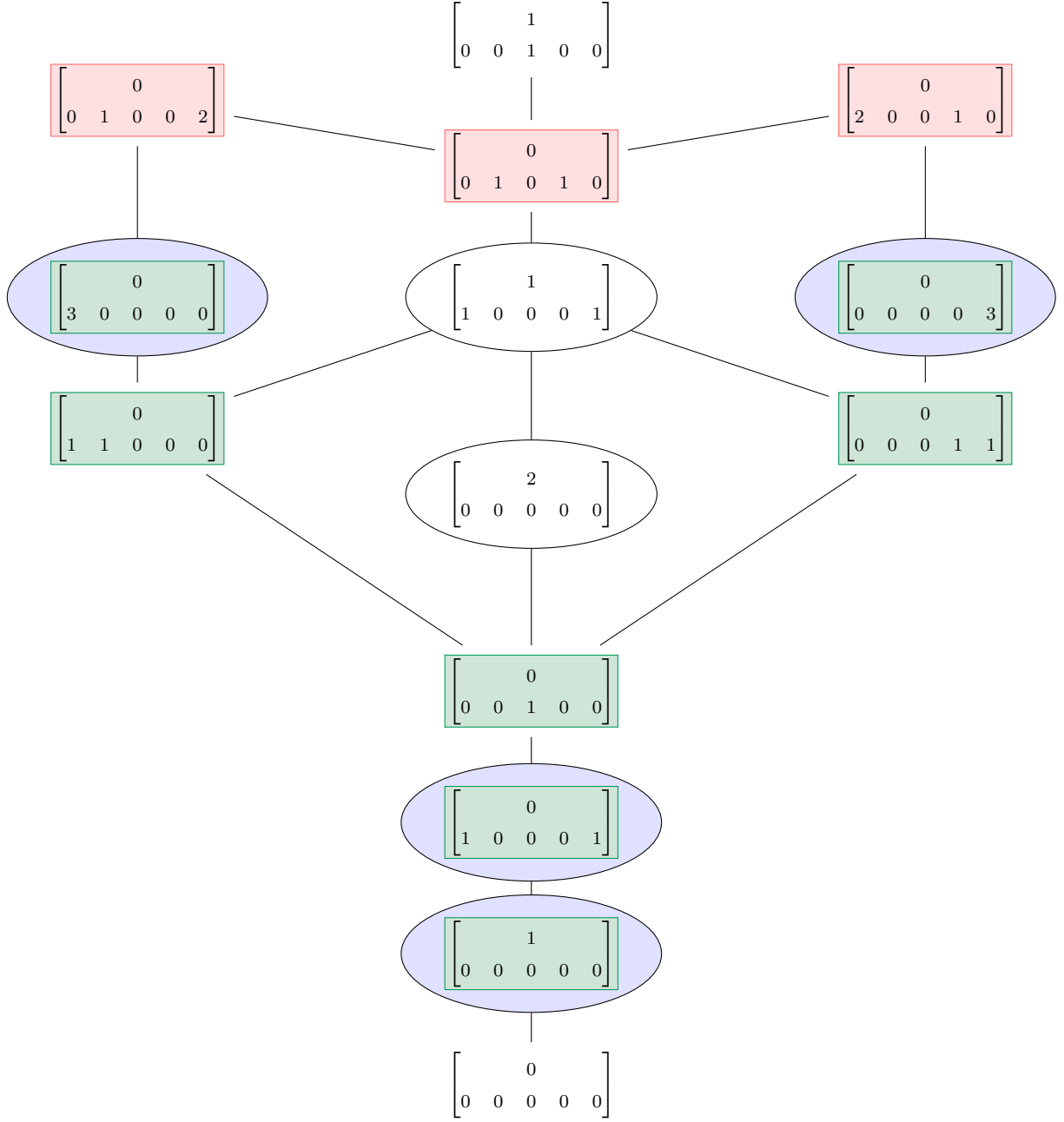
\begin{figure}
    \centering
 \begin{tikzpicture}[
  circFill/.style={draw=black, fill=blue!12, ellipse, inner sep=2.2pt},
  circNoFill/.style={draw=black, ellipse, inner sep=2.2pt},
  innerSqGreen/.style={draw=ForestGreen, fill=ForestGreen!18, rectangle, inner sep=1.2pt},
  innerSqPink/.style={draw=red!60, fill=red!12, rectangle, inner sep=1.2pt},
]

 \node (n0) at (0,0) {\scriptsize$\begin{bmatrix}
      &&0&&\\
      0&0&0&0&0
  \end{bmatrix}$};

  \node[circFill] (n1) at (0,2) {\tikz[baseline]{\node[innerSqGreen]{\scriptsize$\begin{bmatrix}
      &&1&&\\
      0&0&0&0&0
  \end{bmatrix}$};}};
  \node[circFill] (n2) at (0,4) {\tikz[baseline]{\node[innerSqGreen]{\scriptsize$\begin{bmatrix}
      &&0&&\\
      1&0&0&0&1
  \end{bmatrix}$};}};

  \node (n3) at (0,6) {\tikz[baseline]{\node[innerSqGreen]{\scriptsize$\begin{bmatrix}
      &&0&&\\
      0&0&1&0&0
  \end{bmatrix}$};}};

  \node[circNoFill] (n4) at (0,9) {\scriptsize$\begin{bmatrix}
      &&2&&\\
      0&0&0&0&0
  \end{bmatrix}$};
  \node[circNoFill] (n5) at (0,12) {\scriptsize$\begin{bmatrix}
      &&1&&\\
      1&0&0&0&1
  \end{bmatrix}$};

  \node (n4a) at (-6,10) {\tikz[baseline]{\node[innerSqGreen]{\scriptsize$\begin{bmatrix}
      &&0&&\\
      1&1&0&0&0
  \end{bmatrix}$};}}; 

  \node[circFill] (n5a) at (-6,12) {\tikz[baseline]{\node[innerSqGreen]{\scriptsize$\begin{bmatrix}
      &&0&&\\
      3&0&0&0&0
  \end{bmatrix}$};}};

  \node (n4b) at (6,10) {\tikz[baseline]{\node[innerSqGreen]{\scriptsize$\begin{bmatrix}
      &&0&&\\
      0&0&0&1&1
  \end{bmatrix}$};}};

  \node[circFill] (n5b) at (6,12) {\tikz[baseline]{\node[innerSqGreen]{\scriptsize$\begin{bmatrix}
      &&0&&\\
      0&0&0&0&3
  \end{bmatrix}$};}};

  \node (n6a) at (-6,15) {\tikz[baseline]{\node[innerSqPink]{\scriptsize$\begin{bmatrix}
      &&0&&\\
      0&1&0&0&2
  \end{bmatrix}$};}};
  \node (n6) at (0,14) {\tikz[baseline]{\node[innerSqPink]{\scriptsize$\begin{bmatrix}
      &&0&&\\
      0&1&0&1&0
  \end{bmatrix}$};}};
  \node (n6b) at (6,15) {\tikz[baseline]{\node[innerSqPink]{\scriptsize$\begin{bmatrix}
      &&0&&\\
      2&0&0&1&0
  \end{bmatrix}$};}};

  \node (n7) at (0,16) {\scriptsize$\begin{bmatrix}
      &&1&&\\
      0&0&1&0&0
  \end{bmatrix}$};

  \draw (n0) -- (n1);
  \draw (n1) -- (n2);
  \draw (n2) -- (n3);

  \draw (n3) -- (n4a);
  \draw (n4a) -- (n5a);

  \draw (n3) -- (n4b);
  \draw (n4b) -- (n5b);

  \draw (n3) -- (n4);
  \draw (n4) -- (n5);

  \draw (n5) -- (n4a);
  \draw (n4b) -- (n5);

  \draw (n5) -- (n6);
  \draw (n5a) -- (n6a);
  \draw (n5b) -- (n6b);

  \draw (n6) -- (n6a);
  \draw (n6) -- (n6b);
  \draw (n6) -- (n7);

\end{tikzpicture}
      \caption{Lowest portion of the poset of root lattice dominant coweights for the $E_6$ Lie algebra. We follow the conventions for shapes and colors of \Cref{fig:summaryWeightsA4}. We are interested in constructing the threefold associated with the coweight $\tiny{\begin{bmatrix}
            &&0&&\\
            1&1&0&0&0
        \end{bmatrix}}$, that appears at the bottom of the leftmost column. }
      \label{fig:redverteces}
\end{figure}
 In Figure \ref{fig:redverteces}, we organized the root lattice dominant coweights of $E_6$ according to a Hasse diagram, with edges indicating that a certain coweight can be reached from the other by adding a certain positive root. It is immediate to glean from Figure \ref{fig:redverteces} that there are two coweights in the set $M_{\lambda}$ defined in \eqref{red weights algorithm}, given our choice of $\lambda$. Such coweights are: 
 \begin{equation}
     \mu_{1} = \scalemath{0.8}{\begin{bmatrix}
            &&0&&\\
            3&0&0&0&0
        \end{bmatrix}}, \quad \mu_{2} = \scalemath{0.8}{\begin{bmatrix}
            &&1&&\\
            1&0&0&0&1
        \end{bmatrix}}
 \end{equation} 
        The monomials associated to them according to \Cref{table:generators} are: 
        \begin{equation}
            \label{eq:monomialse6example}
            w_{\mu_{1}} = w_3^{E_6} = x + i z^2, \qquad  w_{\mu_{2}} = w_1^{E_6} w_2^{E_6} = y z, 
        \end{equation}
        where, to construct the second monomial, we used that
        \begin{equation}
            \label{eq:additionwieghtse6example}
            \mu_{2} = 
            \scalemath{0.8}{\begin{bmatrix}
            &&0&&\\
            1&0&0&0&1
        \end{bmatrix} }+ \scalemath{0.8}{\begin{bmatrix}
            &&1&&\\
            0&0&0&0&0
        \end{bmatrix}},
        \end{equation}
        and hence $w_{\mu_{2}} = w_1^{E_6} w_2^{E_6}$. 
        
        Then, the threefold associated with $\mu = \tiny{\begin{bmatrix}
            &&0&&\\
            1&1&0&0&0
        \end{bmatrix}}$ is, employing \eqref{red ideal algorithm}, 
        \begin{equation}
            \label{eq:threefoldexample}
            \begin{cases}
            x^2 + y^3 + z^4 = 0,\\
            u v = c_1 w_{\mu_{1}} +   c_2 w_{\mu_{2}} = c_1(x + i z^2) + c_2 yz, \end{cases}
        \end{equation}
        with $c_1$ and $c_2$ arbitrary coefficients.
        Notice that other irrelevant terms can be added to the second equation of \eqref{eq:threefoldexample} keeping the singularity (and the associated 5d SCFT) unchanged. As an example of what can be added in such fashion, note that also the monomial associated with
        $\mu_3 = \tiny{\begin{bmatrix}
            &&1&&\\
            0&0&1&0&0
        \end{bmatrix}}$ can be added to \eqref{eq:threefoldexample} but it is not necessary to construct the threefold. Indeed, $\mu_3$ satisfies $\mu_3 > \lambda$, and so it can be included in the r.h.s.\ of the second equation of \eqref{eq:threefoldexample}. However, since it is not part of the set $M_{\lambda}$ (as $\mu_2\leq \mu_3$) it is not necessary to include the associated monomial once $w_{2}$ is present. 

In the remainder of this Section, we will construct explicitly the resolution of the considered threefold, we will reconstruct the low-energy quiver $\mathsf{Q}_{\mu}$, and we will prove that adding the monomial of the $\mu_3$ to the second equation of \eqref{eq:threefoldexample} does not change the theory.

From a IIA perspective, the addition of the other terms, denoted by the dots in \eqref{eq:threefoldexample} just changes the detailed shape of the non-compact D6 branes appearing in the IIA limit of M-theory on \eqref{eq:threefoldexample}, without changing how they intersect the other curves of the $E_6$ singularity appearing in the IIA limit. Changing the shape of the non-compact D6 branes, without changing their intersection properties, can be seen as the analog of a Hanany-Witten move in the IIA setup \cite{Bourget:2023wlb}. Consequently, the Dynkin quiver phase, and so its associated UV 5d SCFT, is exclusively dictated by the coweights in \eqref{red weights algorithm}.\footnote{Even among the set \eqref{red weights algorithm} there can appear monomials which do not affect the geometry. The clarity of the presentation of the algorithm benefits from ignoring this fact, not affecting our conclusions.} To visualize this concretely, let us add the aforementioned monomial
        \begin{equation}
        \label{eq:subleadintt}
            w_{\mu_{3}} = x z, \qquad \mu_{3} = \small{\begin{bmatrix}
            &&1&&\\
            0&0&1&0&0
        \end{bmatrix}}
        \end{equation}
         in the explicit example \eqref{eq:threefoldexample}: 
             \begin{equation}
            \label{eq:threefoldexampleredundant}
            \begin{cases}
            x^2 + y^3 + z^4 = 0,\\
            u v = c_1 w_{\mu_{1}} +   c_2 w_{\mu_{2}} + \boxed{c_3 w_{\mu_{3}}}= c_1(x + i z^2) + c_2 yz + \boxed{c_3 xz},
            \end{cases}
        \end{equation}
where we highlighted in a box the new, subleading term \eqref{eq:subleadintt}.
        
We then resolve the first equation of \eqref{eq:threefoldexampleredundant}, by introducing the following maps (shown in full generality in Appendix \ref{app:DuValres}):
\begin{equation}
\label{Eq:blowupe6example}
     x = d_1 d_4 d_5^2 d_6^4 d_7^6,\quad
  y = d_2 d_4 d_5^2 d_6^3 d_7^4, \quad
  z= d_3 d_4 d_5 d_6^2 d_7^3,
\end{equation}
that describe the partially resolved threefold \eqref{eq:threefoldexampleredundant} as the following hypersurface in toric ambient space:
\begin{equation}
\label{eq:e6resolved}
\scalemath{0.95}{
\begin{cases}
    d_1^2+d_4d_5^2d_6d_2^3+d_4^2d_3^4 = 0 \\
    u v = \left(c_1 (x + i z^2) + c_2 yz + \boxed{c_3 xz}\right)\big{\rvert}_{\eqref{Eq:blowupe6example}}
\end{cases} \quad \subset \quad 
\renewcommand{\arraystretch}{1.3}
    \begin{array}{ccccccccc}
    d_1   & d_2  & d_3 & 
    d_4  & d_5 & d_6 & d_7  &u&v \\
    \hline
     1 &  1 & 1   & -1 & 0 & 0 & 0 & 0& 0 \\
    1 &  1 & 0  & 1 & -1 & 0 & 0& 0& 0 \\
    1 & 0 & 0  & 1 & 1 & -1 & 0 & 0& 0\\
    1 &  0 & 0  & 1 & 0 & 1 & -1 & 0& 0\\
    \end{array}
    }
\end{equation}
where we are substituting $x,y,z$ with their expressions in \eqref{Eq:blowupe6example}. 

Reducing to IIA along the $\mathbb C^*$ acting on $u,v$ with opposite charges: 
\begin{equation}
    u \to \xi u, \qquad v \to \frac{v}{\xi},
\end{equation}
we see that the D6 brane locus is 
\begin{equation}
    \label{eq:branelocuse6example}
    \left(c_1 (x + i z^2) + c_2 yz + \boxed{c_3 xz}\right)\big{\rvert}_{\eqref{Eq:blowupe6example}} = d_4 d_5^2 d_6^4 d_7^6 \left[d_3 d_4 d_5 d_6 d_7 \left(c_2 d_2 + \boxed{c_3 d_1 d_6 d_7^2}\right) + c_1 \left(d_1 + i d_3^2 d_4\right)\right],
\end{equation}
inside the following ambient space: 
        \begin{equation}
\label{eq:e6surfacemaintext}
    d_1^2+d_4d_5^2d_6d_2^3+d_4^2d_3^4 = 0 
    \qquad \subset \qquad 
\renewcommand{\arraystretch}{1.3}
    \begin{array}{ccccccc}
    d_1   & d_2  & d_3 & 
    d_4  & d_5 & d_6 & d_7  \\
    \hline
    1 &  1 & 1   & -1 & 0 & 0 & 0  \\
    1 &  1 & 0  & 1 & -1 & 0 & 0 \\
   1 & 0 & 0  & 1 & 1 & -1 & 0\\
   1 &  0 & 0  & 1 & 0 & 1 & -1\\
    \end{array}
\end{equation}
that models the \textit{resolved} $E_6$ surface singularity \cite{Collinucci:2020jqd}.

Let us locate the D6 branes that appear in \eqref{eq:branelocuse6example}. The curves of \eqref{eq:e6surfacemaintext} are
\begin{equation}\label{eq:curvesE6maintext}
\begin{array}{l}
    \mathcal{C}_1 =(d_5,d_1+id_4d_3^2) \\
    \mathcal{C}_2 =(d_6,d_1+id_4d_3^2) \\
    \mathcal{C}_3 =(d_7)\\
    \mathcal{C}_4 = (d_6,d_1-id_4d_3^2)\\ 
    \mathcal{C}_5 =(d_5,d_1-id_4d_3^2) \\
    \mathcal{C}_6 =(d_4,d_1) \\
\end{array} \quad\quad
    \scalebox{0.65}{\begin{tikzpicture}
        \draw[thick] (0,0) circle (0.65);
        \node at (0,0) {\small$\mathcal{C}_3$};
        \draw[thick] (0.7,0)--(1.3,0);
        \draw[thick] (2,0) circle (0.65);
        \node at (2,0) {\small$\mathcal{C}_4$};
        \draw[thick] (-2.7,0)--(-3.3,0);
        \draw[thick] (-0.7,0)--(-1.3,0);
        \draw[thick] (-4,0) circle (0.65);
        \node at (-4,0) {\small$\mathcal{C}_1$};
        \draw[thick] (-2,0) circle (0.65);
        \node at (-2,0) {\small$\mathcal{C}_2$};
        \draw[thick] (0,0.7)--(0,1.3);
        \draw[thick] (0,2) circle (0.65);
        \node at (0,2) {\small$\mathcal{C}_6$};
        \draw[thick] (2.7,0)--(3.3,0);
        \draw[thick] (4,0) circle (0.65);
        \node at (4,0) {\small$\mathcal{C}_5$};
        \end{tikzpicture}}
\end{equation}
and one can explicitly check that, \textit{for each choice of $c_3 \in \mathbb C$}, \eqref{eq:branelocuse6example} has a zero of order 
\begin{equation}
    \label{eq:gaugevector}
    r_j  = C_{\mathfrak g}^{-1}\cdot \small{\begin{bmatrix}
            &&0&&\\
            1&1&0&0&0
        \end{bmatrix}} = (3,5,6,4,2,3)_j
\end{equation}
on the respective curve in \eqref{eq:curvesE6maintext} (employing the notation of Section \ref{sec: special unitary}). There is also a zero of order one on the zero-locus of the factor in square brackets in \eqref{eq:branelocuse6example}. This zero locus intersect just $\mathcal C_1$ and $\mathcal C_2$, as expected for the coweight $\tiny{\begin{bmatrix}
            &&0&&\\
            1&1&0&0&0
        \end{bmatrix}}$. Such vanishing orders coincide with the number of D6 branes wrapped on the corresponding compact\footnote{We recall that the vanishing orders on the right-hand side of \eqref{eq:threefoldexample} are those associated with compact curves, and reproduce the gauge labels. The vanishing orders on the non-compact curves reproduce the flavor labels.} and non-compact curves, and reproduce correctly the gauge and flavor labels of $\tiny{\begin{bmatrix}
            &&0&&\\
            1&1&0&0&0
        \end{bmatrix}}$. Since the result we obtained is independent of the choice of $c_3$, we conclude that its value does not change the physics of M-theory on \eqref{eq:threefoldexample}.

\section{$S^1$-compactification to 4d and Class-$\mathcal{S}$}\label{sec:class S}

It is argued in \cite{DeMarco:2025ugw} that $\mathfrak{g}$-type 5d CM atoms can be directly compactified to $\mathfrak{g}$-type 4d $\mathcal{N}=2$ class-$\mathcal{S}$ fixtures (with all punctures assumed to be regular), but that 5d CM molecules cannot. In \cite{DeMarco:2025ugw} not all 5d CM atoms were known, and 5d CM hybrids were not explored. Nevertheless, the statement that only 5d CM atoms can directly be compactified to 4d $\mathcal{N}=2$ class-$\mathcal{S}$ fixtures seems to hold, as we show in this section.

We make the following claim, summarised in Figure \ref{fig:SmallRepAtomFixture}:
\begin{enumerate}
    \item A small coweight $\mu$ of $\mathfrak{g}$ determining a 5d CM atom $\mathcal{T}_{\mu}$ via a low-energy Dynkin quiver $\mathsf{Q}_{\mu}$ also determines a nilpotent orbit $\mathcal{O}$ of $\mathfrak{g}$, such that the $\mathfrak{g}$-type class-$\mathcal{S}$ fixture with two maximal punctures and an $\mathcal{O}$ puncture is the circle compactification of $\mathcal{T}_{\mu}$.
    \item The nilpotent orbit $\mathcal{O}$ is obtained from the small coweight $\mu$ via the work of Achar and Henderson \cite{achar2013geometric}, see Figures \ref{fig:AHforD5}-\ref{fig:AHforE8} in Appendix \ref{app:acharhend}.\footnote{These figures are adopted from \cite[Table 6]{achar2013geometric}; supplemented with elementary slice data computed using \cite{2003math......5095M,Braverman:2016pwk,Bourget:2021siw} for leaves in the affine Grassmannian, and taken from \cite{fu2017generic} for exceptional nilpotent orbits.} The coweight $\mu$ (being dominant and lying on the coroot lattice) defines an affine Grassmannian slice $\overline{\mathcal{W}}^{\,\mu}_{\,0}$, which incidentally is the 3d Coulomb branch of $\mathsf{Q}^{\mathrm{u}}_{\mu}$ \cite{Braverman:2016pwk}. For $\mu$ a small coweight, there is some nilpotent orbit $\mathcal{O}$, such that there exists a map \cite{achar2013geometric}
    \begin{equation}
        \overline{\mathcal{W}}^{\,\mu}_{\,0}\rightarrow \overline{\mathcal{O}}
    \end{equation}
    which is either an isomorphism, cover, or normalisation. In the case of a cover, the nilpotent orbit is the top one in the special piece.
\end{enumerate}
For coweights other than small ones, there is no map from the associated affine Grassmannian slice to a nilpotent orbit. Hence our method to produce the class-$\mathcal{S}$ fixture from an atom cannot be extended to hybrids or molecules.

While molecules made purely from atoms have a class-$\mathcal{S}$ description after turning on certain mass deformations in 5d first, as explained in \cite{DeMarco:2025ugw}, we do not know how to relate hybrids (and molecules involving hybrids) to class-$\mathcal{S}$ at all. We leave this question for future work.

We conclude this section with a comment. We note that the conformal matter atoms are exactly those threefolds for which $w_{\mu}$  can be chosen as a polynomial containing at least a linear term in the coordinates of the $\mathbb C^3_{x,y,z}$ inside which we write $Y_{\mathfrak g}$. In particular, this means that the second equation of \eqref{CY3 CM intro} can be eliminated, and the threefold is a hypersurface of $\mathbb C^4$. We can formulate this statement in a way that is  coordinate independent, saying that the threefolds corresponding to atoms are the only ones in this work whose minimal embedding dimension\footnote{Given an affine threefold singularity, the minimal embedding dimension is the smallest $N$ such that the threefold can be defined as the zero-locus of a set of equations in $\mathbb C^N$.} is four, while we observe that for the threefolds of this work that do not correspond to atoms, the minimal embedding dimension is five in all checked examples.\footnote{The tangent space to a complete intersection at a point is the orthogonal space to the span of the gradients of the equations. At singular points, where the rank of the matrix whose rows are the gradient of the equations of the variety is non-maximal, the dimension of the tangent space (that now becomes the normal cone of the singular cone) increases, and equals the dimension of the ambient space minus the rank of such matrix. A necessary condition to embed a threefold singularity in $\mathbb C^4$ is that the dimension of the tangent cone is at most four. We have checked that this holds true for atoms, while for molecules and hybrids the normal cone at the maximal singular point has dimension 5.} It is interesting to notice that class-$\mathcal{S}$ theories with just regular punctures can be realized \cite{Xie:2015rpa} as IIB string theory on a non-isolated singularity with minimal embedding dimension four. Consequently,   such minimal embedding dimension seems to be preserved passing from the M-theory realization of the 5d SCFT to the IIB construction of its KK reduction.

\section{Higgs branch and dynamical complex deformations of 5d Conformal Matter}\label{sec: Higgs branch}

In this Section we translate the Hasse diagrams relating dominant coweights, including the portions shown in detail in Figures \ref{fig:AHforD5}--\ref{fig:AHforE8}, into geometric and physical language. As we have shown, to each point in the Hasse diagram there corresponds a singular CY3, specified by \eqref{CY3 CM}. The Hasse diagram encodes an ordering "$>$" of the weights, and if $\mu_0 \rightsquigarrow \mu_1$,\footnote{We redirect the reader to \Cref{sec:atomeshybridsmolecules} for the definitions of $>$ and $\rightsquigarrow$.} then we draw an oriented edge from $\mu_0$ to $\mu_1$. An oriented edge connecting two nodes corresponds to a RG flow relating two 5d SCFTs, triggered by a Higgs branch deformation of the singular geometry dual to the starting node. Hence, given some parent theory corresponding to $\mu_0$, one can recover all the geometries corresponding to the descendant theories encoded by $\mu_i$, with $\mu_0 > \mu_i$, by turning on a suitable deformation in the theory $\mu_0$.\\
\indent Recall that, according to the geometric engineering dictionary, normalizable complex structure deformations of the threefold are in relation with Higgs Branch modes of the corresponding 5d SCFT. In general, proving that a deformation is normalizable is an intricate task (see e.g.\ \cite{Acharya:2024bnt} for recent progress in some 5d SCFTs), that we do not address directly in this work. We rather employ an effective definition of ``dynamical'' complex structure deformations of a Calabi-Yau threefold, introduced in \cite{Collinucci:2020jqd} and grounded on a physical argument. We leave a direct proof of their normalizability for the future\footnote{In general, normalizable means that the corresponding kinetic term is not infinite, and hence the deformation is dynamical. This can be seen by thinking about the large volume limit of the CY, where we can use 11d supergravity to distinguish dynamical and non-dynamical modes.}. We further show that the ``dynamical'' deformations studied in this work precisely correspond to Higgs Branch modes, thanks to their relation with the 5d low-energy quiver phases and the class-$\mathcal{S}$ construction obtained after dimensional reduction. We refer to \cite{Chacaltana:2010ks,Chacaltana:2011ze,Chacaltana:2012zy,Chacaltana:2012ch,Chacaltana:2013oka,Chacaltana:2014jba,Chacaltana:2015bna,Chacaltana:2016shw,Chacaltana:2017boe,Chacaltana:2018vhp} for the full details concerning the class-$\mathcal{S}$ fixtures.\\

In general, the dynamical deformations we have just outlined can be written as
\begin{equation}\label{general equation}
    \begin{cases}
        Y_{\mathfrak{g}}(x,y,z) = 0\\
        uv = w(x,y,z) + \text{def}(x,y,z)
    \end{cases},
\end{equation}
where "def" is a polynomial in $(x,y,z)$, with arbitrary coefficients, each valued in $\mathbb{C}$. "def" depends on $\mathfrak{g}$ and $w(x,y,z)$, and we will return to its specific shape momentarily. 
There are $\nu$ independent deformations in $\text{def}(x,y,z)$ each contributing one quaternionic dimension to the HB of the 5d SCFT engineered by \eqref{CY3 CM}. Note that $\nu$, as we will  show in \Cref{sec:E6RGflowexample}, does not in general correspond to the number of monomials appearing on the right-hand side of \eqref{general equation}. The total UV dimension of the Higgs branch is
\begin{equation}\label{UV HB resolution}
    \text{dim}(HB)_{UV} = \nu + \text{dim}(\mathfrak{g}).
\end{equation}
As we have argued in \cite{DeMarco:2025ugw}, we expect that the deformations yielding the $\text{dim}(\mathfrak{g})$ contribution arise as deformations of the first equation in \eqref{general equation}. These trigger RG flows that lead \textit{outside} the Hasse diagrams explored in this work. For this reason, we leave their detailed analysis for future exploration. Conversely, the deformations in def$(x,y,z)$ correspond to the (partial) closure of the puncture $\mathcal{O}$ in the class-$\mathcal{S}$ construction reviewed in Section \ref{sec:class S}. For the case of atoms, expression \eqref{UV HB resolution} precisely coincides with the Higgs Branch dimension predicted from the class-$\mathcal{S}$ point of view \cite{Chacaltana:2012zy}: $\text{dim}(\mathfrak{g})$ is supplied by the two maximal punctures, and $\nu$ by the third puncture. 

In the remainder of this Section, we show how to explicitly compute the dynamical complex structure deformations $\text{def}(x,y,z)$.\\

As the singularities in the CY3 geometries at hand are along non-compact complex lines, deformation theory is not rigorously defined. Nonetheless, heavily relying on the tools developed in \cite{Collinucci:2020jqd}, we are able to identify the (finite number of) genuine Higgs branch deformations, reducing our setups to a Type IIA setting.

Let us summarize the argument of \cite{Collinucci:2020jqd}, applying it to our general setting: the procedure kicks off by performing the resolution \eqref{pi resolution}. On top of each exceptional $\mathbb{P}^1$, a singularity of type $A$ is still present. This resolution data yields a low-energy special unitary quiver description of the 5d SCFT engineered by the singular threefold \eqref{CY3 CM}. In accordance to the previous sections, we denote the quiver by $\mathsf{Q}_{\mu}$, where $\mu$ is the coweight related to the singular threefold. We recall that the partially resolved phase reads:
\begin{equation}\label{partial resolution 3}
    \widetilde{X}: \quad \begin{cases}
    \widetilde{Y}_{\mathfrak{g}}(x,y,z)\big|_{x = x(d_i),y = y(d_i),z = z(d_i)} = 0\\
        uv = w(x,y,z)\big|_{x = x(d_i),y = y(d_i),z = z(d_i)} \equiv \Delta_{D6}(d_i)
    \end{cases},
\end{equation}
with $\Delta_{D6}(d_i)$ the D6-brane locus.
We would now like to define dynamical complex deformations of the second equation in \eqref{partial resolution 3}.
From the perspective of the Type IIA reduction, \cite{Collinucci:2020jqd} argues that the correct Higgs branch deformation counting is produced if one discards deformations that act on a non-compact stack of D6-branes, as they require infinite energy and are thus non-dynamical. As far as the brane locus $\Delta_{D6}(d_i)$ is concerned, this is reflected in requiring that one only adds to the second equation of \eqref{partial resolution 3} deformations that satisfy the following requirements:
    \begin{itemize}
        \item The deformations $\lambda_{k}$ must be invariant under all the $\mathbb{C}^*$-actions enjoyed by the toric ambient space.\footnote{This comes from the fact that the combination $uv$, and hence the brane locus, must have zero GLSM charge, in the GLSM description of the toric ambient space.} This can be equivalently restated as restricting the deformations to be polynomial functions of $x,y,z$.
        Thus:
        \begin{equation}\label{deformation}
            \lambda_k = \lambda_k(x,y,z).
        \end{equation}
        \item Inserting the blow-up map into \eqref{deformation}, the allowed deformations produce a brane locus which is \textit{not} proportional to the brane locus $\Delta_{D6}(d_i)$ in \eqref{partial resolution 3}, up to the first equation of \eqref{partial resolution 3}. Physically, if the deformation \textit{were} proportional to the brane locus, it would be acting on some non-compact stack of D6-branes, and therefore it should not be taken into account. Hence, this requirement implies that all the allowed monomial deformations should produce a low-energy quiver phase where \textit{at least} one gauge label is lower than the corresponding gauge label in $\mathsf{Q}_{\mu}$. In general, not all monomial deformations produced in this way are independent, as some of their polynomial combinations might be proportional to the brane locus. In order to elude such subtleties altogether, one can compute the independent deformations developing the commutative algebra method introduced in footnote \ref{footnote ideals}\footnote{Concretely, the partial resolution defines a ring homomorphism $\phi: \hspace{0.1cm} S\rightarrow R$, where $S = \mathbb{C}[x,y,z]$. Moreover, we can pull back $M_{j}^{p_j}$ back to $R$, obtaining $C_j^{(p_j)}\equiv (C_j^{p_j}R_{C_j}\cap R)$. We employ $C_j^{(p_j)}$ in lieu of $C_j^{p_j}$ in order to discard the ideals corresponding to the intersection points between different curves. Then, the monomials that specify a quiver with at least one gauge label greater or equal than the corresponding one dictated by $\Delta_{D6}(d_i)$ on the curve $C_j$ are the ones belonging to $C_j^{(p_j)}$. We can pull them back via the homomorphism $\phi$, defining:
        \begin{equation*}
            I_j = \phi^{-1}(C_j^{(p_j)}).
        \end{equation*}
        The monomials above the bound for all curves $C_j$ are then given by the intersection:
        \begin{equation*}
            I_{\text{bound}} = \bigcap_j I_j.
        \end{equation*}
        Finally, the independent dynamical deformations are obtained through a choice of basis for the ideal $S/I_{\text{bound}}$. There are exactly $\nu$ independent elements in the basis. Once again, the software implementation of this algorithm is straightforward, and it automatically takes care of discarding combinations of monomials that produce a quiver with labels greater or equal than $\Delta_{D6}(d_i)$. See the ancillary Macaulay2 code for an implementation.
        \label{footnote ideals 2}}.
        \end{itemize}
All in all, one obtains the deformed 5d conformal matter equation as:
\begin{equation}\label{3fold deformation}
        \begin{cases}
        Y_{\mathfrak{g}}(x,y,z)= 0\\
        UV = w(x,y,z) + \sum_{k=1}^{\nu} c_k\lambda_k(x,y,z) \\
    \end{cases},
\end{equation}
with the $c_k$ complex coefficients and $k = 1,\ldots,\nu$ the number of allowed dynamical complex deformations. 
Therefore, the total Higgs branch dimension in the UV can be neatly computed according to expression \eqref{UV HB resolution}.
Notice that \eqref{UV HB resolution} precisely coincides with the UV Higgs branch dimension expected from the low-energy quiver description, provided that
\begin{equation}\label{nu}
    \nu = n_H-n_V -\text{rank}(\mathfrak{g}).
\end{equation}
The validity of \eqref{nu} can be explicitly checked for \textit{all} the 5d conformal matter theories explored in this work.
As we have mentioned, as many deformations as the dimension of $\mathfrak{g}$ must be added to $\nu$ in \eqref{nu} in order to obtain the full UV Higgs branch dimension, as our procedure has selected a specific partial resolution that freezes precisely that amount of wrapped M2-brane modes, as previously argued in \cite{DeMarco:2025ugw}.

Crucially, tuning one or more of the coefficients $c_k$ in the deformed equation \eqref{3fold deformation} to be non-vanishing (with all the others set to zero) triggers a path along the Higgs branch RG flow, ending up in a new conformal matter theory:
\begin{equation}\label{transition geometry}
        \begin{cases}
         Y_{\mathfrak{g}}(x,y,z)= 0\\
     UV=  w(x,y,z)
    \end{cases} \quad \xrightarrow{Higgsing} \quad \begin{cases}
        Y_{\mathfrak{g}}(x,y,z)= 0\\
        UV = w(x,y,z)+\sum_{l=1}^{\nu'}\lambda_l(x,y,z) \equiv w'(x,y,z) \\
    \end{cases},
\end{equation}
with $(1,\ldots,\nu')$ a subset (upon possible reordering) of $(1,\ldots,\nu)$. The partially resolved phase of the theory at the end of the RG flow, on the right of \eqref{transition geometry}, can be systematically read off analyzing the new brane locus $w'(x,y,z) = \Delta'_{D6}(x,y,z)$. In Lie-theoretic language, the RG flow is interpreted as an edge (or a concatenation of edges) between the weight $\mu_0$ corresponding to $w(x,y,z)$, and the weight $\mu_i$ related to $w'(x,y,z)$.

In the next paragraph, we exhibit how the procedure outlined above to compute Higgs branch deformations from the geometry of the Calabi-Yau threefold works in an explicit example.

\subsection{Example 1: $D_6$ conformal matter molecule}
We consider the following threefold, which engineers a 5d conformal matter molecule of type $D_6$:
\begin{equation}
\begin{cases}\label{D6 atom}
    x^2 + z y^2+z^5 = 0 \\
   uv = y^2 \\
 \end{cases}.
\end{equation}
 In the previous section we have reviewed that, in order to count Higgs branch deformations of \eqref{D6 atom}, one must perform a resolution of the Du Val $D_6$ singularity in a suitably chosen toric ambient space. We have summarized this class of resolutions for the $A_n,D_5,D_6,E_6,E_7,E_8$ singularities in Appendix \ref{app:DuValres}. 
Using these notations, the Type IIA brane locus is
\begin{equation}\label{brane locus y2}
    \Delta(d_j) = y^{2}\big|_{y=y(d_j)} = (d_5 d_4 d_6^2 d_7^2 d_3)^2.
\end{equation}
The low-energy quiver phase encoded by the brane-locus \eqref{brane locus y2} reads:
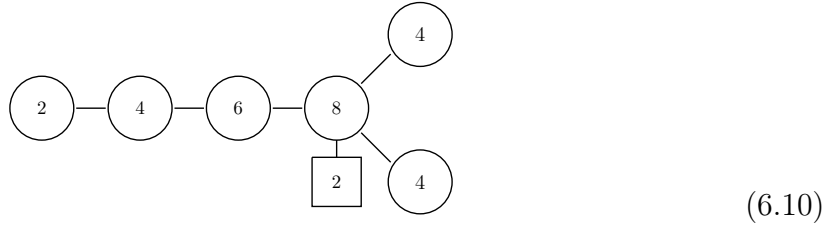
\begin{equation}\label{quiver y2}
       \scalebox{0.65}{\begin{tikzpicture}
        \draw[thick] (0,0) circle (0.65);
        \node at (0,0) {\small$4$};
        \draw[thick] (0.7,0)--(1.3,0);
        \draw[thick] (2,0) circle (0.65);
        \node at (2,0) {\small$6$};
        \draw[thick] (-0.7,0)--(-1.3,0);
        \draw[thick] (-2,0) circle (0.65);
        \node at (-2,0) {\small$2$};
        \draw[thick] (2.7,0)--(3.3,0);
        \draw[thick] (4,0) circle (0.65);
        \node at (4,0) {\small$8$};
        \draw[thick] (5.7,-1.5) circle (0.65);
         \node at (5.7,-1.5) {$4$};
          \draw[thick] (5.7,1.5) circle (0.65);
         \node at (5.7,1.5) {$4$};
         \draw[thick] (4.5,0.5)--(5.1,1.1);
         \draw[thick] (4.5,-0.5)--(5.1,-1.1);
         \draw[thick] (3.5,-1)--(4.5,-1)--(4.5,-2)--(3.5,-2)--(3.5,-1);
         \draw[thick] (4,-0.65)--(4,-1);
         \node at (4,-1.5) {\small$2$};
        \end{tikzpicture}}
\end{equation}
Let us now show how to explicitly identify the Higgs branch deformations. Dynamical deformations in the partially resolved phase are those that \textit{are not} proportional (up to the first equation of \eqref{proper D6}) to the brane locus. Only allowing dynamical deformations, one finds that the deformed threefold is:
\begin{equation}
    \begin{cases}
    x^2 + z y^2+z^5 = 0 \\
   uv =  y^2+c_1+c_2 x+c_3 y+c_4 z+c_5 z^2+c_6 z^3+c_7yz+c_8 xz\\
 \end{cases}.
\end{equation}
Hence $\nu = 8$, which perfectly matches the result that can be obtained from the low-energy quiver phase \eqref{quiver y2}.
All in all, recalling \eqref{UV HB resolution} and applying it in this specific example, we obtain that the dimension of the Higgs branch at the UV fixed point is:
\begin{equation}
\text{dim}(HB)_{UV}= \nu + \text{dim}(\mathfrak{g})  = 8+\text{dim}(D_6) = 74.
\end{equation} 

\subsection{Example 2: $E_6$ conformal matter molecule}
\label{sec:E6RGflowexample}
We now briefly present an example involving a conformal matter molecule of type $E_6$, showcasing the care that should be taken in order to correctly account for the genuine Higgs branch deformations.
The molecule at hand is:
\begin{equation}\label{E6 molecule}
\begin{cases}
    x^2+y^3+z^4 = 0\\
    uv = y(x+iz^2)\\
\end{cases}    
\end{equation}
Proceeding as in the previous example, one completely resolves the $E_6$ singularity, by means of the maps in Appendix \ref{app:DuValres}, and computes the D6-brane locus. This yields the following low-energy quiver description:
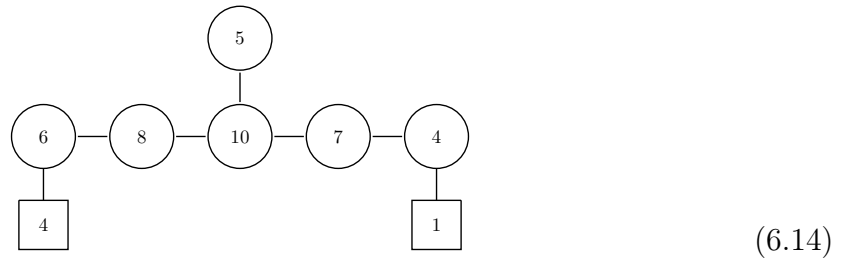
\begin{equation}\label{E6 molecule quiver}
     \scalebox{0.65}{\begin{tikzpicture}
        \draw[thick] (0,0) circle (0.65);
        \node at (0,0) {\small$10$};
        \draw[thick] (0.7,0)--(1.3,0);
        \draw[thick] (2,0) circle (0.65);
        \node at (2,0) {\small$7$};
        \draw[thick] (-2.7,0)--(-3.3,0);
        \draw[thick] (-0.7,0)--(-1.3,0);
        \draw[thick] (-4,0) circle (0.65);
        \node at (-4,0) {\small$6$};
        \draw[thick] (-2,0) circle (0.65);
        \node at (-2,0) {\small$8$};
        \draw[thick] (0,0.7)--(0,1.3);
        \draw[thick] (0,2) circle (0.65);
        \node at (0,2) {\small$5$};
        \draw[thick] (2.7,0)--(3.3,0);
        \draw[thick] (4,0) circle (0.65);
        \node at (4,0) {\small$4$};
         \draw[thick] (-4.5,-1.3)--(-3.5,-1.3)--(-3.5,-2.3)--(-4.5,-2.3)--(-4.5,-1.3);
         \draw[thick] (-4,-0.65)--(-4,-1.3);
         \node at (-4,-1.8) {\small$4$};
             \draw[thick] (4.5,-1.3)--(3.5,-1.3)--(3.5,-2.3)--(4.5,-2.3)--(4.5,-1.3);
         \draw[thick] (4,-0.65)--(4,-1.3);
         \node at (4,-1.8) {\small$1$};
        \end{tikzpicture}}
\end{equation}
Following the recipe of \cite{Collinucci:2020jqd}, the dynamical deformations of the second equation of \eqref{E6 molecule} read:
\begin{equation}\label{E6 molecule def}
\begin{cases}
    x^2+y^3+z^4 = 0\\
   \begin{split} uv &= y(x+iz^2)+c_1 x z^3+c_2 yz^3+c_3 z^5+c_4z^4+c_5xz^2+c_6yz^2+c_7xyz+c_8x^2z+c_9y^2z+\\
&+c_{10}z^3+c_{11}xz+c_{12}xy+c_{13}yz+c_{14}x^2+c_{15}y^2+c_{16}z^2+c_{17}x+c_{18}y+c_{19}z+c_{20}\\
\end{split}
\end{cases}.
\end{equation}
Apparently, there are 20 dynamical deformations. However, it is easy to check from the quiver description in \eqref{E6 molecule quiver} that we expect $\nu = 14$. The crucial point is that not all the deformations in \eqref{E6 molecule def} give rise to genuine Higgs Branch degrees of freedom, since some combinations of the deforming monomials still produce a quiver with entries greater or equal than \eqref{E6 molecule quiver}, and hence should be discarded. This can be seen as follows (here we carefully show how to discard the excess combinations by hand, but we remark that this can be automatically implemented via software based on a commutative algebra approach, as explained in footnote \ref{footnote ideals 2}):
\begin{itemize}
    \item the terms $c_{14}x^2+c_5xz^2$ can be rewritten as:
    \begin{equation}
        c_{14}x^2+c_5xz^2 = -ic_{5}x(x+iz^2) +(c_{14}+ic_{5})x^2. 
    \end{equation}
    Notice that the first term, which is proportional to $x(x+iz^2)$, produces a brane locus that is \textit{proportional} to the one of the undeformed threefold \eqref{E6 molecule}. Hence, $c_{5}$ \textit{does not} produce a dynamical deformation. Only the combination $(c_{14}+ic_{5})$ does. Furthermore, thanks to the $E_6$ Du Val equation, $(c_{14}+ic_{5})x^2$ can be exchanged with terms involving the powers $y^3$ and $z^4$. Namely, we can reabsorb $(c_{14}+ic_{5})x^2$ in a redefinition of the coefficients of $c_4z^4$ (as $y^3$ is proportional to the brane locus of the undeformed threefold, and hence irrelevant). All in all, $c_{14}x^2+c_5xz^2$ does not contribute to the genuine UV Higgs branch dimension.
    \item A similar reasoning can be carried out for the terms $c_{12}xy+c_6yz^2$, as:
    \begin{equation}\label{def 1}
        c_{12}xy+c_6yz^2 = c_{12}y(x+iz^2)+(c_6-ic_{12})yz^2.
    \end{equation}
    The first term in \eqref{def 1} is proportional to the undeformed brane locus of \eqref{E6 molecule}. Hence only the combination $(c_6-ic_{12})$ is a genuine dynamical deformation.
    \item Analogously for the term $c_7xyz+c_2 yz^3$, since:
    \begin{equation}\label{def 2}
       c_7xyz+c_2 yz^3 = c_{7}yz(x+iz^2)+(c_2-ic_{7})yz^3.
    \end{equation}
    The first term in \eqref{def 2} is proportional to the undeformed brane locus of \eqref{E6 molecule}. Therefore only the combination $(c_2-ic_{7})$ is a genuine dynamical deformation. An identical reasoning shows that only one combination of the coefficients in $c_1 x z^3+c_8x^2z$ is a dynamical deformation.
    \item The term $c_8x^2z$ can be exchanged for $c_8(y^3z+z^5)$. Hence it can be discarded via a redefinition of $c_3z^5$.
\end{itemize}
Carefully tracking the caveats above, it is easy to convince oneself that, out of the 20 candidate deformations displayed in \eqref{E6 molecule def}, only 14 are genuine dynamical deformations. This precisely coincides with the expectation from the low-energy quiver \eqref{E6 molecule quiver}. Hence:
\begin{equation}
    \text{dim}(HB)_{UV} = \nu + \text{dim}(E_6) = 14 + 78 = 92.
\end{equation}
The computation of dynamical deformations, and therefore of the UV Higgs Branch dimension, for all other molecules and hybrids studied in this work proceeds employing exactly the same techniques as the ones reviewed in the examples we have presented.

\section{Extended Coulomb branch RG flows and CS levels}\label{sec: CB RG flow}
In this Section, we analyze a second type of RG flow that hierarchically connects the conformal matter theories presented in this paper. Such an RG flow is triggered by taking large CB vevs and large masses, and corresponds to sending to infinity the volumes of specific curves and divisors in a resolution of the threefold. 
We  discuss three aspects of such RG flows. First, in \Cref{sec:extendedcbrg}, we connect some of the theories we presented in the previous sections with such extended Coulomb branch (ECB) RG flows, highlighting the limits of such procedure with respect to the HB RG flows presented in \Cref{sec: Higgs branch}. Then, in \Cref{sec:ECBRGCS}, we  discuss how ECB RG flows can produce threefolds whose associated 5d Dynkin quiver phase has non-zero Chern-Simons (CS) levels. Finally, in \Cref{sec:genericgeomertryCSAn} we  present the \textit{resolved} Calabi-Yau threefolds that correspond to a 5d quiver with generic gauge labels $\mathbf{r}$, flavor labels $\lambda \in \Gamma_w$, and CS levels $\mathbf{k}$. This will give us a glimpse into the 5d physical interpretation of coweights that do not belong to the coroot lattice.  
\subsection{Extended Coulomb branch RG flows}
\label{sec:extendedcbrg}
We wish to comment on a second possible type of RG flows: the (extended) CB RG flow. Reaching the Dynkin shaped Lagrangian phase for the 5d CM corresponds to turning on a diagonal mass vev for the $\mathfrak g \times \oplus g$ flavor symmetry \cite{DeMarco:2025ugw}, and hence to moving along the extended Coulomb branch (ECB) of the 5d SCFT. On the generic point of the ECB of the quiver phase we have families of \textit{resolved} $A$-type singularities fibered over the compact curves (the exceptional $\mathbb P^1$s) of a resolved $Y_{\mathfrak g}$ singularity.  We now examine 5d CB RG-flows obtained by taking the limit of large CB vevs, first in terms of the M-theory geometry, then from the $(p,q)$-web perspective, and finally at a low-energy field theory level.

From the perspective of the threefold, to reach a Dynkin quiver with lower rank special unitary nodes, we send to infinity the volumes of some of the $\mathbb P^1$s resolving the aforementioned curves of $A_n$ singularities. As the curves of $A_n$ singularity are themselves compact, this operation amounts to sending to infinity the volumes of some divisors within the CY3, decoupling the M5 branes wrapped on them.

In the $(p,q)$ web picture, the aforementioned operation corresponds to sending to infinity some compact D5 branes moving just along the $x_5, x_6$-plane, and decoupling the D3 branes that fill the area of the parallelogram that the D5s define in the 5-6 directions. 

At a level of field theory, these operations translate into moving towards the boundaries of the ECB, giving large vevs to some CB operators\footnote{Such vevs geometrically coincide with the volumes of the aforementioned $\mathbb P^1$s.} (with fixed background masses). This procedure, as a byproduct, gives infinite mass to a certain number of BPS magnetic strings that are realized in M-theory by the aforementioned M5 branes, and in the $(p,q)$ web by the D3 branes.

\begin{figure}
\begin{center}
\begin{tikzpicture}

    \draw (0,0) -- (1,0);
    \draw (1,0) -- (2,0);
    \draw (2,0) -- (2.3,0);
    \draw (4,0) -- (3.7,0);

    \draw (0,1) -- (1,1);
    \draw (1,1) -- (2,1);
    \draw (2,1) -- (2.3,1);
    \draw (4,1) -- (3.7,1);

    \draw (0,3) -- (1,3);
    \draw (1,3) -- (2,3);
    \draw (2,3) -- (2.3,3);
    \draw (4,3) -- (3.7,3);

    \draw (0,4) -- (1,4);
    \draw (1,4) -- (2,4);
    \draw (2,4) -- (2.3,4);
    \draw (4,4) -- (3.7,4);

    \draw (0,0) -- (0,1);
    \draw (0,1) -- (0,1.3);
    \draw (0,3) -- (0,2.7);

    \draw (1,0) -- (1,1);
    \draw (1,1) -- (1,1.3);
    \draw (1,3) -- (1,2.7);

    \draw (2,0) -- (2,1);
    \draw (2,1) -- (2,1.3);
    \draw (2,3) -- (2,2.7);

    \draw (4,0) -- (4,1);
    \draw (4,1) -- (4,1.3);
    \draw (4,3) -- (4,2.7);

    \draw (0,3) -- (0,4);
    \draw (1,3) -- (1,4);
    \draw (2,3) -- (2,4);
    \draw (4,3) -- (4,4);

    \node at (3,0) {$\dots$};
    \node at (3,1) {$\dots$};
    \node at (3,2) {$\dots$};
    \node at (3,3) {$\dots$};
    \node at (3,4) {$\dots$};

    \node at (0,2) {$\vdots$};
    \node at (1,2) {$\vdots$};
    \node at (2,2) {$\vdots$};
    \node at (4,2) {$\vdots$};

    \filldraw (0,0) circle (2pt);
    \filldraw (1,0) circle (2pt);
    \filldraw (2,0) circle (2pt);
    \filldraw (4,0) circle (2pt);

    \filldraw (0,1) circle (2pt);
    \filldraw (1,1) circle (2pt);
    \filldraw (2,1) circle (2pt);
    \filldraw (4,1) circle (2pt);

    \filldraw (0,3) circle (2pt);
    \filldraw (1,3) circle (2pt);
    \filldraw (2,3) circle (2pt);
    \filldraw (4,3) circle (2pt);

    \filldraw (0,4) circle (2pt);
    \filldraw (1,4) circle (2pt);
    \filldraw (2,4) circle (2pt);
    \filldraw (4,4) circle (2pt);

\end{tikzpicture}
\end{center}
\caption{Toric diagram for the \eqref{eq:nmrectangle} threefold.}
\label{fig:rectangles}
\end{figure}
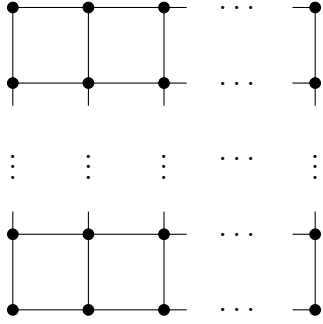

Let us now visualize this ECB RG-flow in an explicit example, by considering the $X_{A_{n}}^{z^{m+1}}$ singularity: 
\begin{equation}
\label{eq:nmrectangle}
    x y = z^{n+1}, \qquad u v = z^{m+1}, 
\end{equation}
whose associated toric diagram is depicted in \Cref{fig:rectangles}.

First, we perform the base-change resolution of \cite{DeMarco:2023irn} on the first equation of \eqref{eq:nmrectangle}, with blowup map
\begin{equation}
    \label{eq:firstblowup}
    x = z_1^{n+1} e_1^n \dots e_n, \qquad y = e_1 \dots e_n^n z_2^{n+1}, \qquad z = z_1 e_1 \dots e_n z_2,
\end{equation}
obtaining the following hypersurface in toric ambient space: 
\begin{equation}
\label{eq:nmrectanglefirstres}
\begin{cases}
u v = z^{m+1}, \\
z_1 e_1 \dots e_{n} z_2 = z,
\end{cases}  \qquad \subset \qquad 
\renewcommand{\arraystretch}{1.3}
\begin{array}{c|ccccccccc|ccc|c}
  & z_1 & e_1 & e_2 & e_3  & \dots &  z_2 & u & v & z& \text{FI}\\
\hline
 \mathbb C^*_1 & 1 & -2 & 1 & 0 & \dots & 0 & 0 &0 &0 &\xi_1  \\
 \mathbb C^*_2 & 0 & 1 & -2 & 1 & \dots & 0 & 0 &0 & 0 &\xi_2 \\
  \vdots & \vdots & \vdots& \vdots& \vdots& \vdots& \vdots & \vdots  & \vdots & \vdots & \vdots \\
   \mathbb C^*_n & 0 & 0 & 0 & 0 & \dots & 1 &  0  & 0 &0&\xi_n \\
\end{array}
\end{equation}
Looking at the two equations on the left of \eqref{eq:nmrectanglefirstres}, we see that the partially resolved threefold is again a base-change of, now, an $A_{m}$ singularity by the monomial $z_1 e_1 \dots e_n z_2$\footnote{This comes from the fact that the $A$-type singularities appearing in the resolution, or equivalently the ranks of the nodes of the Dynkin quiver associated with the phase \eqref{eq:nmrectanglefirstres}, are all the same in this specific threefold.}. We can then iterate the base-change resolution procedure, now focusing on the variables $(u,v,z)$. The second blowup map reads
\begin{equation}
\label{eq:secondblowup}
    u = \tilde{z}_1^{m+1} \tilde{e}_1^m \dots \tilde{e}_m, \qquad v = \tilde{e}_1 \dots \tilde{e}_m^m \tilde{z}_2^{m+1} , \qquad z = \tilde{z}_1 \tilde{e}_1 \dots \tilde{e}_m \tilde{z}_2,
\end{equation}
and automatically satisfies the first equation of \eqref{eq:nmrectanglefirstres}. The second equation of \eqref{eq:nmrectanglefirstres} instead is not automatically satisfied, and after \eqref{eq:secondblowup} the threefold is the set of solutions of
\begin{equation}
    \label{eq:nmrectanglesecond}
z_1 e_1 \dots e_{n} z_2 =  \tilde{z}_1 \tilde{e}_1 \dots \tilde{e}_m \tilde{z}_2,
\end{equation}
in the following ambient space: 
\begin{equation}
\label{eq:nmrectanglesecondambient}
    \renewcommand{\arraystretch}{1.3}
\begin{array}{c|cccccc|cccccccccccc|c}
  & z_1 & e_1 & e_2 & e_3  & \dots &  z_2 & \tilde{z}_1 & \tilde{e}_1 & \tilde{e}_2 & \tilde{e}_3  & \dots &  \tilde{z}_2  & \text{FI}\\
\hline
 \mathbb C^*_1 & 1 & -2 & 1 & 0 & \dots & 0 & 0 & 0 & 0 &0 &0 &0 & \xi_1  \\
 \mathbb C^*_2 & 0 & 1 & -2 & 1 & \dots & 0 &0 &0 &0 &0 &0 &  0 &\xi_2 \\
  \vdots & \vdots & \vdots& \vdots& \vdots& \vdots& \vdots &  \vdots& \vdots& \vdots& \vdots & \vdots& \vdots& \vdots&\\
   \mathbb C^*_n & 0 & 0 & 0 & 0 & \dots & 1 &  0 &  0 &  0 &  0 &  0 &  0 &  \xi_n \\
   \hline
    \tilde{\mathbb C}^*_1 & 0 & 0 & 0 & 0 & \dots & 0 & 1 & -2 & 1 &0 & \dots &0 & \tilde{\xi}_1  \\
 \tilde{\mathbb C}^*_2& 0 & 0 & 0 & 0 & \dots & 0 &0 &1 &-2 &1 & \dots &  0 &\tilde{\xi}_2 \\
  \vdots & \vdots & \vdots& \vdots& \vdots& \vdots& \vdots &  \vdots& \vdots& \vdots& \vdots & \vdots& \vdots& \vdots&\\
 \tilde{\mathbb C}^*_m& 0 & 0 & 0 & 0 & \dots & 0 &  0 &  0 &  0 &  0 &  \dots &  1 &  \tilde{\xi}_m \\
\end{array}
\end{equation}
Let us describe the geometry of \eqref{eq:nmrectanglesecond}, \eqref{eq:nmrectanglesecondambient}: we have $n$ compact curves of resolved $A_m$ singularities, over the loci $e_j = 0$, with $e_j = 1,...,n$. The curve, on the generic point, is fully resolved, with the $\mathbb P^1_i$-bundle of the resolution of the curve over $e_j =0$  (that we call $\mathbb{D}_{i,j}$) being described by the following equation
\begin{equation}
\label{eq:pijbundle}
    e_j = \tilde{e_i} = 0.
\end{equation}
Furthermore, we have two non-compact curves of resolved $A_m$ singularities, over $z_1 = 0$ and $z_2 = 0$. At the intersection of the various $\mathbb P^1$ bundles, e.g.
\begin{equation}
    \label{eq:conifoldpointij}
    e_j = e_{j+1} = \tilde{e}_i = \tilde{e}_{i+1} = 0
\end{equation}
we have some unresolved conifold points, that can be visualized by fixing to one, inside \eqref{eq:nmrectanglesecond}, all the variables that do not appear in \eqref{eq:conifoldpointij}. 

Let us now describe the ECB RG flow, namely the decompactification of some of the $\mathbb{D}_{i,j}$-bundles, for all $j = 0,...,n+1$ (here $\mathbb{D}_{i,0}$ is the non-compact $\mathbb P^1$ bundle over $z_1 = 0$, and  $\mathbb{D}_{i,n+1}$ the one over $z_2 = 0$). First, let us recall that the volume of $\mathbb{D}_{i,j}$ are described by the F.I. parameters of  the GLSM. In particular, we are interested to the D-terms of the actions $\mathbb C^*_j$ and $\mathbb C^*_i$: 
\begin{equation}
    |e_{j+1}|^2 + |e_{j-1}|^2 -2|e_{j}|^2 = \xi_j, \qquad   |\tilde{e}_{i+1}|^2 + |\tilde{e}_{i-1}|^2 -2|\tilde{e}_{i}|^2 = \tilde{\xi}_i.
\end{equation}
Imposing \eqref{eq:pijbundle}, we see that the variables $[e_{j-1}:e_{j+1}]$ (resp. $[\tilde{e}_{i-1}:\tilde{e}_{i+1}]$) describe respectively the base $e_j = 0$ and one of the fibers $\tilde{e}_i = 0$ of the corresponding $\mathbb{D}_{i,j}$-bundles. The K\"ahler volume of the base curve $e_j =0$ (resp., of the fibral curve $\tilde{e}_i =0$) are $\xi_j$ and $\tilde{\xi}_j$, that can be used as some of the 5d CB coordinates. 
The various ECB RG flows we are interested in are obtained for (some of the) $\xi_i \to \infty$.

Let us take e.g.,
\begin{equation}
\label{eq:deccompexample}
    \tilde{\xi_1} \to \infty,
\end{equation}
a limit that sends to infinity the tension of the M5s wrapped over $\mathbb P^1_{j,1}$, for all $j = 0, ..., n+1$, or, equivalently, of the corresponding BPS strings of the 5d SCFT. 

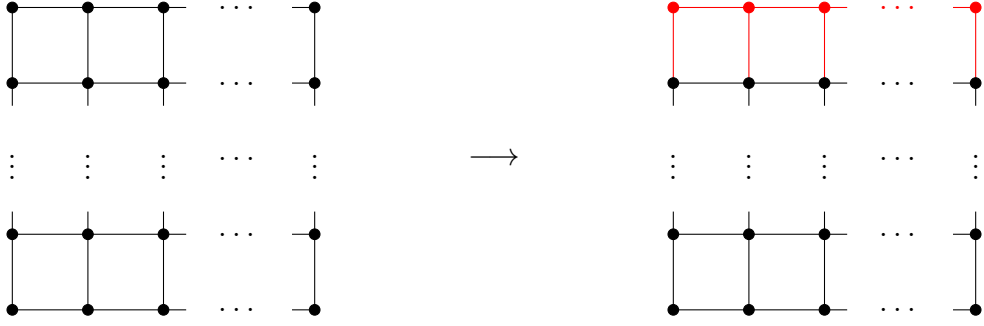
\begin{figure}[H]
\centering

\begin{minipage}{0.44\textwidth}
\centering
\begin{tikzpicture}

    \draw (0,0) -- (1,0);
    \draw (1,0) -- (2,0);
    \draw (2,0) -- (2.3,0);
    \draw (4,0) -- (3.7,0);

    \draw (0,1) -- (1,1);
    \draw (1,1) -- (2,1);
    \draw (2,1) -- (2.3,1);
    \draw (4,1) -- (3.7,1);

    \draw (0,3) -- (1,3);
    \draw (1,3) -- (2,3);
    \draw (2,3) -- (2.3,3);
    \draw (4,3) -- (3.7,3);

    \draw (0,4) -- (1,4);
    \draw (1,4) -- (2,4);
    \draw (2,4) -- (2.3,4);
    \draw (4,4) -- (3.7,4);

    \draw (0,0) -- (0,1);
    \draw (0,1) -- (0,1.3);
    \draw (0,3) -- (0,2.7);

    \draw (1,0) -- (1,1);
    \draw (1,1) -- (1,1.3);
    \draw (1,3) -- (1,2.7);

    \draw (2,0) -- (2,1);
    \draw (2,1) -- (2,1.3);
    \draw (2,3) -- (2,2.7);

    \draw (4,0) -- (4,1);
    \draw (4,1) -- (4,1.3);
    \draw (4,3) -- (4,2.7);

    \draw (0,3) -- (0,4);
    \draw (1,3) -- (1,4);
    \draw (2,3) -- (2,4);
    \draw (4,3) -- (4,4);

    \node at (3,0) {$\dots$};
    \node at (3,1) {$\dots$};
    \node at (3,2) {$\dots$};
    \node at (3,3) {$\dots$};
    \node at (3,4) {$\dots$};

    \node at (0,2) {$\vdots$};
    \node at (1,2) {$\vdots$};
    \node at (2,2) {$\vdots$};
    \node at (4,2) {$\vdots$};

    \filldraw (0,0) circle (2pt);
    \filldraw (1,0) circle (2pt);
    \filldraw (2,0) circle (2pt);
    \filldraw (4,0) circle (2pt);

    \filldraw (0,1) circle (2pt);
    \filldraw (1,1) circle (2pt);
    \filldraw (2,1) circle (2pt);
    \filldraw (4,1) circle (2pt);

    \filldraw (0,3) circle (2pt);
    \filldraw (1,3) circle (2pt);
    \filldraw (2,3) circle (2pt);
    \filldraw (4,3) circle (2pt);

    \filldraw (0,4) circle (2pt);
    \filldraw (1,4) circle (2pt);
    \filldraw (2,4) circle (2pt);
    \filldraw (4,4) circle (2pt);

\end{tikzpicture}
\end{minipage}
\hfill
\begin{minipage}{0.08\textwidth}
\centering
$\longrightarrow$
\end{minipage}
\hfill
\begin{minipage}{0.44\textwidth}
\centering
\begin{tikzpicture}


    \draw (0,0) -- (1,0);
    \draw (1,0) -- (2,0);
    \draw (2,0) -- (2.3,0);
    \draw (4,0) -- (3.7,0);

    \draw (0,1) -- (1,1);
    \draw (1,1) -- (2,1);
    \draw (2,1) -- (2.3,1);
    \draw (4,1) -- (3.7,1);

    \draw (0,3) -- (1,3);
    \draw (1,3) -- (2,3);
    \draw (2,3) -- (2.3,3);
    \draw (4,3) -- (3.7,3);

    \draw[red] (0,4) -- (1,4);
    \draw[red] (1,4) -- (2,4);
    \draw[red] (2,4) -- (2.3,4);
    \draw[red] (4,4) -- (3.7,4);

    \draw (0,0) -- (0,1);
    \draw (0,1) -- (0,1.3);
    \draw (0,3) -- (0,2.7);

    \draw (1,0) -- (1,1);
    \draw (1,1) -- (1,1.3);
    \draw (1,3) -- (1,2.7);

    \draw (2,0) -- (2,1);
    \draw (2,1) -- (2,1.3);
    \draw (2,3) -- (2,2.7);

    \draw (4,0) -- (4,1);
    \draw (4,1) -- (4,1.3);
    \draw (4,3) -- (4,2.7);

    \draw[red] (0,3) -- (0,4);
    \draw[red] (1,3) -- (1,4);
    \draw[red] (2,3) -- (2,4);
    \draw[red] (4,3) -- (4,4);

    \node at (3,0) {$\dots$};
    \node at (3,1) {$\dots$};
    \node at (3,2) {$\dots$};
    \node at (3,3) {$\dots$};
    \node[red] at (3,4) {$\dots$};

    \node at (0,2) {$\vdots$};
    \node at (1,2) {$\vdots$};
    \node at (2,2) {$\vdots$};
    \node at (4,2) {$\vdots$};

    \filldraw (0,0) circle (2pt);
    \filldraw (1,0) circle (2pt);
    \filldraw (2,0) circle (2pt);
    \filldraw (4,0) circle (2pt);

    \filldraw (0,1) circle (2pt);
    \filldraw (1,1) circle (2pt);
    \filldraw (2,1) circle (2pt);
    \filldraw (4,1) circle (2pt);

    \filldraw (0,3) circle (2pt);
    \filldraw (1,3) circle (2pt);
    \filldraw (2,3) circle (2pt);
    \filldraw (4,3) circle (2pt);

    \filldraw[red] (0,4) circle (2pt);
    \filldraw[red] (1,4) circle (2pt);
    \filldraw[red] (2,4) circle (2pt);
    \filldraw[red] (4,4) circle (2pt);

\end{tikzpicture}
\end{minipage}

\caption{Description of the ECB RG flow at the level of the toric diagram of \eqref{eq:nmrectangle}. The red dots in the top row are removed in the rightmost toric diagram.}
\label{fig:rectanglesRG}
\end{figure}

At the level of the GIT quotient underlying the GLSM construction, the decompactification limit is realized simply by removing the point $[\tilde{z_1}:\tilde{e}_2] = [0:\tilde{e}_2]$ from all the $\mathbb P^1_{j,1}$. This is achieved by gauge fixing 
\begin{equation}
    \label{eq:gaugefixingexample}
    \tilde{z}_1 = 1,
\end{equation}
acting with a $\tilde{\mathbb C}^*_1$ with gauge parameter $\tilde{\lambda}_1 = \frac{1}{\tilde{z}_1}$. This produces the following new ambient space coordinates:
\begin{equation}
    \hat{z}_1 = \tilde{e}_1 \tilde{z}_1^2, \qquad \hat{e}_1 = \frac{\tilde{e}_2}{\tilde{z}_1}, 
\end{equation}
while leaving the other ambient space coordinates in \eqref{eq:nmrectanglesecondambient} untouched. The new ambient space,\footnote{The shortcut to see that the charges under the other $\mathbb C^*$ actions are left untouched by the gauge-fixing \eqref{eq:gaugefixingexample} is to note that $\tilde{z}_1$ is charged just under $\tilde{\mathbb C}^*_1$.} after this decompactification, is  
\begin{equation}
\label{eq:nmrectangledecompambient}
    \renewcommand{\arraystretch}{1.3}
\begin{array}{c|cccccc|cccccccccccc|c}
  & z_1 & e_1 & e_2 & e_3  & \dots &  z_2 & \hat{z}_1 & \hat{e}_1 & \tilde{e}_3 & \tilde{e}_4  & \dots &  \tilde{z}_2  & \text{FI}\\
\hline
\mathbb C^*_1 & 1& -2 & 1 & 0 & \dots & 0 &0 &0 &0 &0 &0 &  0 &\xi_1 \\
 \mathbb C^*_2 & 0 & 1 & -2 & 1 & \dots & 0 &0 &0 &0 &0 &0 &  0 &\xi_2 \\
  \vdots & \vdots & \vdots& \vdots& \vdots& \vdots& \vdots &  \vdots& \vdots& \vdots& \vdots & \vdots& \vdots& \vdots&\\
   \mathbb C^*_n & 0 & 0 & 0 & 0 & \dots & 1 &  0 &  0 &  0 &  0 &  0 &  0 &  \xi_n \\
   \hline
 \tilde{\mathbb C}^*_2& 0 & 0 & 0 & 0 & \dots & 0 &1&-2 &1 &0 & \dots &  0 &\tilde{\xi}_2 \\
  \vdots & \vdots & \vdots& \vdots& \vdots& \vdots& \vdots &  \vdots& \vdots& \vdots& \vdots & \vdots& \vdots& \vdots&\\
 \tilde{\mathbb C}^*_m& 0 & 0 & 0 & 0 & \dots & 0 &  0 &  0 &  0 &  0 &  \dots &  1 &  \tilde{\xi}_m \\
\end{array}
\end{equation}
We can now prove that \eqref{eq:deccompexample}, \eqref{eq:gaugefixingexample} realize the following RG flow: 

\begin{equation}
\label{eq:RGflowCB}
    X_{A_n}^{z^{m+1}}  \quad \twoheadrightarrow \qquad  X_{A_n}^{z^{m}}.
\end{equation}
Indeed, after expressing \eqref{eq:nmrectanglesecond} in terms of  $\hat{z}_1, \hat{e}_1$,
\begin{equation}
\label{eq:newconifold}
    z_1 e_1 \dots e_n z_2 = \hat{z}_1 \hat{e}_1 \dots \tilde{e}_m \tilde{z}_2
\end{equation}
and  renaming
\begin{equation}
    \hat{z}_1 \to \tilde{z}_{1}, \qquad \hat{e}_1 \to \tilde{e}_1, \qquad \tilde{e}_i \to \tilde{e}_{i-1}, \qquad  \tilde{\xi}_i \to \widetilde{\xi}_{i-1},
\end{equation}
for $i = 2,...,m$, we see that \eqref{eq:newconifold} and \eqref{eq:nmrectangledecompambient} coincide with \eqref{eq:nmrectanglesecond} and \eqref{eq:nmrectanglesecondambient}, with $m' = m-1$. Consequently, the new threefold contracts to
\begin{equation}
    x y = z^{n+1}, \qquad uv = z^m,
\end{equation}
proving \eqref{eq:RGflowCB}. We depicted graphically, at the level of the toric diagram, the RG we just described in \Cref{fig:rectanglesRG}.

For the other SCFTs considered in this work, one can apply a similar procedure. However, the computation is more cumbersome, as we cannot use anymore the iterated base-change resolution of \eqref{eq:firstblowup}, \eqref{eq:secondblowup}. Nevertheless, we can resolve each line of $A$-type singularities appearing after reaching the Dynkin quiver phase. Then, after iteratively performing a gauge fixing of the form \eqref{eq:deccompexample}, \eqref{eq:gaugefixingexample}, we can reach the GLSM associated to a new Dynkin quiver, realizing in this way the ECB RG-flow between the corresponding conformal matter theories.

We conclude by noticing that, in contrast with the HB RG flow described at the beginning of this Section, if we start with the theory associated
to a given coweight  $\mu_0$, we cannot reach by 
ECB RG flow all the theories with $\mu < \mu_0$. Indeed, as explained in \Cref{sec:atomeshybridsmolecules},  the Dynkin labels of a coweight correspond to the flavor labels of $\mathsf{Q}_{\mu}$. As the associated flavor groups correspond to non-compact curves of A-type singularities, the partial resolution associated with the ECB RG flow can only lower their rank,  constraining in such a way the set of weights that we can reach with this kind of flow. Summing up: 

\textit{A coweight $\mu \leq \mu_0$ can be reached from $\mu_0$ by ECB RG flow if and only if all the Dynkin labels of $\mu$ are smaller or equal to those of $\mu_0$.}

\subsection{ECB RG flow and CS levels}
\label{sec:ECBRGCS}
Let us now discuss how ECB RG flows allow us to obtain ADE special unitary quivers with non-zero Chern-Simons (CS) terms. 

We can show the procedure already in the example \eqref{eq:nmrectangle}. 
First, let us perform again the resolution \eqref{eq:firstblowup}, \eqref{eq:nmrectanglefirstres}. As we saw in the previous Section, we have two non-compact curves of $A_{m}$ singularities over $z_1 = 0$ (resp., $z_2 = 0$) and $n$ compact curves of $A_m$ singularities on $e_i = 0$. The former curves give special-unitary $\mathfrak{su}(m+1)$ flavor groups, while the latter give special unitary $\mathfrak{su}(m+1)$ gauge groups of a linear balanced $A_n$-shaped quiver. 

Let us perform ECB RG flow by taking a large vev, say, for the adjoint scalar of the $\mathfrak{su}(m+1)$ gauge node associated with $e_i = 0$. We do that in such a way that the gauge group is higgsed as 
\begin{equation}
\label{eq:higgsingchain}
    \mathfrak{su}(m+1) \to \mathfrak{su}(m) \oplus \mathfrak{u}(1) \to \mathfrak{su}(m) 
\end{equation}
where in the last step we have sent to infinity the vev of the adjoint scalar of the $\mathfrak{u}(1)$ group. Such procedure corresponds, in the $(p,q)$ web, to sending to infinity one compact D5 brane. In geometry, this is equivalent to performing a partial resolution of the corresponding  compact curve of $A_m$ singularities: 
\begin{equation}
\label{eq:blowupdecoupling}
    A_{m} \to A_{m-1} \oplus \mathbb P^1 \to A_{m-1}
\end{equation}
where $A_{m-1} \oplus \mathbb P^1 $ denote the fact that after a partial resolution of the $A_m$ curve, we have a residual $A_{m-1}$ singularity on the exceptional $\mathbb P^1$.  The last arrow of \eqref{eq:blowupdecoupling} mimics the corresponding arrow of \eqref{eq:higgsingchain}.\footnote{This can be indeed made more precise applying the M-theory geometric engineering dictionary: a singular $A_{m-1}$ gives $\mathfrak{su}(m)$ SYM, while a resolved $\mathbb P^1$ with a $\mathcal O(-2) \oplus \mathcal O$ normal bundle (as in the current example), gives just an  $\mathfrak u(1)$ IR photon: the M2 branes wrapping the $\mathbb P^1$ that would enhance $\mathfrak{u}(1)$ to $\mathfrak{su}(2)$ get a large mass, and decouple from the IR theory.}

Let us realize \eqref{eq:blowupdecoupling} quantitatively: to perform the partial resolution, we blowup 
\begin{equation}
\label{eq:partialblowup}
    u = \lambda u', \qquad e_i = \lambda e_i', 
\end{equation}
introducing new GLSM variables $u',e_i', \lambda$ and a new GLSM action with charges 
\begin{equation}
\label{eq:chargespartialblowup}
    q_{u'} = q_{e_i'} = 1, \qquad q_{\lambda} = -1,
\end{equation}
and GLSM D-term
\begin{equation}
\label{eq:partialblowupFI}
     \qquad |u'|^2 + |e_1'|^2 - 2 |\lambda|^2 = \xi_{bl} > 0. 
\end{equation}
The action \eqref{eq:chargespartialblowup} has to be added to the row of the GLSM description in \eqref{eq:nmrectanglefirstres}. 
Plugging in \eqref{eq:partialblowup} into \eqref{eq:nmrectanglefirstres}, we get 
\begin{equation}
\label{eq:nmrectangleCS}
\scalemath{0.95}{uv = (z_1 e_1 \dots e_i' \dots z_2)^{m+1} \lambda^m \qquad \subset \qquad 
\renewcommand{\arraystretch}{1.3}
\begin{array}{c|ccccccccc|ccc|c}
  & z_1 & e_1 & \dots & e_i'  & \dots &  z_2 & u & v & \lambda& \text{FI}\\
\hline
 \mathbb C^*_1 & 1 & -2 & \dots & 0 & \dots & 0 & 0 &0 &0 &\xi_1  \\
 \mathbb C^*_2 & 0 & 1 & \dots & 0 & \dots & 0 & 0 &0 & 0 &\xi_2 \\
  \vdots & \vdots & \vdots& \vdots& \vdots& \vdots& \vdots & \vdots  & \vdots & \vdots & \vdots \\
  \mathbb C^*_{i-1} & 0 & 0 & \dots & 1 & \dots & 0 &  0  & 0 &0&\xi_{i-1} \\
     \mathbb C^*_i & 0 & 0 & \dots & -2 & \dots & 0 &  0  & 0 &0&\xi_i \\
   \mathbb C^*_{i+1} & 0 & 0 & \dots & 1 & \dots & 0 &  0  & 0 &0&\xi_{i+1} \\
   \vdots & \vdots & \vdots& \vdots& \vdots& \vdots& \vdots & \vdots  & \vdots & \vdots & \vdots \\
      \mathbb C^*_{bl} & 0 & 0 & 0 & 1 & \dots & 0 &  1  & 0 &-1&\xi_{bl} \\
\end{array}
}
\end{equation}
Let us comment on the singularities of \eqref{eq:nmrectangleCS}, we again have the same singularities of \eqref{eq:nmrectanglefirstres} outside the locus $e_i' = 0$. On $e_i' = 0$ the space is now smooth, as $u' = e_i' = 0$ is excluded by \eqref{eq:partialblowupFI}. Instead, we have a new compact curve of $A_{m-1}$ singularities on $\lambda = u' = v = 0$. We have then realized the first arrow of the transition \eqref{eq:blowupdecoupling}. To fully decouple the branes wrapped on the $\mathbb P^1$, and realize the ECB RG flow to a theory with CS terms, we need to send to infinity the volume of the $\mathbb P^1$. This can be achieved with a procedure analogous to the one outlined below \eqref{eq:deccompexample}, zooming on the region $\lambda = 0$, acting with $\mathbb C^{*}_{bl}$ as follows\footnote{Again, we note that the transformation is well defined as, by \eqref{eq:partialblowupFI}, $\lambda$ and $e_i'$ can not be both zero.}
\begin{equation}
\label{eq:newcordinates}
    u' \to \hat{u} \equiv \frac{u'}{e_i'}, \qquad e_i' \to 1, \qquad \lambda \to \hat{e}_i \equiv e_i' \lambda
\end{equation}
\begin{equation}
\label{eq:nmrectangleCSfinal}
\scalemath{0.95}{
uv = (z_1 e_1 \dots e_{i-1} e_{i+1} \dots z_2)^{m+1} \hat{e}_i^m \qquad \subset \qquad 
\renewcommand{\arraystretch}{1.3}
\begin{array}{c|cccccccc|cccc}
  & z_1 & e_1 & \dots & \hat{e}_i  & \dots &  z_2 & \hat{u} & v & \text{FI}\\
\hline
 \mathbb C^*_1 & 1 & -2 & \dots & 0 & \dots & 0 & 0 &0  &\xi_1  \\
 \mathbb C^*_2 & 0 & 1 & \dots & 0 & \dots & 0 & 0 &0  &\xi_2 \\
  \vdots & \vdots & \vdots& \vdots& \vdots& \vdots& \vdots & \vdots  & \vdots &  \vdots \\
  \mathbb C^*_{i-1} & 0 & 0 & \dots & 1 & \dots & 0 &  -1  & 0 &\xi_{i-1} \\
     \mathbb C^*_i & 0 & 0 & \dots & -2 & \dots & 0 &  2  & 0 &\xi_i \\
   \mathbb C^*_{i+1} & 0 & 0 & \dots & 1 & \dots & 0 &  -1  & 0 &\xi_{i+1} \\
   \vdots & \vdots & \vdots& \vdots& \vdots& \vdots& \vdots & 0 & \vdots & \vdots  \\
\end{array}
}
\end{equation}
\eqref{eq:nmrectangleCSfinal} differs from \eqref{eq:nmrectanglefirstres} by the following aspects: 
\begin{itemize}
    \item the hypersurface equation shows $A_{m}$ singularities for all $e_j = 0$, with $e_j \neq \hat{e}_{i}$ and for $z_1 =0 $ (resp. $z_2 = 0$). For $\hat{e}_i = 0$, we get an $A_{m-1}$ singularity. 
\item Now, $\hat{u}$ has GLSM charges $(C_{\mathfrak{g}})_{ij}$  w.r.t $\mathbb C^*_j$, and the action $\mathbb C^*_{bl}$ disappeared due to the gauge-fixing.  
\end{itemize}
The GLSM charges of the combination $uv$ can be identified with the CS levels, taking the IIA limit using the action \eqref{eq:cstarreductionfirst}, that we recall here for the reader's convenience: 
\begin{equation}
\label{eq:cstarreduction}
\hat{u} \to\lambda \hat{u} , \qquad v \to \frac{v}{\lambda }, \qquad \lambda \in \mathbb C^*,
\end{equation}
 with the CS levels of the quiver \cite{Closset_2019}.\footnote{We can see \eqref{eq:cstarreduction} as a new row of the GLSM charge matrix of the threefold.} 
Let us review the proof of this fact, following \cite{Closset_2019}:  M-theory on the threefolds considered in this work reduces, in IIA, to systems of D6 branes wrapped on the resolved $Y_{\mathfrak g}$ by using the action \eqref{eq:cstarreduction}. Such D6 branes stacks correspond to the special unitary groups of the $Q_{\mu}$. The CS levels of the gauge theory phase are related to the slopes of the K\"ahler parameters of $Y_{\mathfrak g}$ over a transverse real coordinate \begin{equation}
    x_9 = |u|^2 - |v|^2,
\end{equation}
that is identified with the FI parameter of \eqref{eq:cstarreduction}. 

For $x_9 > 0,$ we can reduce along \eqref{eq:cstarreduction} by fixing $u = 1$. Each $\mathbb C^*_j$ action then becomes  
 \begin{equation}
 \label{eq:chargetransf}
  \mathbb C^*_j \to \mathbb C^*_j - q_M(u)q_j(u) \mathbb C^*_M = \mathbb C^*_{j} + (C_{\mathfrak g})_{ij} \mathbb C^*_M, 
 \end{equation}
 with $\mathbb C^*_M$ the gauge action in \eqref{eq:cstarreduction}.\footnote{A shortcut to see that \eqref{eq:chargetransf} is the correct transformation is to note that, after the gauge fixing, $u \to 1$, and hence the column of the GLSM charge matrix corresponding to $u$ must be a column of zeros.} The corresponding FI parameter then acquires a $x_9$-dependence
 \begin{equation*}
     \xi_j \to \xi_j + (C_{\mathfrak g})_{ij}x_9, 
 \end{equation*}
 that fibers non-trivially $Y_{\mathfrak g}$ on $x_9$. 
 
 On the other hand, if $x_9 < 0$,  the sign of $x_9$ binds us to gauge-fix \eqref{eq:cstarreduction}  $v = 1$. Then, we have 
 \begin{equation}
     \mathbb C^*_j \to \mathbb C^*_j - q_M(v)q_j(v) \mathbb C^*_M = \mathbb C^*_j, \qquad \xi_j \to \xi_j - q_M(v)q_j(v) x_9 = \xi_j, 
 \end{equation}
 where we have used that $q_j(v) = 0$ for all $j$.
 
 Summing up, the K\"ahler volumes  $\xi_j$ of the $j$-th $\mathbb P^1$ resolving $Y_{\mathfrak g}$ get, after the IIA limit, a piecewise linear dependence on $x_9$. Then, following \cite{Closset_2019}, we can compute the 5d CS levels as the average of the slopes of the K\"ahler parameter for $x_9 > 0$ and $x_9 < 0$. The result that we get from the ECB RG flow \eqref{eq:higgsingchain} on the $i$-th node of the $\mathfrak g$ Dynkin diagram is 
\begin{equation}
\label{eq:CSlevelsECBinduced}
    \Delta k_j = \frac{(C_{\mathfrak g})_{ij}}{2}. 
\end{equation}
We observe that the same procedure can be applied to reduce the rank of a flavor node, increasing by $1/2$ the CS level of the gauge node connected to the flavor node. Finally, we observe that not all such RG-flows lead to CY3 that can be contracted to a full singular phase. For the toric case, the criteria amount to the field theory requiring the corresponding toric diagram to be convex. We leave the analysis of such criteria for the $D, E$ cases for future work. 
\subsection{General coweights and quivers with CS levels: the $A$-case}
\label{sec:genericgeomertryCSAn}
It is tempting to take a quick look at the coweights that do no belong to the coroot lattice. In what follows, we briefly conclude the discussion on CS levels by giving a recipe to compute the partially resolved CY3 geometry for a 5d $A_n$-shaped special unitary quiver with CS levels $\vec{k}$, gauge labels $\vec{r}$, and flavor labels $\lambda$, with $\lambda$ a \textit{generic} coweight of $\mathfrak g$. In particular, we \textit{do not} impose \eqref{eq:gRLDWdef}.

It is a well-known fact that the toric GLSM charges capture the intersection between the cycles of a toric variety. In the case of the $\mathsf{Y}_{A_n}$ singularity,\footnote{See \Cref{sec:appendixAnres} for an explicit description of the resolution of the $\mathsf{Y}_{A_n}$ singularity.} a curve that intersects just the $i-$th $\mathbb P^1$ of the resolved $\mathsf{Y}_{A_n}$ is given by the zero-locus of a polynomial\footnote{In general, these polynomials are not monomial, and the threefold corresponding to the generic coweight is non-toric also for the $A_n$ case.} $\eta_{i}$ in the homogeneous coordinates $(z_1, e_1, ..., e_n, z_2)$, with GLSM charges
\begin{equation}
\label{eq:deltasection}
    q_k(\eta_i)  = \delta^{i}_k. 
\end{equation}
We can construct these $\eta_i$ using some of the partial GLSM gauge invariants, obtained "blowing down" all the $\mathbb P^1$ on the left (resp. right) of the considered $\mathbb P^1$: 
\begin{equation}
\label{eq:etadef}
    \eta_i = a_{left} z_{1}^{i} e_{1}^{i-1} \dots e_{i-1} + a_{right} e_{i+1}\dots e_n^{j-i}z_2^{j-i+1},
\end{equation}
where we take both $a_{left}$ and $a_{right}$ different from zero. 

Each flavor node connected just with the $i$-th gauge node of the  quiver  corresponds to a non-compact two-cycle in the resolved $\mathsf{Y}_{A_n}$ that intersects just the $i$-th $\mathbb P^1$ \cite{DeMarco:2023irn} and hence corresponds to zero locus of one of such $\eta_i$; if the flavor label is $\lambda_i$, we take the polynomial $\eta_i^{\lambda_{i}}$.\footnote{We can also consider $\prod_{i=s}^{n+1} (a_{left,s} z_{1}^{i} e_{1}^{i-1} \dots e_{i-1} + a_{right,s} e_{i+1}\dots z_2^{j-i+1}\eta_{i}^{n+1})$, that correspond to a generic non-zero vev for a complex-valued background mass parameter supported on the $SU(n+1)$ flavor node.} Applying this statement to each flavor node, we can  associate to $\mathbf{r}$ and $ \lambda$, the following equation, in the partially resolved phase analogous to the one presented in \Cref{sec: special unitary}: 
\begin{equation}
\label{eq:branelocusgeneral}
    u v = e_{1}^{\mathbf{r}_{\lambda,1}}e_{2}^{\mathbf{r}_{\lambda,2}}\dots e_{n}^{\mathbf{r}_{\lambda,n}} P_{\lambda}(z_1,e_1,...,e_n,z_2), 
\end{equation}
with $P_{\lambda}(z_1,e_1,...,e_n,z_2) = \eta_{1}^{\lambda_1}...\eta_{n}^{\lambda_n}$. We take \eqref{eq:branelocusgeneral} inside the following toric ambient space
\begin{equation}
\label{eq:ambientspacecs}
  \begin{array}{cccccccccc|c}
      z_1   & e_1  & e_2 & 
    \cdots &  & \cdots & e_n & z_2  & u & v & \text{FI}\\
    \hline
      1 & -2  & 1 & 0 & \cdots & & & 0 & q_{1}(u) & q_1(v) & \xi_1 \\
      0 & 1 & -2 & 1 & 0 & \cdots & & 0 & q_{2}(u) & q_2(v) & \xi_2\\
      \vdots &  &  & &  &  &  & \vdots & \vdots & \vdots & \vdots  \\
      0 & \cdots & &  &  \cdots& 1 & -2 & 1 & q_{n}(u) & q_n(v) & \xi_n\\
    \end{array}
    \end{equation}
We note that to fully specify the threefold, and the ambient space, we need to give the GLSM charges $\mathbf{q}(u), \mathbf{q}(v)$; let us show that this datum corresponds to specify the CS levels $\mathbf{k}$. 

By construction $P_{\lambda}$ has GLSM charges\footnote{We can equivalently say, using the relation between line bundles and coweights recalled below \eqref{eq:coweightlatticeasrelativehomology}, that the polynomial $P_{\lambda}$ is a section of the line bundle $\mathcal O(\lambda)$ on the resolved $\mathsf{Y}_{A_n}$.} 
\begin{equation}
\label{eq:linebundlecharges}
    q(P_{\lambda})_i = \lambda_i,
\end{equation}  
and we have that 
\begin{equation}
\label{eq:chargescompact}
    q_{i}(e_{1}^{\mathbf{r}_{\mu,1}}e_{2}^{\mathbf{r}_{\mu,2}}\dots e_{j}^{\mathbf{r}_{\mu,j}}) = \sum_{j=1}^{\text{rank}\mathfrak (g)}(Q_{red,ij}) \textbf{r}^{j}_\mu = \sum_{j=1}^{\text{rank}\mathfrak (g)}(-C_{\mathfrak g})_{ij} \textbf{r}^{j}_\mu,
\end{equation}
with $Q_{red,{ij}}$ being the charge matrix \eqref{eq:ambientspacecs} with the columns of $z_1, z_2, u, v$ removed.

Adding \eqref{eq:linebundlecharges} and \eqref{eq:chargescompact} we get that 
\begin{equation}
    q_i(u) + q_i(v) = \lambda_i -\sum_j( C_{\mathfrak g})_{ij} \mathbf{r}^j, 
\end{equation}
and hence to specify the geometry we just need to specify $\textbf{q}(u) - \textbf{q}(v)$. In deriving \eqref{eq:CSlevelsECBinduced}, we showed that the K\"ahler volumes in the IIA limit of \cite{Closset_2019} are respectively, for $x_9 = |u|^2 - |v|^2$ greater than zero or less than zero,
\begin{equation}
    \xi_{i,+} = \xi_i -q_i(u) x_9, \qquad \xi_{i,-} = \xi_i + q_i(v) x_9. 
\end{equation}
Consequently, the CS levels are \cite{Closset_2019}
\begin{equation}
    \mathbf{k}_i = \frac{\partial_{x_9}\xi_{i,+} + \partial_{x_9}\xi_{i,-}}{2} = \frac{-q_i(u) + q_i(v)}{2},
\end{equation}
fixing the geometry of the threefold \eqref{eq:branelocusgeneral}, \eqref{eq:ambientspacecs} in terms of the triple $(\textbf{r},\lambda,\textbf{k})$, and viceversa. 
We notice that taking $q_i(u) = 0$, we have\footnote{An analogous argument holds for $q_i(v) = 0$.} 
\begin{equation}  \label{eq:CSlvelsaturation}
\mathbf{k}_i = \textbf{q}_i(v)  = \textbf{q}_i(u v) = \lambda_i -\sum_j( C_{\mathfrak g})_{ij} \mathbf{r}^j,  
\end{equation}
that, for $q_i(v) > 0$, reproduces the flavor enhancement condition \cite{Yonekura:2015ksa}. Furthermore, imposing as in the previous sections $\mathbf{q}(u) = \mathbf{q}(v) = 0$, (and hence, in particular, that $\mathbf{k} = 0$) we see that \eqref{eq:CSlvelsaturation} reduces to \eqref{eq:gRLDWdef}. 

We notice that not all the CS $\mathfrak g$-shaped Dynkin quivers are expected to be UV completed by a 5d SCFT. Indeed, this corresponds to the fact that nothing guarantees that the threefold specified by \eqref{eq:branelocusgeneral}, \eqref{eq:ambientspacecs}  can be crepantly contracted to a canonical singularity. 

We conclude by noticing that the  intersection form coincides with minus the Cartan matrix also for the other Du Val singularities. Hence, even if in the non-toric Du Val singularities we do not have a neat GLSM description, one might expect a similar procedure to be the key to obtain systematically quiver with CS levels also for those cases.\footnote{In a sense, however, a similar result is reasonably expected by the following argument: we can always resolve $\mathsf{Y}_{\mathfrak{g}}$ with blowups in toric ambient spaces, whose $\mathbb C^*$ charges label the line bundles over the resolved $\mathsf{Y}_{\mathfrak{g}}$. This can be achieved by blowing up each time a non-Cartier divisor of the Du Val singularity, instead of a point inside of it. A compact divisor on $\mathsf{Y}_{\mathfrak{g}}$ is uniquely identified by its intersection with all the other divisors, as the intersection form is non-degenerate. Consequently, the $\mathbb C^*$ charges of the sections of a certain line bundle over $\mathsf{Y}_{\mathfrak{g}}$ are again determined by multiplying by the intersection form (namely, minus the Cartan matrix). In particular, the analogue of the $\eta_j$ would be again obtained starting from a full resolution of $\mathsf{Y}_{\mathfrak g}$, and blowing down, passing to the corresponding partial GLSM invariants, all the $\mathbb P^1$'s but the $j$-th one.}

\section{6d origin of 5d CM}\label{sec: 6d origin}
We can now note an interesting consequence of our work. A conjecture formulated in \cite{Jefferson:2017ahm} asserts that every 5d SCFT has a 6d origin. Available mechanisms to obtain 5d SCFTs in such fashion are the circle reduction of 6d $\mathcal{N}=(1,0)$ theories, possibly followed by mass deformations, as well as integrating out BPS strings \cite{Bhardwaj:2019xeg}.\footnote{Compactifications of 6d conformal matter were also exploited to produce 4d theories with 8 \cite{Ohmori:2015pua,Ohmori:2015pia} and 4 supercharges  \cite{Razamat:2016dpl,Kim:2017toz,Kim:2018bpg,Kim:2018lfo}.} In what follows, we will obtain all the 5d conformal matter theories presented in this work as circle reduction of 6d conformal matter, followed by an Higgs branch RG flow.

\paragraph{HB RG flows and 6d origin of 5d conformal matter}
To understand the 6d origin of 5d conformal matter, we first show how, for each $\mathfrak g$, there are  5d CM molecules that directly descend from circle reduction (without background for the flavor symmetries on the circle) of 6d CM molecules. Then, starting from such molecules, we reach any other kind of 5d CM theory by the HB deformations followed by an RG flow. 

 6d conformal matter can be constructed via F-theory on an elliptically fibered CY3 $X_{ell}$:
\begin{equation}
\label{eq:ellfibrationgen}
\begin{tikzcd}
\mathbb E \arrow[r, hook] & X_{ell} \arrow[d, "\pi"] \\
& B
\end{tikzcd}
\end{equation}
where $\mathbb E$ is the elliptic curve, fibered over a complex surface $B$. $X_{ell}$ is typically presented in the form of a Weierstrass model: 
\begin{equation}
\label{eq:weierstrass}
    x^2  = y^3 + f(u,v) y + g(u,v), \qquad (u,v) \in B, \qquad (x,y) \in \mathbb C^2.
\end{equation}
It is important to notice that \eqref{eq:weierstrass} only represents a patch of $\mathbb E$, that covers all of $\mathbb E$ but a point $p_{\infty}$; the full $\mathbb E$ is obtained by compactifying \eqref{eq:weierstrass} inside $\mathbb P_{231}$. In the F-theory setup, the elliptic curve does not represent spacetime directions, but rather the value of the axio-dilaton of a non-perturbative IIB background. Reducing the 6d conformal matter theory on a circle is equivalent to take a decompactification limit of $\mathbb E$, \cite{Bhardwaj:2019fzv}, with different kinds of decompactifications corresponding to different backgrounds, on the circle, for line operators charged under the global symmetries of the 6d theory.
If we do not want to turn on any background for the flavor symmetries, the decompactification limit simply amounts to forgetting about $p_{\infty} \in \mathbb E$, working in the affine patch \eqref{eq:weierstrass}.

We can see that some of the threefolds that we used to construct 5d conformal matter are elliptically fibered, and \textit{hence} can be thought as the outcome of a decompactification of an F-theory model, or equivalently as a circle reduction without Wilson line backgrounds of a 6d  conformal matter (1,0) theory. An example of this is $X_{E_6}^z$: 
\begin{equation}
\label{eq:e6zcut}
    x^2 +y^3 + u^4 v^4=0, 
\end{equation}
that corresponds to a Weierstrass model with $f = 0$, and, up to performing the coordinate change $x \to i x$, $g = u^4 v^4$ and $B = \mathbb C^2 \ni (u,v).$ This model, if considered in the F-theory setup, engineers the 6d conformal matter of type $E_6$ \cite{DelZotto:2014hpa}. We can then conclude that the 5d conformal matter atom obtained considering M-theory on \eqref{eq:e6zcut} is  the circle reduction of the corresponding 6d conformal matter theory, as it arises as a special decompactification of the latter.\\ 
As pointed out in \cite{DeMarco:2023irn}, the same statement applies to the  $X_{E_6}^{z^j}$,$X_{E_7}^{z^j}$,$X_{E_8}^{z^j}$.

Consider now the threefolds $X_{D_k}^{z^j}$
\begin{equation}
    \label{eq:dkftheory}
    x^2 + z y^2 + z^{k-1} = 0, \quad z^j = u v.
\end{equation}
We now first add a new (irrelevant) term to the first equation of \eqref{eq:dkftheory}:

\begin{equation}
\label{eq:irrelevantshift}
    x^2 + z y^2 + z^{k-1} = 0, \qquad \longrightarrow \qquad x^2 + z y^2 + \boxed{y^3} +  z^{k-1} = 0.
\end{equation}
The boxed term does not change the leading behaviour of the threefold nearby the locus where the 5d SCFT localizes, and then preserves the 5d theory.  
Then, we perform the following coordinate redefinition
\begin{equation}
    y\to \frac{1}{3} (y-z),\qquad x\to \frac{x}{3 \sqrt{3}}.
\end{equation}
that allows us to rewrite the  threefold in the following form: 
\begin{equation}
\label{eq:weiermoleculedk}
    x^2+y^3-3 y z^2+2 z^3 + 27 z^{k-1} = 0, \qquad z^j = u v. 
\end{equation}
We can explicitly see that, fixing $z,u,v$ in \eqref{eq:weiermoleculedk}, we get an elliptic curve (described by varying $x,y$), and furthermore the threefold is globally in the Weierstrass form. In the notation of \eqref{eq:ellfibrationgen}, the base $B$ is the zero locus of $z^j - u v$ in $\mathbb C^3 \ni (u,v,z)$, and the projection $\pi$ is realized by fixing $(u,v,z)$ to specific numerical values satisfying the aforementioned equation. 

We then conclude that the 5d SCFT engineered via M-theory on \eqref{eq:weiermoleculedk} is the circle reduction of the 6d conformal matter one. To go back to 6d,  we  homogenize the first equation of \eqref{eq:weiermoleculedk} inside $\mathbb P_{231}$, adding in this way the point at infinity of the elliptic curve:
\begin{equation}
    \label{eq:homogenizedweiermoleculedk}
      \begin{cases}
          X^2+Y^3-3 T^4 Y z^2+(2 z^3 + 27 z^{k-1}) T^6 = 0,\\
          z^j = u v
      \end{cases} 
      \qquad [X:Y:T] \in \mathbb P_{231}, \quad (u,v,z) \in \mathbb C^3.  
\end{equation}

This corresponds to taking the F-theory limit of M-theory on \eqref{eq:weiermoleculedk}, making an infinite tower of KK states light, and producing in this way the 6d (1,0) $D_k$ conformal matter molecule obtained by the fusion of $j$ conformal matter atoms. One can also proceed in the opposite direction, starting from \eqref{eq:homogenizedweiermoleculedk} and then restricting to the patch $T \neq 0$, obtaining \eqref{eq:weiermoleculedk} and circle reducing the 6d (1,0) theory to 5d.
We will adopt this perspective for the remaining part of this section. 

Finally, we briefly mention that the same argument also applies to $X_{A_n}^{z^j}$, simply adding a cubic term as follows: 
\begin{equation}
    \begin{cases}
    x^2 + y^2 + z^n = 0, \\
    uv = z^j
    \end{cases} \qquad \longrightarrow \qquad  \begin{cases}
    x^2 + y^2 + \boxed{y^3} +z^n = 0, \\
    uv = z^j,
    \end{cases}
\end{equation}
and homogenizing the equation within $\mathbb P_{231}$. 

To sum up, we have shown that, for each $\mathfrak g$, the molecule $X_{\mathfrak g}^{z^j}$ descends from circle reduction of the corresponding 6d CM matter molecule. We can now exploit this fact to describe the 6d origin of \textit{any} conformal matter theory of $\mathfrak{g}$-type in 5d. 

More specifically, let us now show that all the 5d conformal matter theories of type $\mathfrak{g}$ can be reached with HB RG-flows from $X_{\mathfrak{g}}^{z^j}$, for $j$ sufficiently large. We can prove this as follows: given a 5d CM theory of $X_{\mathfrak{g}}^{z^j}$ type
\begin{equation}\label{Dk zj 3fold}
\begin{cases}
    Y_{\mathfrak{g}}(x,y,z) =0 \\
uv = z^j
    \end{cases},
\end{equation}
we have shown in Section \eqref{sec: special unitary} a choice of partial resolution of the threefold \eqref{Dk zj 3fold}, that produces a low-energy Lagrangian special unitary quiver phase, with gauge nodes arranged in the shape of the $\mathfrak{g}$ Dynkin diagram. As reviewed  in Section \ref{sec: Higgs branch}, the dynamical complex structure deformations of this phase can be characterized via the techniques of \cite{Collinucci:2020jqd}, relying on a type IIA setup with D6-branes wrapped on the singular $\mathbb{P}^1$'s of the partially resolved threefold obtained from \eqref{Dk zj 3fold}. The dynamical deformations modify the second equation in \eqref{Dk zj 3fold}:
\begin{equation}\label{Dk zj 3fold def}
\begin{cases}
    Y_{\mathfrak{g}}(x,y,z) =0 \\
uv = z^j + \text{def}(x,y,z)
    \end{cases},
\end{equation}
where $\text{def}(x,y,z)$ can be explicitly computed, and comprises a \textit{finite} number of monomials.\\
\indent With these tools at hand, it is straightforward to prove that \textit{all} 5d CM theories of type $\mathfrak{g}$ can be obtained via Higgsing from $X_{\mathfrak{g}}^{z^j}$, for some sufficiently large $j$. Consider a generic 5d CM theory of type $\mathfrak{g}$:
\begin{equation}\label{generic Dk CM}
  \begin{cases}
    Y_{\mathfrak{g}}(x,y,z) =0 \\
uv = w(x,y,z)
    \end{cases}  ,
\end{equation}
with $w(x,y,z)$ a combination of the generators of type $\mathfrak{g}$ in Table \ref{table:generators}. Then, there always exists a choice of $j$ such that \eqref{Dk zj 3fold} admits a dynamical deformation that triggers a flow to \eqref{generic Dk CM}:
\begin{equation}
    \begin{cases}
       Y_{\mathfrak{g}}(x,y,z) =0 \\
uv = z^j 
    \end{cases}
 \quad \longrightarrow \quad\quad
 \begin{cases}
         Y_{\mathfrak{g}}(x,y,z) =0 \\
uv = z^j + w(x,y,z)
    \end{cases}.
\end{equation}
This is possible since, for $j$ large enough, the polynomial $w(x,y,z)$ is obtained as a tuning of the dynamical deformations $\text{def}(x,y,z)$ appearing in \eqref{Dk zj 3fold def}.

Geometrically, this means that any 5d conformal matter threefold, even the \textit{non-elliptically fibered ones}, can be obtained by first performing an  appropriate complex deformation of \eqref{Dk zj 3fold}, and then by decoupling irrelevant monomials.

We can therefore conclude what follows:\\

\textit{Each 5d conformal matter theory of $\mathfrak g$-type can be obtained from a $\mathfrak g$-type 6d conformal matter molecule by circle-reduction and HB RG flow.}\\

It is then an open question to understand whether the trinions and tetraons predicted in \cite{DeMarco:2023irn}, which are engineered by M-theory on CY3 which are constructed in a similar fashion as the 5d bifundamental CM studied in this paper, do have a 6d origin \cite{DeMarcoDelZottoGraffeoSangiovanni:WIP}.
\section*{Acknowledgements}
We would like to thank Amihay Hanany, Rudolph Kalveks and Zhenghao Zhong for discussions on small representations, their associated affine Grassmannian slices and nilpotent orbits. Furthermore, we would like to thank Andr\'es Collinucci, Roberto Valandro and Shani Meynet for discussions.

JFG is supported by the EPSRC Open Fellowship (Schafer-Nameki) EP/X01276X/1 and the ``Simons Collaboration on Special Holonomy in Geometry, Analysis and Physics''. JFG is grateful for the hospitality of the Centre for Geometry and Physics at Uppsala University, where part of this project was completed.
The research of M.D.M. is funded through an ARC advanced project, and further supported by IISN-Belgium (convention 4.4503.15).
The research of AS is funded by the VR Centre for Geometry and Physics (VR grant No.\ 2022-06593) at Uppsala University. AS wishes to thank the IPhT in Saclay for hospitality during the initial stage of this work, and the the IRN:QFS network that supported the visit. AS also acknowledges the ERC grant No. 101171852-HIGH and the VR project grant No. 2023-05590 for partial support. The work of MDZ is supported by the European Research Council (ERC) under the European Union’s Horizon Europe research and innovation program (grant agreement No. 101171852) and by the VR project grant
No. 2023-05590. MDZ also acknowledges the VR Centre for Geometry and Physics
(VR grant No. 2022-06593) and the Simons Foundation International Grant for the
Simons Collaboration on Global Categorical Symmetries.

\appendix

\section{Toric ambient space resolution of Du Val singularities}
\label{app:DuValres}

In this appendix we schematically collect the data that encodes the crepant resolution of Du Val singularities in a suitable toric ambient space. We list:
\begin{itemize}
    \item the Du Val singularity;
    \item the toric ambient space coordinates and the corresponding $\mathbb{C}^*$ actions;
    \item the resolved Du Val singularity and the blowup map;
    \item the loci of the resolved curves.
\end{itemize}

\subsection{$\boldsymbol{A_n}$}
\label{sec:appendixAnres}
\noindent The $A_n$ singularities are:
\begin{equation}\label{An Du Val}
xy=z^{n+1}.
\end{equation}
One can perform a resolution of \eqref{An Du Val} in the toric ambient space characterized by the weights:
\begin{equation}\label{toric action A4}
    \begin{array}{cccccccc}
      z_1   & e_1  & e_2 & 
    \cdots &  & \cdots & e_n & z_2  \\
    \hline
      1 & -2  & 1 & 0 & \cdots & & & 0 \\
      0 & 1 & -2 & 1 & 0 & \cdots & & 0 \\
      \vdots &  &  & &  &  &  & \vdots  \\
      0 & \cdots & &  &  \cdots& 1 & -2 & 1 \\
    \end{array}
\end{equation}
The blowup map is:
\begin{equation}\label{An blowdown}
\begin{split}
    &x =e_1e_2^2\cdot\ldots\cdot e_n^nz_2^{n+1},\\
    &y = z_1^{n+1}e_1^n e_2^{n-1}\cdot\ldots\cdot e_n, \\
    & z = z_1e_1e_2\cdot\ldots\cdot e_n z_2,
    \end{split}
\end{equation}

\noindent The exceptional curves are given by the ideals:
\begin{equation}\label{curves A}
\begin{array}{l}
    \mathcal{C}_1 =(e_1) \\
    \mathcal{C}_2 =(e_2) \\
    \hspace{0.2cm}\vdots\\ 
    \mathcal{C}_n =(e_n) \\
\end{array} \quad\quad
    \scalebox{0.65}{\begin{tikzpicture}
        \draw[thick] (0,0) circle (0.65);
        \node at (0,0) {\small$\mathcal{C}_2$};
        \draw[thick] (0.7,0)--(1.3,0);
        \node at (2,0) {\small$\cdots$};
        \draw[thick] (-0.7,0)--(-1.3,0);
        \draw[thick] (-2,0) circle (0.65);
        \node at (-2,0) {\small$\mathcal{C}_1$};
        \draw[thick] (2.7,0)--(3.3,0);
        \draw[thick] (4,0) circle (0.65);
        \node at (4,0) {\small$\mathcal{C}_n$};
        \end{tikzpicture}}
\end{equation}

\subsection{$\boldsymbol{D_5}$}
The $D_5$ Du Val singularity reads:
\begin{equation}\label{D5 du val}
    x^2+zy^2+z^4 = 0
\end{equation}
The resolution of the Du Val singularity is embedded in a toric ambient space with coordinates $d_i$ and toric action:
\begin{equation}\label{toric action D5}
    \begin{array}{ccccccc}
      d_1   & d_2  & d_3 & 
    d_4  & d_5 & d_6 & d_7  \\
    \hline
      1 & 1  & 1 & -1 & 0 & 0 & 0 \\
      1 & 1 & 0 & 1 & -1 & 0 & 0 \\
      1 & 0 & 1 & 1 & 0 & -1 & 0 \\
      1 & 0 & 0 & 1 & 0 & 0 & -1\\
    \end{array}
\end{equation}
The proper transform of the Du Val singularity \eqref{D5 du val} is:
\begin{equation}
d_1^2 d_7+d_3^2 d_2 d_4
   d_6+d_2^4 d_4^2 d_5^4 d_7 = 0.
\end{equation}
The Stanley-Reisner ideal implies that the following loci are not part of the resolved variety:
\begin{equation}
\begin{split}
    & L_1 = \{d_2 d_5,d_3 d_6,d_1
   d_5 d_6 d_7\}=0,\\
  & L_2 = \{d_2,d_1
   d_6 d_7,d_4 d_6 d_7\}=0,\\
   & L_3 = \{d_1 d_7,d_4 d_7,d_3\}=0,\\
   & L_4 = \{d_1,d_4\}=0. \\
   \end{split}
\end{equation}
The blow-down map acts on the starting coordinates $x,y,z$ as:
\begin{equation}\label{D5 blowdown}
\begin{split}
  &  x = d_4 d_1d_5^2d_6^2d_7^2,\\
  & y = d_4d_3d_5d_6^2d_7, \\
  & z= d_4d_2d_5^2d_6d_7.
    \end{split}
\end{equation}
There are $\text{rank}(D_5)$ exceptional curves leftover after the blow-up, arranged exactly like the $D_5$ Dynkin diagram. The exceptional curves are given by the ideals:
\begin{equation}\label{curves D5}
\begin{array}{l}
    \mathcal{C}_1 = (d_5,d_7d_1^2+d_4d_2d_3^2d_6)\\ 
       \mathcal{C}_2 =(d_7,d_4)\\
    \mathcal{C}_3 =(d_7,d_6) \\
     \mathcal{C}_4 =(d_6,d_1-id_4d_2^2d_5^2) \\   
    \mathcal{C}_5 =(d_6,d_1+id_4d_2^2d_5^2) \\
 
\end{array} \quad\quad
   \scalebox{0.65}{\begin{tikzpicture}
        \draw[thick] (0,0) circle (0.65);
        \node at (0,0) {\small$\mathcal{C}_1$};
        \draw[thick] (0.7,0)--(1.3,0);
        \draw[thick] (2,0) circle (0.65);
        \node at (2,0) {\small$\mathcal{C}_2$};
        \draw[thick] (2.7,0)--(3.3,0);
        \draw[thick] (4,0) circle (0.65);
        \node at (4,0) {\small$\mathcal{C}_3$};
        \draw[thick] (5.7,-1.5) circle (0.65);
         \node at (5.7,-1.5) {$\mathcal{C}_4$};
          \draw[thick] (5.7,1.5) circle (0.65);
         \node at (5.7,1.5) {$\mathcal{C}_5$};
         \draw[thick] (4.5,0.5)--(5.1,1.1);
         \draw[thick] (4.5,-0.5)--(5.1,-1.1);
        \end{tikzpicture}}
\end{equation}

\subsection{$\boldsymbol{D_6}$}

The $D_6$ Du Val singularity reads:
\begin{equation}\label{D6 du val}
    x^2+zy^2+z^5 = 0
\end{equation}
The resolution of the Du Val singularity is embedded in a toric ambient space with coordinates $d_i$ and toric action:

\begin{equation}\label{toric action D6}
    \begin{array}{ccccccc}
    d_1 & d_2 & d_3 & d_4 & d_5 & d_6 & d_7 \\
\hline
 1 & 1 & 1 & -1 & 0 & 0 & 0 \\
 1 & 1 & 0 & 1 & -1 & 0 & 0 \\
 1 & 0 & 1 & 1 & 0 & -1 & 0 \\
 1 & 0 & 0 & 0 & 0 & 1 & -1 \\
    \end{array}
\end{equation}
The proper transform of the Du Val singularity \eqref{D6 du val} is:
\begin{equation}\label{proper D6}
d_1^2 d_7+d_2 d_3^2 d_4 d_6+d_2^5 d_4^3 d_5^6 d_6= 0
\end{equation}
The Stanley-Reisner ideal implies that the following loci are not part of the resolved variety:
\begin{equation}
\begin{split}
    & L_1 = \left\{d_2 d_5,d_6 d_3 d_7,d_6 d_1 d_5 d_7\right\}=0,\\
   & L_2 = \left\{d_2,d_6 d_1 d_7,d_4 d_6 d_7\right\}=0,\\
   & L_3 = \{d_1 d_7,d_4,d_3\}=0,\\
   & L_4 = \{d_1,d_6\} = 0.
   \end{split}
\end{equation}
The blow-down map acts on the starting coordinates $x,y,z$ as:
\begin{equation}\label{D6 blowdown}
\begin{split}
   & x = d_5^2 d_4 d_6^2 d_7^3 d_1, \\
   & y = d_5 d_4 d_6^2 d_7^2 d_3,\\
   & z = d_5^2 d_4 d_6 d_7 d_2.\\
    \end{split}
\end{equation}
There are $\text{rank}(D_6)$ exceptional curves leftover after the blow-up, arranged exactly like the $D_6$ Dynkin diagram. The exceptional curves are given by the ideals:
\begin{equation}\label{curves D6}
\begin{array}{l}
\mathcal{C}_1 =(d_5,d_7d_1^2+d_4 d_6 d_2d_3^2)\\
\mathcal{C}_2 =(d_4,d_1) \\
\mathcal{C}_3 =(d_7,d_4) \\ 
\mathcal{C}_4 =(d_7,d_6) \\
\mathcal{C}_5 =(d_7,d_3-i d_5^3 d_4 d_2^2) \\
\mathcal{C}_6 =(d_7,d_3+i d_5^3 d_4 d_2^2)\\
 \\
\end{array} \quad\quad
    \scalebox{0.65}{\begin{tikzpicture}
        \draw[thick] (0,0) circle (0.65);
        \node at (0,0) {\small$\mathcal{C}_2$};
        \draw[thick] (0.7,0)--(1.3,0);
        \draw[thick] (2,0) circle (0.65);
        \node at (2,0) {\small$\mathcal{C}_3$};
        \draw[thick] (-0.7,0)--(-1.3,0);
        \draw[thick] (-2,0) circle (0.65);
        \node at (-2,0) {\small$\mathcal{C}_1$};
        \draw[thick] (2.7,0)--(3.3,0);
        \draw[thick] (4,0) circle (0.65);
        \node at (4,0) {\small$\mathcal{C}_4$};
        \draw[thick] (5.7,-1.5) circle (0.65);
         \node at (5.7,-1.5) {$\mathcal{C}_5$};
          \draw[thick] (5.7,1.5) circle (0.65);
         \node at (5.7,1.5) {$\mathcal{C}_6$};
         \draw[thick] (4.5,0.5)--(5.1,1.1);
         \draw[thick] (4.5,-0.5)--(5.1,-1.1);
        \end{tikzpicture}}
\end{equation}

\subsection{$\boldsymbol{E_6}$}
\label{sec:rese6}
We perform a resolution of the $E_6$ Du Val singularity 
\begin{equation}\label{E6 du val}
    x^2+y^3+z^4 = 0
\end{equation}
in a toric ambient space, specified by the weights:
\begin{equation}\label{toric action E6}
    \begin{array}{ccccccc}
      d_1   & d_2  & d_3 & 
    d_4  & d_5 & d_6 & d_7   \\
    \hline
    1 &  1 & 1   & -1 & 0 & 0 & 0 \\
    1 &  1 & 0  & 1 & -1 & 0 & 0 \\
    1 & 0 & 0  & 1 & 1 & -1 & 0 \\
    1 &  0 & 0  & 1 & 0 & 1 & -1 \\
    \end{array}
\end{equation}
The Stanley-Reisner ideal implies that the following loci are not part of the resolved variety:
\begin{equation}
\begin{split}
    & L_1 =\left\{d_5 d_7 d_6
   d_2,d_3,d_5 d_7 d_6
   d_1\right\}=0,\\
   & L_2 = \left\{d_2,d_7 d_6
   d_1,d_7 d_4 d_6\right\}=0,\\
   & L_3 = \{d_7
   d_1,d_7
   d_4,d_5\}=0,\\
   & L_4 = \{d_1,d_4,d_6\}=0.
   \end{split}
\end{equation}
The blow-down map acts on the starting coordinates $x,y,z$ as:
\begin{equation}\label{E6 blowdown}
\begin{split}
  &  x = d_1 d_4 d_5^2 d_6^4 d_7^6,\\
  & y = d_2 d_4 d_5^2 d_6^3 d_7^4, \\
  & z= d_3 d_4 d_5 d_6^2 d_7^3.
    \end{split}
\end{equation}
The proper transform of the Du Val singularity \eqref{E6 du val} is then:
\begin{equation}
d_1^2+d_2^3d_4d_5^2d_6+d_3^4d_4^2 = 0.
\end{equation}
There are $\text{rank}(E_6)$ exceptional curves leftover after the blow-up, arranged exactly like the $E_6$ Dynkin diagram. The exceptional curves are given by the ideals:
\begin{equation}\label{curves E6}
\begin{array}{l}
    \mathcal{C}_1 =(d_5,d_1+id_4d_3^2) \\
    \mathcal{C}_2 =(d_6,d_1+id_4d_3^2) \\
    \mathcal{C}_3 =(d_7,d_1^2+d_4d_5^2d_6d_2^3+d_4^2d_3^4)\\
    \mathcal{C}_4 = (d_6,d_1-id_4d_3^2)\\ 
    \mathcal{C}_5 =(d_5,d_1-id_4d_3^2) \\
    \mathcal{C}_6 =(d_4,d_1) \\
\end{array} \quad\quad
    \scalebox{0.65}{\begin{tikzpicture}
        \draw[thick] (0,0) circle (0.65);
        \node at (0,0) {\small$\mathcal{C}_3$};
        \draw[thick] (0.7,0)--(1.3,0);
        \draw[thick] (2,0) circle (0.65);
        \node at (2,0) {\small$\mathcal{C}_4$};
        \draw[thick] (-2.7,0)--(-3.3,0);
        \draw[thick] (-0.7,0)--(-1.3,0);
        \draw[thick] (-4,0) circle (0.65);
        \node at (-4,0) {\small$\mathcal{C}_1$};
        \draw[thick] (-2,0) circle (0.65);
        \node at (-2,0) {\small$\mathcal{C}_2$};
        \draw[thick] (0,0.7)--(0,1.3);
        \draw[thick] (0,2) circle (0.65);
        \node at (0,2) {\small$\mathcal{C}_6$};
        \draw[thick] (2.7,0)--(3.3,0);
        \draw[thick] (4,0) circle (0.65);
        \node at (4,0) {\small$\mathcal{C}_5$};
        \end{tikzpicture}}
\end{equation}

\subsection{$\boldsymbol{E_7}$}
We perform a resolution of the $E_7$ Du Val singularity 
\begin{equation}\label{E7 du val}
    x^2+y^3+y z^3 = 0
\end{equation}
in a toric ambient space, specified by the weights:
\begin{equation}\label{toric action E7}
    \begin{array}{cccccccc}
      d_1   & d_2  & d_3 & 
    d_4  & d_5 & d_6 & d_7 & d_8  \\
    \hline
      1 & 1  & 1 & -1 & 0 & 0 & 0 & 0\\
      1 & 1 & 0 & 1 & -1 & 0 & 0 & 0\\
      1 & 1 & 0 & 0 & 1 & -1 & 0 & 0\\
      1 & 0 & 0 & 1 & 1 & 0 & -1 & 0\\
      1 & 0 & 0 & 0 & 0 & 0 & 1 & -1\\
    \end{array}
\end{equation}
The proper transform of the Du Val singularity \eqref{E7 du val} is then:
\begin{equation}
d_1^2 d_8+d_4 d_5 d_7 d_2
   \left(d_4 d_3^3+d_5 d_2^2 d_6^3\right) = 0.
\end{equation}
The Stanley-Reisner ideal implies that the following loci are not part of the resolved variety:
\begin{equation}
\begin{split}
    & L_1 =\left\{d_5 d_7 d_2 d_6 d_8,d_3,d_5 d_7 d_1 d_6
   d_8\right\}=0,\\
   & L_2 = \left\{d_2 d_6,d_7
   d_1 d_6 d_8,d_4 d_7
   d_8\right\}=0,\\
   & L_3 = \left\{d_7 d_1 d_8,d_2,d_5 d_7 d_8\right\}=0,\\
   & L_4 = \{d_4,d_5,d_1 d_8\}=0,\\
   & L_5 = \{d_1,d_7\}=0.
   \end{split}
\end{equation}
The blow-down map acts on the starting coordinates $x,y,z$ as:
\begin{equation}\label{E7 blowdown}
\begin{split}
  &  x = d_6^3 d_4 d_5^2 d_8^5 d_7^4 d_1,\\
  & y = d_6^3 d_4 d_5^2 d_8^3 d_7^3 d_2, \\
  & z= d_6 d_4 d_5 d_8^2 d_7^2 d_3.
    \end{split}
\end{equation}
There are $\text{rank}(E_7)$ exceptional curves leftover after the blow-up, arranged exactly like the $E_7$ Dynkin diagram. The exceptional curves are given by the ideals:
\begin{equation}\label{curves E7}
\begin{array}{l}
    \mathcal{C}_1 =(d_4,d_1) \\
    \mathcal{C}_2 =(d_4,d_8) \\
    \mathcal{C}_3 =(d_7,d_8)\\
    \mathcal{C}_4 = (d_5,d_8)\\ 
    \mathcal{C}_5 =(d_5,d_1) \\
    \mathcal{C}_6 =(d_6,d_4^2 d_5 d_7 d_2 d_3^3+d_8 d_1^2) \\
    \mathcal{C}_7 =(d_8,d_4 d_3^3+d_6^3 d_5 d_2^2) \\
\end{array} \quad\quad
    \scalebox{0.65}{\begin{tikzpicture}
        \draw[thick] (0,0) circle (0.65);
        \node at (0,0) {\small$\mathcal{C}_3$};
        \draw[thick] (0.7,0)--(1.3,0);
        \draw[thick] (2,0) circle (0.65);
        \node at (2,0) {\small$\mathcal{C}_4$};
        \draw[thick] (-2.7,0)--(-3.3,0);
        \draw[thick] (-0.7,0)--(-1.3,0);
        \draw[thick] (-4,0) circle (0.65);
        \node at (-4,0) {\small$\mathcal{C}_1$};
        \draw[thick] (-2,0) circle (0.65);
        \node at (-2,0) {\small$\mathcal{C}_2$};
        \draw[thick] (0,0.7)--(0,1.3);
        \draw[thick] (0,2) circle (0.65);
        \node at (0,2) {\small$\mathcal{C}_7$};
        \draw[thick] (2.7,0)--(3.3,0);
        \draw[thick] (4,0) circle (0.65);
        \node at (4,0) {\small$\mathcal{C}_5$};
        \draw[thick] (4.7,0)--(5.3,0);
        \draw[thick] (6,0) circle (0.65);
        \node at (6,0) {\small$\mathcal{C}_6$};
        \end{tikzpicture}}
\end{equation}

\subsection{$\boldsymbol{E_8}$}
We perform a resolution of the $E_8$ Du Val singularity 
\begin{equation}\label{E8 du val}
    x^2+y^3+z^5 = 0
\end{equation}
in a toric ambient space, specified by the weights:
\begin{equation}\label{toric action E8}
    \begin{array}{ccccccccc}
      d_1   & d_2  & d_3 & 
    d_4  & d_5 & d_6 & d_7 & d_8 & d_9  \\
    \hline
      1 & 1  & 1 & -1 & 0 & 0 & 0 & 0 & 0\\
  1&1&0&1&-1&0&0 &0 &0\\
  1&0&0&1&1&-1&0&0&0 \\
  1&0&0&1& 0& 1&-1& 0&0 \\
  1&0&0&0&1&1&0&-1&0 \\
  1&0&0&0&0&0&0&1&-1 \\
    \end{array}
\end{equation}
The proper transform of the Du Val singularity \eqref{E8 du val} is then:
\begin{equation}
d_1^2 d_9+d_4 d_5 d_6 d_8
   \left(d_5 d_3^3+d_4^2 d_6 d_2^5 d_7^3\right)= 0.
\end{equation}
The Stanley-Reisner ideal implies that the following loci are not part of the resolved variety:
\begin{equation}
\begin{split}
    & L_1 =\left\{d_5 d_6 d_8 d_3 d_7
   d_9,d_2,d_5 d_6 d_8 d_1
  d_7 d_9\right\}=0,\\
   & L_2 = \left\{d_3,d_6
   d_8 d_1 d_7 d_9,d_4
   d_6 d_8 d_7 d_9\right\}=0,\\
   & L_3 = \left\{d_8 d_1 d_7d_9,d_4 d_7,d_5 d_8 d_9\right\} = 0,\\
   & L_4 = \left\{d_8 d_1 d_9,d_4,d_6
   d_8 d_9\right\}=0,\\
   & L_5 = \{d_1 d_9,d_5,d_6\}=0,\\
   & L_6 = \{d_1,d_8\}=0.
   \end{split}
\end{equation}
The blow-down map acts on the starting coordinates $x,y,z$ as:
\begin{equation}\label{E8 blowdown}
\begin{split}
  &  x =d_7^6 d_4 d_9^8 d_5^2 d_6^4 d_8^7 d_1,\\
  & y = d_7^4 d_4 d_9^5 d_5^2 d_6^3 d_8^5 d_3, \\
  & z= d_7^3 d_4 d_9^3 d_5 d_6^2 d_8^3 d_2.
    \end{split}
\end{equation}
There are $\text{rank}(E_8)$ exceptional curves leftover after the blow-up, arranged exactly like the $E_8$ Dynkin diagram. The exceptional curves are given by the ideals:
\begin{equation}\label{curves E8}
\begin{array}{l}
    \mathcal{C}_1 =(d_5,d_1) \\
    \mathcal{C}_2 =(d_5,d_9) \\
    \mathcal{C}_3 =(d_8,d_9)\\
    \mathcal{C}_4 = (d_6,d_9)\\ 
    \mathcal{C}_5 =(d_6,d_1) \\
    \mathcal{C}_6 =(d_7,d_4 d_5^2 d_6 d_8 d_3^3+d_9 d_1^2) \\
    \mathcal{C}_7 =(d_4,d_1) \\
    \mathcal{C}_8 =(d_9,d_7^3 d_4^2 d_6 d_2^5+d_5 d_3^3)\\
\end{array} \quad\quad
    \scalebox{0.65}
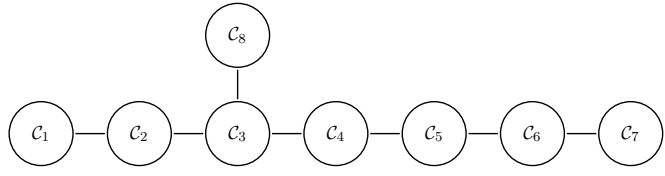
{\begin{tikzpicture}
        \draw[thick] (0,0) circle (0.65);
        \node at (0,0) {\small$\mathcal{C}_3$};
        \draw[thick] (0.7,0)--(1.3,0);
        \draw[thick] (2,0) circle (0.65);
        \node at (2,0) {\small$\mathcal{C}_4$};
        \draw[thick] (-2.7,0)--(-3.3,0);
        \draw[thick] (-0.7,0)--(-1.3,0);
        \draw[thick] (-4,0) circle (0.65);
        \node at (-4,0) {\small$\mathcal{C}_1$};
        \draw[thick] (-2,0) circle (0.65);
        \node at (-2,0) {\small$\mathcal{C}_2$};
        \draw[thick] (0,0.7)--(0,1.3);
        \draw[thick] (0,2) circle (0.65);
        \node at (0,2) {\small$\mathcal{C}_8$};
        \draw[thick] (2.7,0)--(3.3,0);
        \draw[thick] (4,0) circle (0.65);
        \node at (4,0) {\small$\mathcal{C}_5$};
        \draw[thick] (4.7,0)--(5.3,0);
        \draw[thick] (6,0) circle (0.65);
        \node at (6,0) {\small$\mathcal{C}_6$};
         \draw[thick] (6.7,0)--(7.3,0);
        \draw[thick] (8,0) circle (0.65);
        \node at (8,0) {\small$\mathcal{C}_7$};
        \end{tikzpicture}}
\end{equation}

\section{Construction of the basis of distinguished weights}
\label{appendix:proofdistinguished}
In this Appendix, we show that the distinguished coweights listed in \Cref{table:generators} are all the generators, as semigroup, of the set $\tilde{\Gamma}$ defined in \eqref{coweights one parent}.

To prove this, we first show that any element of \eqref{coweights one parent} must have zero Dynkin labels on the inner nodes of the Dynkin diagram. Indeed, assume that one of the Dynkin labels associated with one of the internal nodes of the Dynkin diagram is non-zero. Removing the corresponding node, we get a regular subalgebra $\mathfrak{u}(1) \oplus_{j}^{L} \mathfrak g_{j} \subset \mathfrak g$, with $L$ the number of links of the node with the rest of the Dynkin diagram. The highest root $\theta_i$ of any of the $\mathfrak g_i$ is also a root of $\mathfrak g$, and it has 
\begin{itemize}
    \item Dynkin label minus one on the removed node, 
    \item Dynkin labels zero on all the nodes of $\mathfrak g_j$, with $j \neq i$, 
    \item as it can be easily seen with a case-by-case analysis, non-zero Dynkin labels just on the exterior nodes of the sub-diagram associated to $\mathfrak g_i$. 
\end{itemize}

It follows that we have at least $L$ ways to move a non-zero internal label to one of the outer nodes of the Dynkin diagram by adding one of the $\theta_i$. 

In particular, if $L > 1$ we have at least two ways of doing so, showing that a weight $\lambda$ with a non-zero internal Dynkin label can not belong to $\tilde{\Gamma}$. Indeed, one might object that adding $\theta_i$ is a non-minimal operation (namely, $\theta_i$ is not such that $\theta_i \rightsquigarrow \lambda$). However, since the $\mathfrak g_i$ are orthogonal subalgebras, we always have at least $L$ \textit{distinct} coroots $\widetilde{\theta}_i \in \mathfrak g_i$ that we can add to $\lambda$. Consequently, only the $\lambda$ with $L = 1$ can belong to \eqref{coweights one parent}. 

Now that we have proven that the coweights in $\tilde{\Gamma}$ must have zero Dynkin labels on the interior points of the Dynkin diagram, we can  complete the proof by a case-by-case analysis. We can use that all the coroot lattice dominant coweights are exactly those associated with representations for which the center of the simply connected group associated with $\mathfrak g$ acts trivially. The conditions are (see e.g. \cite{FEGER2015166})
\begin{eqnarray}
\label{eq:centercharges}
    A_n:& \qquad \sum_{j=1}^n j \lambda_j  = 0 \text{ mod}(n+1), \nonumber \\
        D_n:& \qquad \lambda_{n-1}-\lambda_n = 0 \text{ mod}(2), \qquad 2\sum_{k=0}^{\lfloor \frac{n-3}{2}\rfloor}\lambda_{2k+1}+ (n-2)\lambda_{n-1} + n \lambda_n = 0 \text{ mod}(4)\nonumber \\
        E_6:&  \lambda_1 - \lambda_2 + \lambda_4 - \lambda_5 = 0 \text{ mod}(3), \nonumber \\
        E_7: & \lambda_4+ \lambda_6 + \lambda_7 = 0 \text{ mod}(2), 
\end{eqnarray}
whereas in the $E_8$ case all the coweights are coroots, as the center of $E_8$ is trivial. 
One can explicitly check that for the $A_n, E_6, E_7,E_8$ case the $\lambda$ with non-zero labels only on the exterior points of the Dynkin diagrams (namely, the elements of $\tilde{\Gamma}$) can all be written as linear combinations of the coweights in \Cref{table:generators} with positive coefficients.\footnote{Let us consider for example $E_6$. We want coweights whose non-zero Dynkin labels are just  on the outer nodes, hence $\lambda_2 = \lambda_4 = 0$. We then have, by \eqref{eq:centercharges}, $\lambda_1 = - \lambda_5  = 0 \text{ mod}(3)$. The minimal solution with both the labels non-zero is $\lambda_1 = \lambda_5 = 1$, while we can minimally take $(\lambda_1 = 3, \lambda_5 = 0)$, or $(\lambda_1 = 0, \lambda_5 = 3)$. Finally, we can take  $\lambda_1 = \lambda_5 = 0$, while keeping $\lambda_6$ non-zero; the minimal solution is for $\lambda_6 = 1$. This reproduces \Cref{table:generators}.} For the $D_{2j}$ case,   
the first equation is minimally solved by 
\begin{equation}
    \label{eq:centersold2j}
   (\lambda_{n-1} , \lambda_n ) = (2,0), \quad (\lambda_{n-1} , \lambda_n ) = (1,1), \quad (\lambda_{n-1} , \lambda_n ) = (0,2), \quad (\lambda_{n-1} , \lambda_n ) = (0,0), 
\end{equation}
with \eqref{eq:centersold2j} representing the minimal choices. All the other positive solutions can be obtained by taking positive combinations of \eqref{eq:centersold2j}. For $D_{2j}$ the second equation of the D-case in \eqref{eq:centercharges} becomes
\begin{equation}
    \lambda_1 + (j-1)\lambda_{n-1} + j \lambda_{n}  = 0 \text{ mod}(2), \qquad n = 2j.
\end{equation}
Then, plugging in the minimal choices \eqref{eq:centersold2j}, we get respectively, as minimal solutions,  
\begin{equation}
    \lambda_1 = 0, \quad \lambda_1 = 1, \quad \lambda_1 = 0, \quad \lambda_1 = 2,
\end{equation}
where for the last solution we did not take $\lambda_1 = 0$, because in that case the coweight would be zero. These solutions reproduce exactly the coweights in \Cref{table:generators}. 

To conclude, let us consider the $D_{2j+1}$ case, for which the equations in \eqref{eq:centercharges} become 
\begin{equation}
\label{eq:doddfirstmanipulation}
    \lambda_{n-1} - \lambda_n = 0 \text{ mod}(2), \qquad 2(\lambda_1 + j(\lambda_{n-1} + \lambda_n)) -\left(\lambda_{n-1} - \lambda_n\right) = 0 \text{ mod}(4).
\end{equation}
We note that the quantity $\left(\lambda_{n-1} - \lambda_n\right)$ is zero modulo two for the first equation of \eqref{eq:doddfirstmanipulation}, and hence we can split the solutions in 
two cases: 
\begin{equation}
\label{eq:cases}
    \lambda_{n-1} -\lambda_n = 0 \text{ mod (4)}, \qquad \lambda_{n-1} -\lambda_n = 2 \text{ mod (4)},
\end{equation}
In the first case the minimal solution are 
\begin{equation}
\label{eq:doddpartiallist1}
     (\lambda_{n-1} , \lambda_n ) = (4,0), \quad (\lambda_{n-1} , \lambda_n ) = (1,1), \quad (\lambda_{n-1} , \lambda_n ) = (0,4), \quad (\lambda_{n-1} , \lambda_n ) = (0,0), 
\end{equation}
and the second equation of \eqref{eq:doddfirstmanipulation}
becomes
\begin{equation}
\label{eq:reducedeqD2j+1}
       \lambda_{n-1} - \lambda_n = 0 \text{ mod}(2), \qquad \lambda_1 + j(\lambda_{n-1} + \lambda_n)  = 0 \text{ mod}(2),
\end{equation}
that is minimally solved by picking $\lambda_1 = 0$ for the first three solutions in \eqref{eq:doddpartiallist1}, and $\lambda_1 = 2$ for the last one. 
To conclude, let us pass to the second case of \eqref{eq:cases}. We have, minimally, 
\begin{equation}
\label{eq:listsecondcase}
    (\lambda_{n-1},\lambda_n) = (2,0), \qquad  (\lambda_{n-1},\lambda_n) = (0,2)
\end{equation}
with the other minimal solutions satisfying the equation mod four. Using that, for \eqref{eq:listsecondcase}, $\lambda_{n-1} - \lambda_n$ is even, the second equation of \eqref{eq:doddfirstmanipulation} becomes 
\begin{equation}
    \lambda_1 + j(\lambda_{n-1} + \lambda_n) - 1 = 0 \text{ mod(2)}
\end{equation}
that is minimally solved by $\lambda_1 = 1$. 
Summing up, for the $D_{2j+1}$ case we have 
\begin{eqnarray}
\scalemath{1}{\begin{split}
    &(\lambda_1, \lambda_{n-1},\lambda_n) = (0,0,4), \quad  (\lambda_1, \lambda_{n-1},\lambda_n) = (0,4,0),\\
    & (\lambda_1, \lambda_{n-1},\lambda_n) = (0,1,1), \quad  (\lambda_1, \lambda_{n-1},\lambda_n) = (2,0,0),  \nonumber \\
    \end{split}}
\end{eqnarray}
and
\begin{equation}
   (\lambda_1, \lambda_{n-1},\lambda_n) = (1,0,2), \quad 
      (\lambda_1, \lambda_{n-1},\lambda_n) = (1,2,0),
\end{equation}
that coincide with the coweights  in \Cref{table:generators}. 

\section{Tables of 5d conformal matter atoms and hybrids}\label{sec: tables}

\begin{table}[H]
\makebox[\linewidth][c]{

\resizebox{10cm}{!}{
    \begin{tabular}{|c|c|c|c|c|}

        \hline
        Coweight & Dynkin Quiver & Type & $I_{\lambda}$ & Orbit\\
        \hline \hline
        \scriptsize$\AtwoWeight{1}{1}$ & \scriptsize\AtwoRgens{1} & \cellcolor{green!15}Atom & $(x,y)$ & $[2,1]$\\
        \hline
        \scriptsize$\AtwoWeight{3}{0}$ & \scriptsize\AtwoRgens{2} & \cellcolor{green!15}Atom & $(x)$ &$[3]$\\
        \hline
        \scriptsize$\AtwoWeight{0}{3}$ & \scriptsize\AtwoRgens{3} & \cellcolor{green!15}Atom & $(y)$ & $[3]$\\
          \hline

    \end{tabular}
    }
 }
    \caption{Indecomposable balanced $A_2$ Dynkin Quivers constructed from coweights on the coroot lattice of $A_2$. $I_{\lambda}$ defines the CY3 employed to engineer the corresponding 5d SCFTs, according to \eqref{red ideal algorithm}.}
    \label{tab:DynkQuivA2}
\end{table}

\begin{table}[H]
\makebox[\linewidth][c]{

\resizebox{10cm}{!}{
    \begin{tabular}{|c|c|c|c|c|}

        \hline
        Coweight & Dynkin Quiver & Type & $I_{\lambda}$ & Orbit\\
        \hline \hline
        \scriptsize$\AthreeWeight{1}{0}{1}$ & \scriptsize\AthreeRgens{1} & \cellcolor{green!15}Atom & $(z)$ & $[2,1^2]$\\
        \hline
        \scriptsize$\AthreeWeight{0}{2}{0}$ & \scriptsize\AthreeRgens{2} & \cellcolor{green!15}Atom & $(x,y)$ &$[2^2]$\\
        \hline
        \scriptsize$\AthreeWeight{2}{1}{0}$ & \scriptsize\AthreeRgens{3} & \cellcolor{green!15}Atom & $(x,z^2)$ & $[3,1]$\\
          \hline
          \scriptsize$\AthreeWeight{0}{1}{2}$ & \scriptsize\AthreeRgens{4} & \cellcolor{green!15}Atom & $(y,z^2)$ & $[3,1]$\\
          \hline
            \scriptsize$\AthreeWeight{4}{0}{0}$ & \scriptsize\AthreeRgens{5} & \cellcolor{green!15}Atom & $(x)$ & $[4]$\\
          \hline
           \scriptsize$\AthreeWeight{0}{0}{4}$ & \scriptsize\AthreeRgens{6} & \cellcolor{green!15}Atom & $(y)$ & $[4]$\\
          \hline
    \end{tabular}
    }
 }
    \caption{Indecomposable balanced $A_3$ Dynkin Quivers constructed from coweights on the coroot lattice of $A_3$. $I_{\lambda}$ defines the CY3 employed to engineer the corresponding 5d SCFTs, according to \eqref{red ideal algorithm}.}
    \label{tab:DynkQuivA3}
\end{table}

\begin{table}[H]
\makebox[\linewidth][c]{

\resizebox{10cm}{!}{
    \begin{tabular}{|c|c|c|c|c|}

        \hline
        Coweight & Dynkin Quiver & Type & $I_{\lambda}$ & Orbit\\
        \hline \hline
        \scriptsize$\AfourWeight{1}{0}{0}{1}$ & \scriptsize\AfourRgens{1} & \cellcolor{green!15}Atom & $(z)$ & $[2,1^3]$\\
        \hline
        \scriptsize$\AfourWeight{0}{1}{1}{0}$ & \scriptsize\AfourRgens{2} & \cellcolor{green!15}Atom & $(x,y)$ &$[2^2,1]$\\
        \hline
        \scriptsize$\AfourWeight{2}{0}{1}{0}$ & \scriptsize\AfourRgens{3} & \cellcolor{green!15}Atom & $(x,z^2)$ & $[3,1^2]$\\
          \hline
          \scriptsize$\AfourWeight{0}{1}{0}{2}$ & \scriptsize\AfourRgens{9} & \cellcolor{green!15}Atom & $(y,z^2)$ & $[3,1^2]$\\
          \hline
        \scriptsize$\AfourWeight{1}{2}{0}{0}$ & \scriptsize\AfourRgens{4} & \cellcolor{green!15}Atom & $(x,yz)$ & $[3,2]$\\
        \hline
        \scriptsize$\AfourWeight{0}{0}{2}{1}$ & \scriptsize\AfourRgens{10} & \cellcolor{green!15}Atom & $(y,xz)$ & $[3,2]$\\
        \hline
        \scriptsize$\AfourWeight{3}{1}{0}{0}$ & \scriptsize\AfourRgens{5} & \cellcolor{green!15}Atom & $(x,z^3)$ & $[4,1]$\\
        \hline
        \scriptsize$\AfourWeight{0}{0}{1}{3}$ & \scriptsize\AfourRgens{11} & \cellcolor{green!15}Atom & $(y,z^3)$ & $[4,1]$\\
        \hline
        \scriptsize$\AfourWeight{5}{0}{0}{0}$ & \scriptsize\AfourRgens{6} & \cellcolor{green!15}Atom & $(x)$ & $[5]$\\
        \hline
        \scriptsize$\AfourWeight{0}{0}{0}{5}$ & \scriptsize\AfourRgens{12} & \cellcolor{green!15}Atom & $(y)$ & $[5]$\\
        \hline

    \end{tabular}
    }

    \resizebox{10cm}{!}{
    \begin{tabular}{|c|c|c|c|}

        \hline
        Coweight & Dynkin Quiver & Type & $I_{\lambda}$\\
        \hline \hline
        \scriptsize$\AfourWeight{0}{3}{0}{1}$ & \scriptsize\AfourRgens{7} & \cellcolor{purple!15}Hybrid & $(xz,y^2)$ \\
        \hline
        \scriptsize$\AfourWeight{0}{5}{0}{0}$ & \scriptsize\AfourRgens{8} & \cellcolor{purple!15}Hybrid & $(x^2,y^3)$\\
        \hline
        \scriptsize$\AfourWeight{1}{0}{3}{0}$ & \scriptsize\AfourRgens{13} & \cellcolor{purple!15}Hybrid & $(yz,x^2)$ \\
        \hline
        \scriptsize$\AfourWeight{0}{0}{5}{0}$ & \scriptsize\AfourRgens{14} & \cellcolor{purple!15}Hybrid & $(y^2,x^3)$\\
        \hline

    \end{tabular}
    }

    }
    \caption{Indecomposable balanced $A_4$ Dynkin Quivers constructed from coweights on the coroot lattice of $A_4$. $I_{\lambda}$ defines the CY3 employed to engineer the corresponding 5d SCFTs, according to \eqref{red ideal algorithm}.}
    \label{tab:DynkQuivA4}
\end{table}
 
\begin{landscape}
    
\begin{table}
    \makebox[\linewidth][c]{
    \scalebox{0.9}{
    \begin{tabular}{|c|c|c|}
        \hline
        Weights & Number & Orbits \\
        \hline \hline
        $\begin{bmatrix}
            &&&&&0\\
            0&1&0&\dots&0&\\
            &&&&&0
        \end{bmatrix}$, $\begin{bmatrix}
            &&&&&&&0\\
            0&0&0&1&0&\dots&0&\\
            &&&&&&&0
        \end{bmatrix}$,\dots, $\begin{bmatrix}
            &&&&&0\\
            0&\dots&0&1&0&\\
            &&&&&0
        \end{bmatrix}$ & $n-1$ & \cellcolor{green!15} $[2^2,1^{4n-2}]$, $[2^4,1^{4n-6}]$, \dots, $[2^{2n-2},1^{6}]$\\
        \hline
        $\begin{bmatrix}
            &&&&0\\
            2&0&\dots&0&\\
            &&&&0
        \end{bmatrix}$ & $1$ & \cellcolor{green!15} $[3,1^{4n-1}]$\\
        \hline
        $\begin{bmatrix}
            &&&&&&0\\
            0&0&2&0&\dots&0&\\
            &&&&&&0
        \end{bmatrix}$, $\begin{bmatrix}
            &&&&&&&&0\\
            0&0&0&0&2&0&\dots&0&\\
            &&&&&&&&0
        \end{bmatrix}$, \dots, $\begin{bmatrix}
            &&&&0\\
            0&\dots&0&2&\\
            &&&&0
        \end{bmatrix}$ & $n-1$ & \cellcolor{purple!15} not small rep\\
        \hline
        $\begin{bmatrix}
            &&&&&&0\\
            1&0&1&0&\dots&0&\\
            &&&&&&0
        \end{bmatrix}$, $\begin{bmatrix}
            &&&&&&&&0\\
            1&0&0&0&1&0&\dots&0&\\
            &&&&&&&&0
        \end{bmatrix}$, \dots, $\begin{bmatrix}
            &&&&&0\\
            1&0&\dots&0&1&\\
            &&&&&0
        \end{bmatrix}$ & $n-1$ & \cellcolor{green!15} $[3^2,1^{4n-4}]$, $[3^2,2^2,1^{4n-8}]$, \dots, $[3^2,2^{2n-4},1^4]$\\
        \hline
        \begin{tabular}{r}
            $\begin{bmatrix}
                &&&&&&&&0\\
                0&0&1&0&1&0&\dots&0&\\
                &&&&&&&&0
            \end{bmatrix}$, $\left[\begin{array}{*{11}c}
                &&&&&&&&&&0\\
                0&0&1&0&0&0&1&0&\dots&0&\\
                &&&&&&&&&&0
            \end{array}\right]$, \dots, $\begin{bmatrix}
                &&&&&&0\\
                0&0&1&0\dots&0&1&\\
                &&&&&&0
            \end{bmatrix}$,\\
            $\left[\begin{array}{*{11}c}
                &&&&&&&&&&0\\
                0&0&0&0&1&0&1&0&\dots&0&\\
                &&&&&&&&&&0
            \end{array}\right]$, \dots, $\begin{bmatrix}
                &&&&&&&&0\\
                0&0&0&0&1&0\dots&0&1&\\
                &&&&&&&&0
            \end{bmatrix}$,\\
            $\ddots\qquad\qquad\qquad\vdots\qquad\qquad$,\\
            $\begin{bmatrix}
                &&&&&&0\\
                0&\dots&0&1&0&1&\\
                &&&&&&0
            \end{bmatrix}$
        \end{tabular} & $\frac{(n-1)(n-2)}{2}$ & \cellcolor{purple!15} not small rep\\
        \hline
    \end{tabular}
    }
    }
    \caption{Indecomposable coweights for $D_{2n+1}$, with subset of small representations and their corresponding nilpotent orbits (Part 1). For $D_{2n+1}$ there are $\frac{1}{2}(n^2+7n+4)$ indecomposable coweights, out of which $2n+2$ are small representations.}
    \label{tab:DoddSmallReps1}
\end{table}

\end{landscape}
\begin{landscape}

\begin{table}
    \makebox[\linewidth][c]{
    \scalebox{1}{
    \begin{tabular}{|c|c|c|}
        \hline
        Weights & Number & Orbits \\
        \hline \hline
        $\begin{bmatrix}
            &&&1\\
            0&\dots&0&\\
            &&&1
        \end{bmatrix}$ & $1$ & \cellcolor{green!15} $[2^{2n},1^2]$\\
        \hline
        $\begin{bmatrix}
            &&&&2\\
            1&0&\dots&0&\\
            &&&&0
        \end{bmatrix}$, $\begin{bmatrix}
            &&&&0\\
            1&0&\dots&0&\\
            &&&&2
        \end{bmatrix}$ & $2$ & \cellcolor{green!15} $[3^2,2^{2n-2}]$\\
        \hline
        $\begin{bmatrix}
            &&&&&&2\\
            0&0&1&0&\dots&0&\\
            &&&&&&0
        \end{bmatrix}$, $\begin{bmatrix}
            &&&&&&&&2\\
            0&0&0&0&1&0&\dots&0&\\
            &&&&&&&&0
        \end{bmatrix}$, \dots, $\begin{bmatrix}
            &&&&2\\
            0&\dots&0&1&\\
            &&&&0
        \end{bmatrix}$ & $n-1$ & \cellcolor{purple!15} not small rep\\
        \hline
        $\begin{bmatrix}
            &&&&&&0\\
            0&0&1&0&\dots&0&\\
            &&&&&&2
        \end{bmatrix}$, $\begin{bmatrix}
            &&&&&&&&0\\
            0&0&0&0&1&0&\dots&0&\\
            &&&&&&&&2
        \end{bmatrix}$, \dots, $\begin{bmatrix}
            &&&&0\\
            0&\dots&0&1&\\
            &&&&2
        \end{bmatrix}$ & $n-1$ & \cellcolor{purple!15} not small rep\\
        \hline
        $\begin{bmatrix}
            &&&4\\
            0&\dots&0&\\
            &&&0
        \end{bmatrix}$, $\begin{bmatrix}
            &&&0\\
            0&\dots&0&\\
            &&&4
        \end{bmatrix}$ & $2$ & \cellcolor{purple!15} not small rep\\
        \hline
    \end{tabular}
    }

    }
    \caption{Indecomposable coweights for $D_{2n+1}$, with subset of small representations and their corresponding nilpotent orbits (Part 2). For $D_{2n+1}$ there are $\frac{1}{2}(n^2+7n+4)$ indecomposable coweights, out of which $2n+2$ are small representations.}
    \label{tab:DoddSmallReps2}
\end{table}
 
\end{landscape}
\begin{landscape}

\begin{table} 
\makebox[\linewidth][c]{
\resizebox{10cm}{!}{
    \begin{tabular}{|c|c|c|c|c|}
        \hline
        Coweight & Dynkin Quiver & Type & $I_{\lambda}$ & Orbit\\
        \hline \hline
        \scriptsize$\DfiveWeight{0}{1}{0}{0}{0}$ & \scriptsize\DfiveRgens{1} & \cellcolor{green!15}Atom & $(y,z)$ & $[2^2,1^{6}]$\\
        \hline
        \scriptsize$\DfiveWeight{2}{0}{0}{0}{0}$ & \scriptsize\DfiveRgens{2} & \cellcolor{green!15}Atom & $(z)$ & $[3,1^7]$\\
        \hline
        \scriptsize$\DfiveWeight{0}{0}{0}{1}{1}$ & \scriptsize\DfiveRgens{4} & \cellcolor{green!15}Atom & $(y)$ & $[2^4,1^2]$\\
        \hline
        \scriptsize$\DfiveWeight{1}{0}{1}{0}{0}$ & \scriptsize\DfiveRgens{3} & \cellcolor{green!15}Atom & $(x+iz^2,x-iz^2)$ & $[3^2,1^4]$\\
        \hline
        \scriptsize$\DfiveWeight{1}{0}{0}{2}{0}$ & \scriptsize\DfiveRgens{5} & \cellcolor{green!15}Atom & $(x-iz^2)$ & $[3^2,2^2]$\\
        \hline
        \scriptsize$\DfiveWeight{1}{0}{0}{0}{2}$ & \scriptsize\DfiveRgens{11} & \cellcolor{green!15}Atom & $(x+iz^2)$ & $[3^2,2^2] $\\
        \hline

    \end{tabular}
    }

    \resizebox{10cm}{!}{
    \begin{tabular}{|c|c|c|c|}

        \hline
        Coweight & Dynkin Quiver & Type & $I_{\lambda}$\\
        \hline \hline
        \scriptsize$\DfiveWeight{0}{0}{2}{0}{0}$ & \scriptsize\DfiveRgens{10} & \cellcolor{purple!15}Hybrid &$ (y^2,z(x-iz^2),z(x+iz^2))$\\
        \hline
        \scriptsize$\DfiveWeight{0}{0}{1}{2}{0}$ & \scriptsize\DfiveRgens{6} & \cellcolor{purple!15}Hybrid & $(y^2,z(x-iz^2))$ \\
        \hline
        \scriptsize$\DfiveWeight{0}{0}{1}{0}{2}$ & \scriptsize\DfiveRgens{7} & \cellcolor{purple!15}Hybrid & $(y^2,z(x+iz^2))$\\
        \hline
        \scriptsize$\DfiveWeight{0}{0}{0}{4}{0}$ & \scriptsize\DfiveRgens{8} & \cellcolor{purple!15}Hybrid & $(y^2+2iz(x-iz^2))$\\
        \hline
        \scriptsize$\DfiveWeight{0}{0}{0}{0}{4}$ & \scriptsize\DfiveRgens{9} & \cellcolor{purple!15}Hybrid & $(y^2-2iz(x+iz^2))$ \\
        \hline

    \end{tabular}
    }
    }
    \caption{Indecomposable balanced $D_5$ Dynkin Quivers constructed from coweights on the coroot lattice of $D_5$. $I_{\lambda}$ defines the CY3 employed to engineer the corresponding 5d SCFTs, according to \eqref{red ideal algorithm}.}
    \label{tab:DynkQuivD5}
\end{table}

\end{landscape}
\begin{landscape}

\begin{table}
    \makebox[\linewidth][c]{
    \scalebox{0.85}{
    \begin{tabular}{|c|c|c|}
        \hline
        Weights & Number & Orbits \\
        \hline \hline
        $\begin{bmatrix}
            &&&&&0\\
            0&1&0&\dots&0&\\
            &&&&&0
        \end{bmatrix}$, $\begin{bmatrix}
            &&&&&&&0\\
            0&0&0&1&0&\dots&0&\\
            &&&&&&&0
        \end{bmatrix}$,\dots, $\begin{bmatrix}
            &&&&0\\
            0&\dots&0&1&\\
            &&&&0
        \end{bmatrix}$ & $n-1$ & \cellcolor{green!15} $[2^2,1^{4n-4}]$, $[2^4,1^{4n-8}]$, \dots, $[2^{2n-2},1^4]$\\
        \hline
        $\begin{bmatrix}
            &&&&0\\
            2&0&\dots&0&\\
            &&&&0
        \end{bmatrix}$ & $1$ & \cellcolor{green!15} $[3,1^{4n-3}]$\\
        \hline
        $\begin{bmatrix}
            &&&&&&0\\
            0&0&2&0&\dots&0&\\
            &&&&&&0
        \end{bmatrix}$, $\begin{bmatrix}
            &&&&&&&&0\\
            0&0&0&0&2&0&\dots&0&\\
            &&&&&&&&0
        \end{bmatrix}$, \dots, $\begin{bmatrix}
            &&&&&0\\
            0&\dots&0&2&0&\\
            &&&&&0
        \end{bmatrix}$ & $n-2$ & \cellcolor{purple!15} not small rep\\
        \hline
        $\begin{bmatrix}
            &&&&&&0\\
            1&0&1&0&\dots&0&\\
            &&&&&&0
        \end{bmatrix}$, $\begin{bmatrix}
            &&&&&&&&0\\
            1&0&0&0&1&0&\dots&0&\\
            &&&&&&&&0
        \end{bmatrix}$, \dots, $\begin{bmatrix}
            &&&&&&0\\
            1&0&\dots&0&1&0&\\
            &&&&&&0
        \end{bmatrix}$ & $n-2$ & \cellcolor{green!15} $[3^2,1^{4n-6}]$, $[3^2,2^2,1^{4n-10}]$, \dots, $[3^2,2^{2n-6},1^6]$\\
        \hline
        \begin{tabular}{r}
            $\begin{bmatrix}
                &&&&&&&&0\\
                0&0&1&0&1&0&\dots&0&\\
                &&&&&&&&0
            \end{bmatrix}$, $\left[\begin{array}{*{11}c}
                &&&&&&&&&&0\\
                0&0&1&0&0&0&1&0&\dots&0&\\
                &&&&&&&&&&0
            \end{array}\right]$, \dots, $\begin{bmatrix}
                &&&&&&&0\\
                0&0&1&0\dots&0&1&0&\\
                &&&&&&&0
            \end{bmatrix}$,\\
            $\left[\begin{array}{*{11}c}
                &&&&&&&&&&0\\
                0&0&0&0&1&0&1&0&\dots&0&\\
                &&&&&&&&&&0
            \end{array}\right]$, \dots, $\begin{bmatrix}
                &&&&&&&&&0\\
                0&0&0&0&1&0\dots&0&1&0&\\
                &&&&&&&&&0
            \end{bmatrix}$,\\
            $\ddots\qquad\qquad\qquad\vdots\qquad\qquad$,\\
            $\begin{bmatrix}
                &&&&&&&0\\
                0&\dots&0&1&0&1&0&\\
                &&&&&&&0
            \end{bmatrix}$
        \end{tabular} & $\frac{(n-2)(n-3)}{2}$ & \cellcolor{purple!15} not small rep\\
        \hline
    \end{tabular}
    }
    }
    \caption{Indecomposable coweights for $D_{2n}$ with subset of small representations and their corresponding nilpotent orbits (Part 1). For $D_{2n}$ there are $\frac{1}{2}(n^2+3n)$ indecomposable coweights, out of which $2n+1$ are small representations.}
    \label{tab:DevenSmallReps1}
\end{table}

\end{landscape}
\begin{landscape}

\begin{table}
    \makebox[\linewidth][c]{
    \scalebox{1}{
    \begin{tabular}{|c|c|c|}
        \hline
        Weights & Number & Orbits \\
        \hline \hline
        $\begin{bmatrix}
            &&&2\\
            0&\dots&0&\\
            &&&0
        \end{bmatrix}$, $\begin{bmatrix}
            &&&0\\
            0&\dots&0&\\
            &&&2
        \end{bmatrix}$ & $2$ & \cellcolor{green!15} $[2^{2n}]$\\
        \hline
        $\begin{bmatrix}
            &&&&1\\
            1&0&\dots&0&\\
            &&&&1
        \end{bmatrix}$ & $1$ & \cellcolor{green!15} $[3^2,2^{2n-4},1^2]$\\
        \hline
        $\begin{bmatrix}
            &&&&&&1\\
            0&0&1&0&\dots&0&\\
            &&&&&&1
        \end{bmatrix}$, $\begin{bmatrix}
            &&&&&&&&1\\
            0&0&0&0&1&0&\dots&0&\\
            &&&&&&&&1
        \end{bmatrix}$, \dots, $\begin{bmatrix}
            &&&&&1\\
            0&\dots&0&1&0&\\
            &&&&&1
        \end{bmatrix}$ & $n-2$ & \cellcolor{purple!15} not small rep\\
        \hline
    \end{tabular}
    }
    }
    \caption{Indecomposable coweights for $D_{2n}$, with subset of small representations and their corresponding nilpotent orbits (Part 2). For $D_{2n}$ there are $\frac{1}{2}(n^2+3n)$ indecomposable coweights, out of which $2n+1$ are small representations.}
    \label{tab:DevenSmallReps2}
\end{table}
    
\end{landscape}
\begin{landscape}

\begin{table}[H]
\makebox[\linewidth][c]{

\resizebox{10cm}{!}{
    \begin{tabular}{|c|c|c|c|c|}

        \hline
        Coweight & Dynkin Quiver & Type & $I_{\lambda}$ & Orbit\\
        \hline \hline
        \scriptsize$\DSixWeight{0}{1}{0}{0}{0}{0}$ & \scriptsize\DSixRgens{1} & \cellcolor{green!15}Atom & $(y-iz^2,y+iz^2,z)$ & $[2^2,1^8]$\\
        \hline
        \scriptsize$\DSixWeight{2}{0}{0}{0}{0}{0}$ & \scriptsize\DSixRgens{2} & \cellcolor{green!15}Atom & $(z)$ & $[3,1^9]$\\
        \hline
        \scriptsize$\DSixWeight{0}{0}{0}{1}{0}{0}$ & \scriptsize\DSixRgens{3} & \cellcolor{green!15}Atom & $(y-iz^2,y+iz^2)$ & $[2^4,1^4]$\\
        \hline
        \scriptsize$\DSixWeight{1}{0}{1}{0}{0}{0}$ & \scriptsize\DSixRgens{4} & \cellcolor{green!15}Atom & $(x,z^2)$ & $[3^2,1^6]$\\
        \hline
        \scriptsize$\DSixWeight{0}{0}{0}{0}{2}{0}$ & \scriptsize\DSixRgens{5} & \cellcolor{green!15}Atom & $(y-iz^2)$ & $[2^6]$\\
        \hline
        \scriptsize$\DSixWeight{0}{0}{0}{0}{0}{2}$ & \scriptsize\DSixRgens{6} & \cellcolor{green!15}Atom & $(y+iz^2)$ & $[2^6]$\\
        \hline
        \scriptsize$\DSixWeight{1}{0}{0}{0}{1}{1}$ & \scriptsize\DSixRgens{7} & \cellcolor{green!15}Atom & $(x)$ & $[3^2,2^2,1^2]$\\
        \hline

    \end{tabular}
    }

    \resizebox{10cm}{!}{
    \begin{tabular}{|c|c|c|c|}

        \hline
        Coweight & Dynkin Quiver & Type & $I_{\lambda}$\\
        \hline \hline
        \scriptsize$\DSixWeight{0}{0}{2}{0}{0}{0}$ & \scriptsize\DSixRgens{8} & \cellcolor{purple!15}Hybrid & $((y+iz^2)^2,(y-iz^2)^2,(y+iz^2)(y-iz^2),z^3)$ \\
        \hline
        \scriptsize$\DSixWeight{0}{0}{1}{0}{1}{1}$ & \scriptsize\DSixRgens{9} & \cellcolor{purple!15}Hybrid & $((y+iz^2)^2,(y-iz^2)^2,(y+iz^2)(y-iz^2),xz)$ \\
        \hline

    \end{tabular}
    }
    
    }
    \caption{Indecomposable balanced $D_6$ Dynkin Quivers constructed from coweights on the coroot lattice of $D_6$. $I_{\lambda}$ defines the CY3 employed to engineer the corresponding 5d SCFTs, according to \eqref{red ideal algorithm}.}
    \label{tab:DynkQuivD6}
\end{table}

\end{landscape}
\begin{landscape}

\begin{table}[H]
\makebox[\linewidth][c]{

\resizebox{10cm}{!}{
    \begin{tabular}{|c|c|c|c|c|}

        \hline
        Coweight & Dynkin Quiver & Type & $I_{\lambda}$ & Orbit\\
        \hline \hline
        \scriptsize$\eSixWeight{0}{0}{0}{0}{0}{1}$ & \scriptsize\EsixRgens{1} & \cellcolor{green!15}Atom & $(z)$ & $A_1$\\
        \hline
        \scriptsize$\eSixWeight{1}{0}{0}{0}{1}{0}$ & \scriptsize\EsixRgens{2} & \cellcolor{green!15}Atom & $(y)$ & $2A_1$\\
        \hline
        \scriptsize$\eSixWeight{0}{0}{1}{0}{0}{0}$ & \scriptsize\EsixRgens{3} & \cellcolor{green!15}Atom & $(x+iz^2,x-iz^2)$  & $A_2$\\
        \hline
        \scriptsize$\eSixWeight{1}{1}{0}{0}{0}{0}$ & \scriptsize\EsixRgens{4} & \cellcolor{green!15}Atom & $(x+iz^2,yz)$ & $A_2+A_1$\\
        \hline
        \scriptsize$\eSixWeight{3}{0}{0}{0}{0}{0}$ & \scriptsize\EsixRgens{5} & \cellcolor{green!15}Atom & $(x+iz^2)$ & $2A_2$\\
        \hline
        \scriptsize$\eSixWeight{0}{0}{0}{1}{1}{0}$ & \scriptsize\EsixRgens{6} & \cellcolor{green!15}Atom & $(x-iz^2,yz)$ &$A_2+A_1$\\
        \hline
        \scriptsize$\eSixWeight{0}{0}{0}{0}{3}{0}$ & \scriptsize\EsixRgens{7} & \cellcolor{green!15}Atom & $(x-iz^2)$ & $2A_2$ \\
        \hline
    \end{tabular}
    }
    \quad

    \resizebox{10cm}{!}{
    \begin{tabular}{|c|c|c|c|}

        \hline
        Coweight & Dynkin Quiver & Type & $I_{\lambda}$\\
        \hline \hline
        \scriptsize$\eSixWeight{0}{1}{0}{1}{0}{0}$ & \scriptsize\EsixRgens{8} & \cellcolor{purple!15}Hybrid & $(z^3,y^2)$ \\
        \hline
        \scriptsize$\eSixWeight{2}{0}{0}{1}{0}{0}$ & \scriptsize\EsixRgens{9} & \cellcolor{purple!15}Hybrid & $(z \left(x+i z^2\right),y^2)$\\
        \hline
        \scriptsize$\eSixWeight{0}{1}{0}{0}{2}{0}$ & \scriptsize\EsixRgens{10} & \cellcolor{purple!15}Hybrid & $(z \left(x-i z^2\right),y^2)$ \\
        \hline
        \scriptsize$\eSixWeight{1}{0}{0}{2}{0}{0}$ & \scriptsize\EsixRgens{11} & \cellcolor{purple!15}Hybrid & $(y \left(x-i z^2\right),(x+iz^2)^2)$\\
        \hline
        \scriptsize$\eSixWeight{0}{2}{0}{0}{1}{0}$ & \scriptsize\EsixRgens{12} & \cellcolor{purple!15}Hybrid & $(y \left(x+i z^2\right),(x-iz^2)^2)$ \\
        \hline
        \scriptsize$\eSixWeight{0}{3}{0}{0}{0}{0}$ & \scriptsize\EsixRgens{13} & \cellcolor{purple!15}Hybrid & $(\left(x+i z^2\right)^2,z^5)$\\
        \hline
        \scriptsize$\eSixWeight{0}{0}{0}{3}{0}{0}$ & \scriptsize\EsixRgens{14} & \cellcolor{purple!15}Hybrid& $(\left(x-i z^2\right)^2,z^5)$ \\
        \hline
    \end{tabular}
    }
    }
    \caption{Indecomposable balanced $E_6$ Dynkin Quivers constructed from coweights on the coroot lattice of $E_6$. Both atoms and hybrids exist for $E_6$. $I_{\lambda}$ defines the CY3 employed to engineer the corresponding 5d SCFTs, according to \eqref{red ideal algorithm}.}
    \label{tab:DynkQuivE6}
\end{table}

\end{landscape}
\begin{landscape}

\begin{table}
    \centering
    \scalebox{0.65}{\begin{tikzpicture}
            \node[fill=green!15] (1) at (0,0) {\scriptsize$\EsixRgens{1}$};
            \node[fill=green!15] (2) at (0,4) {\scriptsize$\EsixRgens{2}$};
            \node[fill=green!15] (3) at (0,8) {\scriptsize$\EsixRgens{3}$};
            \node[fill=green!15] (4) at (-3,12) {\scriptsize$\EsixRgens{4}$};
            \node[fill=green!15] (5) at (-3,16) {\scriptsize$\EsixRgens{5}$};
            \node[fill=green!15] (6) at (3,12) {\scriptsize$\EsixRgens{6}$};
            \node[fill=green!15] (7) at (3,16) {\scriptsize$\EsixRgens{7}$};
            \draw[<-] (1)--(2);
            \draw[<-] (2)--(3);
            \draw[<-] (3)--(4);
            \draw[<-] (4)--(5);
            \draw[<-] (3)--(6);
            \draw[<-] (6)--(7);

            \def\x{14};
            \def\y{17};
            \node[fill=purple!15] (8) at (0+\x,0+\y) {\scriptsize$\EsixRgens{8}$};
            \node[fill=purple!15] (9) at (-3+\x,4+\y) {\scriptsize$\EsixRgens{9}$};
            \node[fill=purple!15] (10) at (3+\x,4+\y) {\scriptsize$\EsixRgens{10}$};
            \node[fill=orange!15] (8d) at (0+\x,-5+\y) {\scriptsize$\begin{tikzpicture}
                \eSixF{1}{0}{0}{0}{1}{1}
                \eSixG{3}{5}{7}{5}{3}{4}
            \end{tikzpicture}$};
            \draw[<-] (8d)--(8);
            \draw[<-] (8)--(9);
            \draw[<-] (8)--(10);

            \draw[<-] (5) .. controls (-3,19) and (0,19) .. (9);
            \draw[<-] (7) .. controls (3,19) and (9,17) .. (10);

            \def\xx{14};
            \def\yy{6};
            \node[fill=purple!15] (11) at (-3+\xx,0+\yy) {\scriptsize$\EsixRgens{11}$};
            \node[fill=purple!15] (12) at (3+\xx,0+\yy) {\scriptsize$\EsixRgens{12}$};
            \node[fill=orange!15] (11d12d) at (0+\xx,-5+\yy) {\scriptsize$\begin{tikzpicture}
                \eSixF{1}{0}{1}{0}{1}{0}
                \eSixG{4}{7}{10}{7}{4}{5}
            \end{tikzpicture}$};
            \draw[<-] (11d12d)--(11);
            \draw[<-] (11d12d)--(12);

            \def\xxx{26};
            \def\yyy{18};
            \node[fill=purple!15] (13) at (0+\xxx,0+\yyy) {\scriptsize$\EsixRgens{13}$};
            \node[fill=orange!15] (13d) at (0+\xxx,-5+\yyy) {\scriptsize$\begin{tikzpicture}
                \eSixF{1}{1}{1}{0}{0}{0}
                \eSixG{5}{9}{12}{8}{4}{6}
            \end{tikzpicture}$};
            \draw[<-] (13d)--(13);
            
            \node[fill=purple!15] (14) at (0+\xxx,-10+\yyy) {\scriptsize$\EsixRgens{14}$};
            \node[fill=orange!15] (14d) at (0+\xxx,-15+\yyy) {\scriptsize$\begin{tikzpicture}
                \eSixF{0}{0}{1}{1}{1}{0}
                \eSixG{4}{8}{12}{9}{5}{6}
            \end{tikzpicture}$};
            \draw[<-] (14d)--(14);
        \end{tikzpicture}}
    \caption{Higgs branch RG flows (indicated by arrows) between $E_6$ conformal matter theories labelled by their Dynkin quiver phase. Atoms are highlighted in green, hybrids in purple, and molecules in orange. We suppress further Higgsings of molecules. Two hybrids can be directly Higgsed to atoms without Higgsing to a molecule first.}
    \label{fig:AtomE6Higgsings}
\end{table}

\end{landscape}
\begin{landscape}

\begin{table}
    \makebox[\linewidth][c]{
     \resizebox{10cm}{!}{
    \begin{tabular}{|c|c|c|c|c|}
        \hline
        Coweight & Dynkin Quiver & Type & $I_{\lambda}$ & Orbit \\
        \hline \hline
        \scriptsize$\eSevenWeight{1}{0}{0}{0}{0}{0}{0}$ & \scriptsize\EsevenRgens{1} & \cellcolor{green!15}Atom & $(z)$ & $A_1$\\
        \hline
        \scriptsize$\eSevenWeight{0}{0}{0}{0}{1}{0}{0}$ & \scriptsize\EsevenRgens{2} & \cellcolor{green!15}Atom & $(y,z^2)$ & $2A_1$\\
        \hline
         \scriptsize$\eSevenWeight{0}{0}{0}{0}{0}{2}{0}$ & \scriptsize\EsevenRgens{4} & \cellcolor{green!15}Atom & $(y)$ & $3(A_1)''$ \\
        \hline
        \scriptsize$\eSevenWeight{0}{1}{0}{0}{0}{0}{0}$ & \scriptsize\EsevenRgens{3} & \cellcolor{green!15}Atom & $(x,z^2)$ & $A_2$\\
        \hline
          \scriptsize$\eSevenWeight{0}{0}{0}{0}{0}{1}{1}$ & \scriptsize\EsevenRgens{5} & \cellcolor{green!15}Atom & $(x)$ & $A_2+A_1$\\
        \hline
    \end{tabular}
    }
    \quad

     \resizebox{10cm}{!}{
    \begin{tabular}{|c|c|c|c|}
        \hline
        Coweight & Dynkin Quiver & Type & $I_{\lambda}$\\
        \hline \hline
        \scriptsize$\eSevenWeight{0}{0}{1}{0}{0}{0}{0}$ & \scriptsize\EsevenRgens{6} & \cellcolor{purple!15}Hybrid & $\makecell{(y^2, c z^3) \\(c\neq 1)}$ \\
        \hline
        \scriptsize$\eSevenWeight{0}{0}{0}{0}{0}{0}{2}$ & \scriptsize\EsevenRgens{7} & \cellcolor{purple!15}Hybrid & $(y^2 + z^3) $\\
        \hline
        \scriptsize$\eSevenWeight{0}{0}{0}{2}{0}{0}{0}$ & \scriptsize\EsevenRgens{8} & \cellcolor{purple!15}Hybrid & $(z^5,y^3)$ \\
          \hline
        \scriptsize$\eSevenWeight{0}{0}{0}{1}{0}{1}{0}$ & \scriptsize\EsevenRgens{9} & \cellcolor{purple!15}Hybrid & $(y^2,xz)$ \\
          \hline
        \scriptsize$\eSevenWeight{0}{0}{0}{1}{0}{0}{1}$ & \scriptsize\EsevenRgens{10} & \cellcolor{purple!15}Hybrid & $(xy,z^4)$ \\
        \hline
    \end{tabular}
    }
    }
    \caption{Indecomposable balanced $E_7$ Dynkin Quivers constructed from coweights on the coroot lattice of $E_7$. Both atoms and hybrids exist for $E_7$. $I_{\lambda}$ defines the CY3 employed to engineer the corresponding 5d SCFTs, according to \eqref{red ideal algorithm}.}
    \label{tab:DynkQuivE7}
\end{table}
 
\end{landscape}
\begin{landscape}
 
\begin{table}
    \centering
    \scalebox{0.7}{\begin{tikzpicture}
                \node[fill=green!15] (1) at (0,0) {\scriptsize$\EsevenRgens{1}$};
                \node[fill=green!15] (2) at (0,4) {\scriptsize$\EsevenRgens{2}$};
                \node[fill=green!15] (3) at (-4,8) {\scriptsize$\EsevenRgens{3}$};
                \node[fill=green!15] (4) at (4,8) {\scriptsize$\EsevenRgens{4}$};
                \node[fill=green!15] (5) at (0,12) {\scriptsize$\EsevenRgens{5}$};
                \draw[<-] (1)--(2);
                \draw[<-] (2)--(3);
                \draw[<-] (2)--(4);
                \draw[<-] (3)--(5);
                \draw[<-] (4)--(5);

                \def\x{18};
                \def\y{7};
                \node[fill=purple!15] (9) at (-7+\x,4+\y) {\scriptsize$\EsevenRgens{9}$};
                \node[fill=orange!15] (9d) at (-7+\x,-1+\y) {\scriptsize$\begin{tikzpicture}
                    \eSevenF{1}{0}{0}{0}{0}{2}{0}
                    \eSevenG{4}{7}{10}{8}{6}{4}{5}
                \end{tikzpicture}$};
                \node[fill=purple!15] (6) at (0+\x,0+\y) {\scriptsize$\EsevenRgens{6}$};
                \node[fill=purple!15] (7) at (0+\x,4+\y) {\scriptsize$\EsevenRgens{7}$};
                \node[fill=orange!15] (6d) at (0+\x,-5+\y) {\scriptsize$\begin{tikzpicture}
                    \eSevenF{1}{0}{0}{0}{1}{0}{0}
                    \eSevenG{4}{7}{10}{8}{6}{3}{5}
                \end{tikzpicture}$};
                \draw[<-] (6d)--(6);
                \draw[<-] (6)--(7);
                \draw[<-] (6)--(9);
                \draw[<-] (9d)--(9);
                \draw[<-] (9d)--(6d);

                \def\xx{25};
                \def\yy{14};
                \node[fill=purple!15] (8) at (0+\xx,0+\yy) {\scriptsize$\EsevenRgens{8}$};
                \node[fill=orange!15] (8d) at (0+\xx,-5+\yy) {\scriptsize$\begin{tikzpicture}
                    \eSevenF{0}{0}{1}{0}{1}{0}{0}
                    \eSevenG{6}{12}{18}{14}{10}{5}{9}
                \end{tikzpicture}$};
                \draw[<-] (8d)--(8);
                
                \def\xxx{25};
                \def\yyy{3};
                \node[fill=purple!15] (10) at (0+\xxx,0+\yyy) {\scriptsize$\EsevenRgens{10}$};
                \node[fill=orange!15] (10d) at (0+\xxx,-5+\yyy) {\scriptsize$\begin{tikzpicture}
                    \eSevenF{0}{1}{0}{0}{1}{0}{0}
                    \eSevenG{5}{10}{14}{11}{8}{4}{7}
                \end{tikzpicture}$};
                \draw[<-] (10d)--(10);
            \end{tikzpicture}}
    \caption{Higgs branch RG flows (indicated by arrows) between $E_7$ conformal matter theories labelled by their Dynkin quiver phase. Atoms are highlighted in green, hybrids in purple, and molecules in orange. We suppress further Higgsings of molecules.}
    \label{fig:AtomE7Higgsings}
\end{table}

\end{landscape}
\begin{landscape}

\begin{table}[H]
    \makebox[\linewidth][c]{
    \resizebox{11cm}{!}{
    \begin{tabular}{|c|c|c|c|c|}
        \hline
        Coweight & Dynkin Quiver & Type & $I_{\lambda}$ & Orbit\\
        \hline \hline
        \scriptsize$\eEightWeight{0}{0}{0}{0}{0}{0}{1}{0}$ & \scriptsize\EeightRgens{1} & \cellcolor{green!15}Atom & $(z)$ & $A_1$\\
        \hline
        \scriptsize$\eEightWeight{1}{0}{0}{0}{0}{0}{0}{0}$ & \scriptsize\EeightRgens{2} & \cellcolor{green!15}Atom & $(y)$ & $2A_1$\\
        \hline
        \scriptsize$\eEightWeight{0}{0}{0}{0}{0}{1}{0}{0}$ & \scriptsize\EeightRgens{3} & \cellcolor{green!15}Atom & $(x,z^2)$ & $A_2$ \\
        \hline
        \scriptsize$\eEightWeight{0}{0}{0}{0}{0}{0}{0}{1}$ & \scriptsize\EeightRgens{4} & \cellcolor{green!15}Atom & $(x)$ & $A_2+A_1$\\
        \hline
    \end{tabular}
    }
    \quad
       \resizebox{11cm}{!}{
    \begin{tabular}{|c|c|c|c|}
        \hline
        Coweight & Dynkin Quiver & Type & $I_{\lambda}$ \\
        \hline \hline
        \scriptsize$\eEightWeight{0}{0}{0}{0}{1}{0}{0}{0}$ & \scriptsize\EeightRgens{5} & \cellcolor{purple!15}Hybrid & $(z^3,y^2)$\\
        \hline
        \scriptsize$\eEightWeight{0}{1}{0}{0}{0}{0}{0}{0}$ & \scriptsize\EeightRgens{6} & \cellcolor{purple!15}Hybrid & $(zx, y^2)$\\
        \hline
        \scriptsize$\eEightWeight{0}{0}{0}{1}{0}{0}{0}{0}$ & \scriptsize\EeightRgens{7} & \cellcolor{purple!15}Hybrid & $(z^4,xy)$\\
        \hline
        \scriptsize$\eEightWeight{0}{0}{1}{0}{0}{0}{0}{0}$ & \scriptsize\EeightRgens{8} & \cellcolor{purple!15}Hybrid & $(z^5,y^3)$\\
        \hline
    \end{tabular}
    }
    }
    \caption{Indecomposable balanced $E_8$ Dynkin Quivers constructed from coweights on the coroot lattice of $E_8$. Both atoms and hybrids exist for $E_8$. $I_{\lambda}$ defines the CY3 employed to engineer the corresponding 5d SCFTs, according to \eqref{red ideal algorithm}.}
    \label{tab:DynkQuivE8}
\end{table}

\end{landscape}
\begin{landscape}

\begin{table}
    \centering
    \scalebox{0.7}{\begin{tikzpicture}
                \node[fill=green!15] (1) at (0,0) {\scriptsize$\EeightRgens{1}$};
                \node[fill=green!15] (2) at (0,4) {\scriptsize$\EeightRgens{2}$};
                \node[fill=green!15] (3) at (0,8) {\scriptsize$\EeightRgens{3}$};
                \node[fill=green!15] (4) at (0,12) {\scriptsize$\EeightRgens{4}$};
                \draw[<-] (1)--(2);
                \draw[<-] (2)--(3);
                \draw[<-] (3)--(4);

                \def\x{10};
                \def\y{7};
                \node[fill=purple!15] (5) at (0+\x,0+\y) {\scriptsize$\EeightRgens{5}$};
                \node[fill=purple!15] (6) at (0+\x,4+\y) {\scriptsize$\EeightRgens{6}$};
                \node[fill=orange!15] (5d) at (0+\x,-5+\y) {\scriptsize$\begin{tikzpicture}
                    \eEightF{1}{0}{0}{0}{0}{0}{1}{0}
                    \eEightG{6}{11}{16}{13}{10}{7}{4}{8}
                \end{tikzpicture}$};
                \draw[<-] (5d)--(5);
                \draw[<-] (5)--(6);

                \def\xx{20};
                \def\yy{14};
                \node[fill=purple!15] (7) at (0+\xx,0+\yy) {\scriptsize$\EeightRgens{7}$};
                \node[fill=orange!15] (7d) at (0+\xx,-5+\yy) {\scriptsize$\begin{tikzpicture}
                    \eEightF{1}{0}{0}{0}{0}{1}{0}{0}
                    \eEightG{8}{15}{22}{18}{14}{10}{5}{11}
                \end{tikzpicture}$};
                \draw[<-] (7d)--(7);

                \def\xxx{22};
                \def\yyy{4};
                \node[fill=purple!15] (8) at (0+\xxx,0+\yyy) {\scriptsize$\EeightRgens{8}$};
                \node[fill=orange!15] (8d) at (0+\xxx,-5+\yyy) {\scriptsize$\begin{tikzpicture}
                    \eEightF{1}{0}{0}{0}{1}{0}{0}{0}
                    \eEightG{10}{19}{28}{23}{18}{12}{6}{14}
                \end{tikzpicture}$};
                \draw[<-] (8d)--(8);
            \end{tikzpicture}}
    \caption{Higgs branch RG flows (indicated by arrows) between $E_8$ conformal matter theories labelled by their Dynkin quiver phase. Atoms are highlighted in green, hybrids in purple, and molecules in orange. We suppress further Higgsings of molecules.}
    \label{fig:AtomE8Higgsings}
\end{table}
\end{landscape}

\section{Affine Grassmannian and Nilpotent Orbits for Small Coweights}\label{app:acharhend}
\begin{figure}[H]
    \centering
    \scalebox{.9}{
    \begin{tikzpicture}
        \node (1) at (0,-4) {\scriptsize$\begin{bmatrix}
             &&&0\\
            0&0&0&\\
            &&&0
        \end{bmatrix}$};
        \node (2) at (0,-1) {\scriptsize$\begin{bmatrix}
             &&&0\\
            0&1&0&\\
            &&&0
        \end{bmatrix}$};
        \node (3a) at (-2,1) {\scriptsize$\begin{bmatrix}
             &&&1\\
            0&0&0&\\
            &&&1
        \end{bmatrix}$};
        \node (3b) at (2,3) {\scriptsize$\begin{bmatrix}
             &&&0\\
            2&0&0&\\
            &&&0
        \end{bmatrix}$};
        \node (4) at (0,5) {\scriptsize$\begin{bmatrix}
             &&&0\\
            1&0&1&\\
            &&&0
        \end{bmatrix}$};
        \node (5a) at (-2,8) {\scriptsize$\begin{bmatrix}
             &&&0\\
            1&0&0&\\
            &&&2
        \end{bmatrix}$};
        \node (5b) at (2,8) {\scriptsize$\begin{bmatrix}
             &&&2\\
            1&0&0&\\
            &&&0
        \end{bmatrix}$};

        \draw  (1) -- node [left]  {\scriptsize$d_5$} (2);
        \draw  (2) -- node [left]  {\scriptsize$a_3$} (3a);
        \draw  (2) -- node [left, yshift=6pt]  {\scriptsize$A_2$} (3b);
        \draw  (3a) -- node [left]  {\scriptsize$a_3$} (4);
        \draw  (3b) -- node [left]  {\scriptsize$d_4$} (4);
        \draw  (4) -- node [left]  {\scriptsize$A_2$} (5a);
        \draw  (4) -- node [left]  {\scriptsize$A_2$} (5b);

        \def\x{7}
        \node[circle,draw] (n0) at (\x,-4) {$0$};
        \node[circle,draw]  (n1) at (\x,-1) {$[2^2,1^6]$};
        \node[circle,draw]  (n2) at (\x-1,1) {$[2^4,1^2]$};
        \node[circle,draw] (n3) at (\x+2,3) {$[3,1^7]$};
        \node[circle,draw] (n4) at (\x,5) {$[3^2,1^4]$};
        \node[circle,draw] (n5) at (\x,8) {$[3^2,2^2]$};

        \draw  (n0) -- node [left]  {\scriptsize$d_5$} (n1);
        \draw  (n1) -- node [left]  {\scriptsize$a_3$} (n2);
        \draw  (n1) -- node [left]  {\scriptsize$A_2$} (n3);
        \draw  (n2) -- node [left, yshift=6pt]  {\scriptsize$a_3$} (n4);
        \draw  (n3) -- node [left]  {\scriptsize$d_4$} (n4);
        \draw  (n4) -- node [left]  {\scriptsize$A_2$} (n5);

        \draw[dashed,->] (1)--(n0);
        \draw[dashed,->] (2)--(n1);
        \draw[dashed,->] (3a)--(n2);
        \draw[dashed,->] (3b)--(n3);
        \draw[dashed,->] (4)--(n4);
        \draw[dashed,->,blue] (5a) .. controls (2,7) .. (n5);
        \draw[dashed,->,blue] (5b)--(n5);
    \end{tikzpicture}}
    \caption{Left: Hasse diagram of symplectic leaves corresponding to small coweights in the affine Grassmannian of $D_5$. Right: Hasse diagram of related nilpotent orbits in $D_5$. Black dashed arrows denote isomorphisms. Blue dashed arrows denote normalisations.}
    \label{fig:AHforD5}
\end{figure}

\begin{figure}
    \centering
    \begin{tikzpicture}
        \node (1) at (0,-4) {\scriptsize$\begin{bmatrix}
             &&&&0\\
            0&0&0&0&\\
            &&&&0
        \end{bmatrix}$};
        \node (2) at (0,-1) {\scriptsize$\begin{bmatrix}
           &&&&0\\
            0&1&0&0&\\
            &&&&0
        \end{bmatrix}$};
         \node (3) at (-4,2) {\scriptsize$\begin{bmatrix}
           &&&&0\\
            2&0&0&0&\\
            &&&&0
        \end{bmatrix}$};
        \node (4) at (0,4) {\scriptsize$\begin{bmatrix}
           &&&&0\\
            0&0&0&1&\\
            &&&&0
        \end{bmatrix}$};
        \node (5) at (-4,6) {\scriptsize$\begin{bmatrix}
             &&&&0\\
            1&0&1&0&\\
            &&&&0
        \end{bmatrix}$};
        \node (6a) at (0,8) {\scriptsize$\begin{bmatrix}
              &&&&0\\
            0&0&0&0&\\
            &&&&2
        \end{bmatrix}$};
         \node (6b) at (3,8) {\scriptsize$\begin{bmatrix}
              &&&&2\\
            0&0&0&0&\\
            &&&&0
        \end{bmatrix}$};
        \node (7) at (0,11) {\scriptsize$\begin{bmatrix}
              &&&&1\\
            1&0&0&0&\\
            &&&&1
        \end{bmatrix}$};

        \draw  (1) -- node [left]  {\scriptsize$d_6$} (2);
        \draw  (2) -- node [left]  {\scriptsize$A_2$} (3);
        \draw  (2) -- node [left, yshift=6pt]  {\scriptsize$d_4$} (4);
        \draw  (3) -- node [left]  {\scriptsize$d_5$} (5);
        \draw  (4) -- node [left]  {\scriptsize$a_3$} (5);
        \draw  (4) -- node [left, yshift=6pt]  {\scriptsize$A_2$} (6a);
        \draw  (4) -- node [left, yshift=6pt]  {\scriptsize$A_2$} (6b);
        \draw  (5) -- node [left]  {\scriptsize$a_3$} (7);
        \draw  (6a) -- node [left]  {\scriptsize$a_5$} (7);
        \draw  (6b) -- node [left]  {\scriptsize$a_5$} (7);

        \def\x{7}
        \node[circle,draw] (n1) at (\x,-4) {$0$};
        \node[circle,draw]  (n2) at (\x,-1) {$[2^2,1^6]$};
        \node[circle,draw]  (n3) at (\x-1,2) {$[2^4,1^2]$};
        \node[circle,draw] (n4) at (\x+2,4) {$[3,1^7]$};
        \node[circle,draw] (n5) at (\x-1,6) {$[3^2,1^4]$};
        \node[circle,draw] (n6) at (\x+2,8) {$[3^2,2^2]$};
        \node[circle,draw] (n7) at (\x,11) {$[3^2,2^2]$};

        \draw  (n1) -- node [left]  {\scriptsize$d_6$} (n2);
        \draw  (n2) -- node [left]  {\scriptsize$A_2$} (n3);
        \draw  (n2) -- node [left]  {\scriptsize$d_4$} (n4);
        \draw  (n3) -- node [left, yshift=6pt]  {\scriptsize$d_5$} (n5);
        \draw  (n4) -- node [left]  {\scriptsize$a_3$} (n5);
        \draw  (n4) -- node [left]  {\scriptsize$A_2$} (n6);
        \draw  (n5) -- node [left]  {\scriptsize$a_3$} (n7);
        \draw  (n6) -- node [left]  {\scriptsize$a_5$} (n7);

        \draw[dashed,->] (1)--(n1);
        \draw[dashed,->] (2)--(n2);
        \draw[dashed,->] (3)--(n3);
        \draw[dashed,->] (4)--(n4);
        \draw[dashed,->] (5)--(n5);
        \draw[dashed,->,blue] (6a) .. controls (3,7) .. (n6);
        \draw[dashed,->,blue] (6b)--(n6);
        \draw[dashed,->] (7)--(n7);
    \end{tikzpicture}
    \caption{Left: Hasse diagram of symplectic leaves corresponding to small coweights in the affine Grassmannian of $D_6$. Right: Hasse diagram of related nilpotent orbits in $D_6$. Black dashed arrows denote isomorphisms. Blue dashed arrows denote normalisations. }
    \label{fig:AHforD6}
\end{figure}

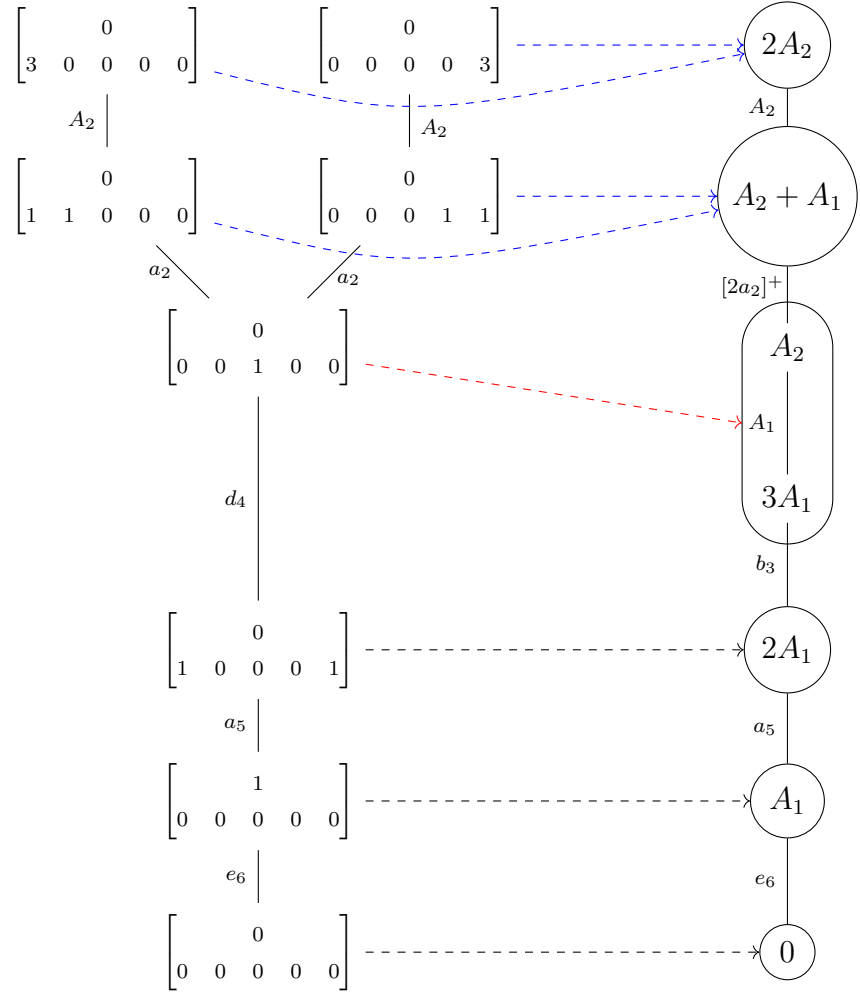
\begin{figure}
    \centering
    \begin{tikzpicture}
        \node (0) at (0,0) {\scriptsize$\begin{bmatrix}
            &&0&&\\
            0&0&0&0&0
        \end{bmatrix}$};
        \node (1) at (0,2) {\scriptsize$\begin{bmatrix}
            &&1&&\\
            0&0&0&0&0
        \end{bmatrix}$};
        \node (2) at (0,4) {\scriptsize$\begin{bmatrix}
            &&0&&\\
            1&0&0&0&1
        \end{bmatrix}$};
        \node (3) at (0,8) {\scriptsize$\begin{bmatrix}
            &&0&&\\
            0&0&1&0&0
        \end{bmatrix}$};
        \node (4a) at (-2,10) {\scriptsize$\begin{bmatrix}
            &&0&&\\
            1&1&0&0&0
        \end{bmatrix}$};
        \node (5a) at (-2,12) {\scriptsize$\begin{bmatrix}
            &&0&&\\
            3&0&0&0&0
        \end{bmatrix}$};
        \node (4b) at (2,10) {\scriptsize$\begin{bmatrix}
            &&0&&\\
            0&0&0&1&1
        \end{bmatrix}$};
        \node (5b) at (2,12) {\scriptsize$\begin{bmatrix}
            &&0&&\\
            0&0&0&0&3
        \end{bmatrix}$};

        \draw  (0) -- node [left]  {\scriptsize$e_6$} (1);
        \draw  (1) -- node [left]  {\scriptsize$a_5$} (2);
        \draw  (2) -- node [left]  {\scriptsize$d_4$} (3);
        \draw  (3) -- node [left]  {\scriptsize$a_2$} (4a);
        \draw  (4a) -- node [left]  {\scriptsize$A_2$} (5a);
        \draw  (3) -- node [right,xshift=-0.1cm,yshift=-0.1cm]  {\scriptsize$a_2$} (4b);
        \draw  (4b) -- node [right,yshift=-0.1cm]  {\scriptsize$A_2$} (5b);

        \def\x{7}
        \node[circle,draw] (n0) at (\x,0) {$0$};
        \node[circle,draw]  (n1) at (\x,2) {$A_1$};
        \node[circle,draw]  (n2) at (\x,4) {$2A_1$};
        \node (n31) at (\x,6) {$3A_1$};
        \node (n32) at (\x,8) {$A_2$};
        \draw \convexpath{n31,n32}{0.6cm};
        \node[circle,draw] (n4) at (\x,10) {$A_2+A_1$};
        \node[circle,draw] (n5) at (\x,12) {$2A_2$};

        \draw  (n0) -- node [left]  {\scriptsize$e_6$} (n1);
        \draw  (n1) -- node [left]  {\scriptsize$a_5$} (n2);
        \draw  (n2) -- node [left]  {\scriptsize$b_3$} (n31);
        \draw  (n31) -- node [left]  {\scriptsize$A_1$} (n32);
        \draw  (n32) -- node [left,xshift=0.1cm,yshift=0.1cm]  {\scriptsize$[2a_2]^+$} (n4);
        \draw  (n4) -- node [left]  {\scriptsize$A_2$} (n5);

        \draw[dashed,->] (0)--(n0);
        \draw[dashed,->] (1)--(n1);
        \draw[dashed,->] (2)--(n2);
        \draw[dashed,->,red] (3)--(6.4,7);
        \draw[dashed,->,blue] (4b)--(n4);
        \draw[dashed,->,blue] (4a) .. controls (2,9) .. (n4);
        \draw[dashed,->,blue] (5b)--(n5);
        \draw[dashed,->,blue] (5a) .. controls (2,11) .. (n5);
    \end{tikzpicture}
    \caption{Left: Hasse diagram of symplectic leaves corresponding to small coweights in the affine Grassmannian of $E_6$. Right: Hasse diagram of related nilpotent orbits in $E_6$. Black dashed arrows denote isomorphisms. Red dashed arrows denote $\mathbb{Z}_2$ covers. Blue dashed arrows denote normalisations. }
    \label{fig:AHforE6}
\end{figure}

\begin{figure}
    \centering
    \begin{tikzpicture}
        \node (0) at (0,0) {\scriptsize$\begin{bmatrix}
            &&0&&&\\
            0&0&0&0&0&0
        \end{bmatrix}$};
        \node (1) at (0,2) {\scriptsize$\begin{bmatrix}
            &&0&&&\\
            1&0&0&0&0&0
        \end{bmatrix}$};
        \node (2) at (0,4) {\scriptsize$\begin{bmatrix}
            &&0&&&\\
            0&0&0&0&1&0
        \end{bmatrix}$};
        \node (3) at (-2,6) {\scriptsize$\begin{bmatrix}
            &&0&&&\\
            0&0&0&0&0&2
        \end{bmatrix}$};
        \node (4) at (2,8) {\scriptsize$\begin{bmatrix}
            &&0&&&\\
            0&1&0&0&0&0
        \end{bmatrix}$};
        \node (5) at (0,10) {\scriptsize$\begin{bmatrix}
            &&1&&&\\
            0&0&0&0&0&1
        \end{bmatrix}$};

        \draw  (0) -- node [left]  {\scriptsize$e_7$} (1);
        \draw  (1) -- node [left]  {\scriptsize$d_6$} (2);
        \draw  (2) -- node [left]  {\scriptsize$A_1$} (3);
        \draw  (3) -- node [left]  {\scriptsize$e_6$} (5);
        \draw  (2) -- node [right,xshift=0.25cm,yshift=0.5cm]  {\scriptsize$d_5$} (4);
        \draw  (4) -- node [right]  {\scriptsize$a_5$} (5);

        \def\x{9}
        \node[circle,draw] (n0) at (\x,0) {$0$};
        \node[circle,draw]  (n1) at (\x,2) {$A_1$};
        \node[circle,draw]  (n2) at (\x,4) {$2A_1$};
        \node[circle,draw]  (n3) at (\x-2,6) {$(3A_1)''$};
        \node (n41) at (\x+2,6) {$(3A_1)'$};
        \node (n42) at (\x+4,8) {$A_2$};
        \draw \convexpath{n41,n42}{0.8cm};
        \node (n51) at (\x,8) {$4A_1$};
        \node (n52) at (\x+2,10) {$A_2+A_1$};
        \draw \convexpath{n51,n52}{0.8cm};

        \draw  (n0) -- node [left]  {\scriptsize$e_7$} (n1);
        \draw  (n1) -- node [left]  {\scriptsize$d_6$} (n2);
        \draw  (n2) -- node [left]  {\scriptsize$A_1$} (n3);
        \draw  (n3) -- node [left,xshift=0.1cm,yshift=0.1cm]  {\scriptsize$f_4$} (n51);
        \draw  (n51) -- node [left]  {\scriptsize$c_3$} (n52);
        \draw  (n52) -- node [right]  {\scriptsize$a_5^+$} (n42);
        \draw  (n42) -- node [right]  {\scriptsize$A_1$} (n41);
        \draw  (n41) -- node [right]  {\scriptsize$b_4$} (n2);
        \draw  (n41) -- node [left,xshift=0.1cm,yshift=-0.1cm]  {\scriptsize$c_3$} (n51);

        \draw[dashed,->] (0)--(n0);
        \draw[dashed,->] (1)--(n1);
        \draw[dashed,->] (2)--(n2);
        \draw[dashed,->] (3)--(n3);
        \draw[dashed,->,red] (4) .. controls (\x+2.5,6) and (\x+1,8) .. (\x+2.4,7.6);
        \draw[dashed,->,red] (5)--(\x+0.4,9.6);
    \end{tikzpicture}
    \caption{Left: Hasse diagram of symplectic leaves corresponding to small coweights in the affine Grassmannian of $E_7$. Right: Hasse diagram of related nilpotent orbits in $E_7$. Black dashed arrows denote isomorphisms. Red dashed arrows denote $\mathbb{Z}_2$ covers. }
    \label{fig:AHforE7}
\end{figure}
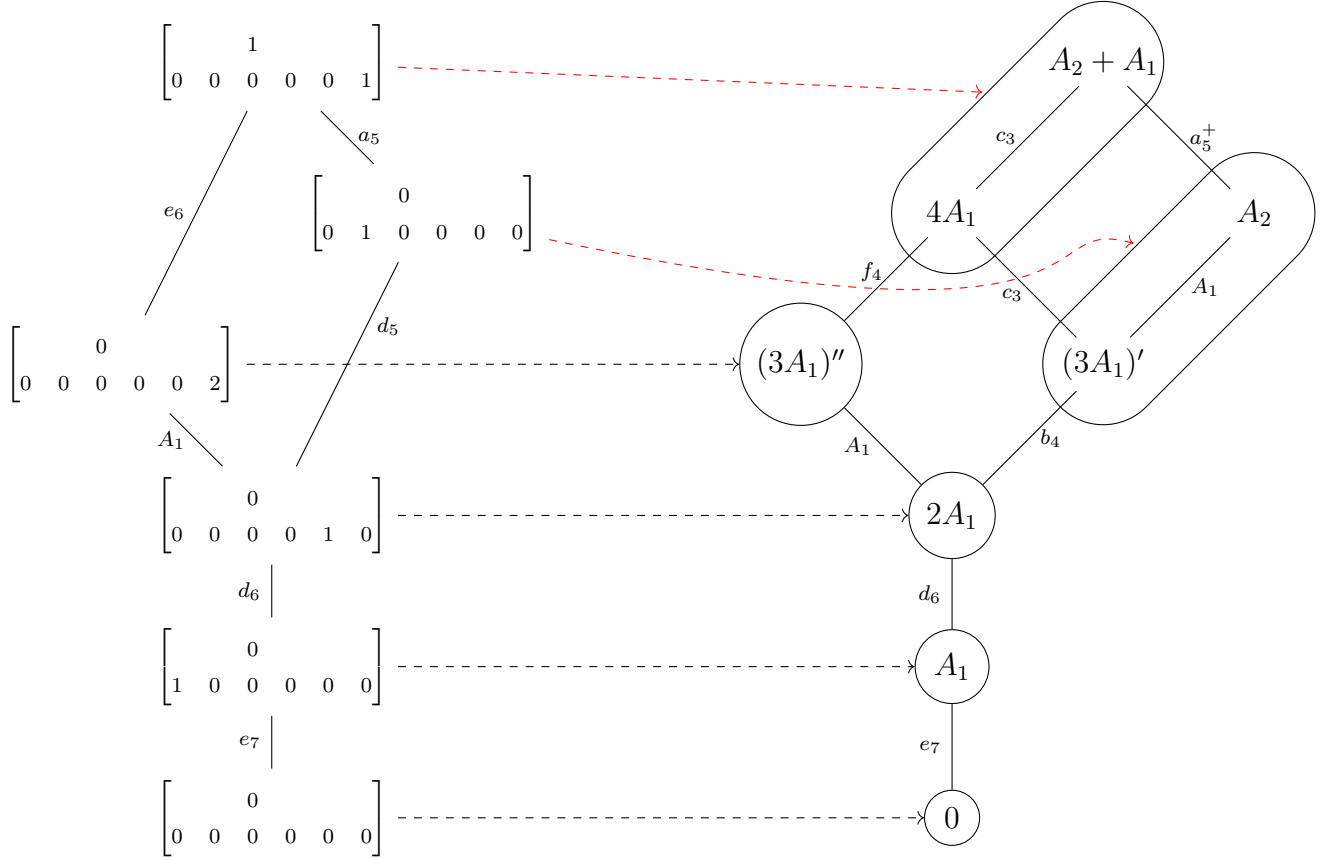

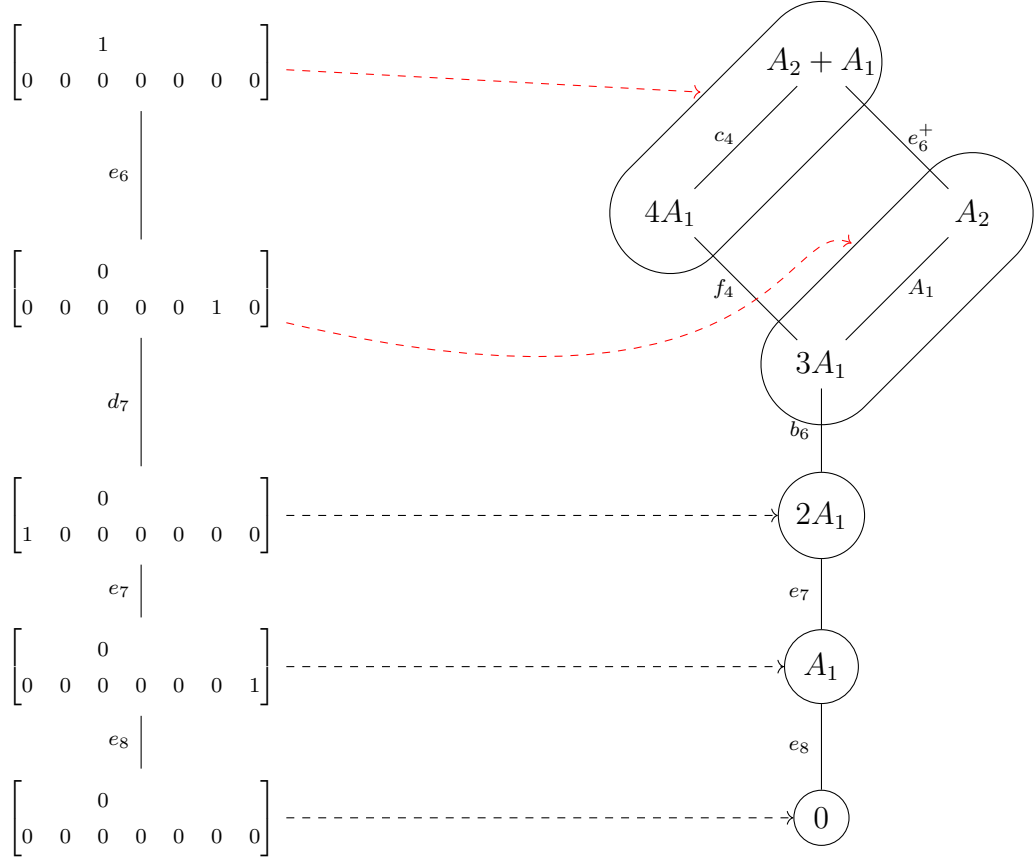
\begin{figure}
    \centering
    \begin{tikzpicture}
        \node (0) at (0,0) {\scriptsize$\begin{bmatrix}
            &&0&&&&\\
            0&0&0&0&0&0&0
        \end{bmatrix}$};
        \node (1) at (0,2) {\scriptsize$\begin{bmatrix}
            &&0&&&&\\
            0&0&0&0&0&0&1
        \end{bmatrix}$};
        \node (2) at (0,4) {\scriptsize$\begin{bmatrix}
            &&0&&&&\\
            1&0&0&0&0&0&0
        \end{bmatrix}$};
        \node (3) at (0,7) {\scriptsize$\begin{bmatrix}
            &&0&&&&\\
            0&0&0&0&0&1&0
        \end{bmatrix}$};
        \node (4) at (0,10) {\scriptsize$\begin{bmatrix}
            &&1&&&&\\
            0&0&0&0&0&0&0
        \end{bmatrix}$};

        \draw  (0) -- node [left]  {\scriptsize$e_8$} (1);
        \draw  (1) -- node [left]  {\scriptsize$e_7$} (2);
        \draw  (2) -- node [left]  {\scriptsize$d_7$} (3);
        \draw  (3) -- node [left]  {\scriptsize$e_6$} (4);

        \def\x{9}
        \node[circle,draw] (n0) at (\x,0) {$0$};
        \node[circle,draw]  (n1) at (\x,2) {$A_1$};
        \node[circle,draw]  (n2) at (\x,4) {$2A_1$};
        \node (n31) at (\x,6) {$3A_1$};
        \node (n32) at (\x+2,8) {$A_2$};
        \draw \convexpath{n31,n32}{0.8cm};
        \node (n41) at (\x-2,8) {$4A_1$};
        \node (n42) at (\x,10) {$A_2+A_1$};
        \draw \convexpath{n41,n42}{0.8cm};

        \draw  (n0) -- node [left]  {\scriptsize$e_8$} (n1);
        \draw  (n1) -- node [left]  {\scriptsize$e_7$} (n2);
        \draw  (n2) -- node [left]  {\scriptsize$b_6$} (n31);
        \draw  (n31) -- node [right]  {\scriptsize$A_1$} (n32);
        \draw  (n32) -- node [right]  {\scriptsize$e_6^+$} (n42);
        \draw  (n42) -- node [left]  {\scriptsize$c_4$} (n41);
        \draw  (n41) -- node [left]  {\scriptsize$f_4$} (n31);

        \draw[dashed,->] (0)--(n0);
        \draw[dashed,->] (1)--(n1);
        \draw[dashed,->] (2)--(n2);
        \draw[dashed,->,red] (3) .. controls (\x-0.5,5) and (\x-0.5,8) .. (\x+0.4,7.6);
        \draw[dashed,->,red] (4)--(\x+0.4-2,9.6);
    \end{tikzpicture}
    \caption{Left: Hasse diagram of symplectic leaves corresponding to small coweights in the affine Grassmannian of $E_8$. Right: Hasse diagram of related nilpotent orbits in $E_8$. Black dashed arrows denote isomorphisms. Red dashed arrows denote $\mathbb{Z}_2$ covers. }
    \label{fig:AHforE8}
\end{figure}

\clearpage

\providecommand{\href}[2]{#2}

\end{document}